\documentclass[a4paper,11pt]{article}
\usepackage{jcappub} 

\arxivnumber{2511.21676} 

\title{Axion Constraints from White Dwarf Cooling in 47 Tucanae}

\author[a,1]{L. Fleury,\note{Corresponding author.}}
\author[a]{A. Obertas,}
\author[a,2]{H. Richer,\note{Deceased.}}
\author[a,1]{and J. Heyl}
\affiliation[a]{Department of Physics and Astronomy, University of British Columbia,\\ Vancouver, BC V6T 1Z1, Canada}

\emailAdd{lfleury@phas.ubc.ca}
\emailAdd{heyl@phas.ubc.ca}

\abstract{%
We analyse the cooling of white dwarfs in the globular cluster 47 Tucanae to look for evidence of axion emission affecting the rate of white dwarf cooling.
If axions exist and couple to electrons, then axions could be produced at an appreciable rate in the electron-degenerate core of a white dwarf through axion bremsstrahlung from electrons.
The emission of these axions would provide an additional cooling mechanism for white dwarfs that would affect the cooling rate, and hints of axions have been suggested based on observations of anomalous cooling reported for white dwarfs in the Galactic disc and halo.
We performed stellar evolution simulations of white dwarf cooling that accounted for the additional energy loss due to axion bremsstrahlung from electrons, producing a suite of white dwarf cooling models for different values of the axion-electron coupling constant, as well as the white dwarf mass and envelope thickness.
These cooling models are compared to observations of white dwarfs in 47 Tucanae from the Hubble Space Telescope through an unbinned likelihood analysis.
The optimal model found by this analysis corresponds to the case of no axion emission with a thick white dwarf envelope, and we find a new bound on the axion-electron coupling of $g_{aee} \leq 0.81 \times 10^{-13}$ at 95\% confidence level, conditional on the adopted white dwarf cooling models and priors.
This bound improves upon the previous white dwarf cooling bound for this coupling and excludes the range of values favoured by the axion hints from the anomalous cooling of Galactic white dwarfs.%
}


\usepackage{main}

\let\citep=\cite
\renewcommand{\citet}{ref.~\cite}
\renewcommand{\Citet}{Ref.~\cite}

\begin{document}
\maketitle
\flushbottom

\section{Introduction} \label{sec:47tuc_axions_intro}

White dwarfs are a popular target of indirect searches for axions and axion-like particles.
For axion models such as the benchmark DFSZ model \citep{Dine:1981,Zhitnitsky:1980} that include a coupling between the axion and electrons, axions can be produced in the very dense, electron-degenerate interior of white dwarfs through axion bremsstrahlung from electrons.
Axions can also in principle be produced in the interiors of white dwarfs through the axion Primakoff effect, but this is a sub-dominant process compared to axion bremsstrahlung for the density and temperature scales typical of white dwarfs.
Furthermore, the typical densities and temperatures of young white dwarfs would lead to high rates of emission for axions produced via axion bremsstrahlung relative to the emission of neutrinos and photons, which are the standard channels through which white dwarfs lose energy over time.

For example, X-ray observations of a hot, magnetic white dwarf have been used to search for the spectroscopic signal of axions that were produced in the core of the white dwarf through axion bremsstrahlung (from electron-ion scattering) and then converted to photons in the strong magnetic field surrounding the white dwarf \citep{2019PhRvL.123f1104D,2022PhRvL.128g1102D}.
As that scenario relies on the axion-photon interaction (to produce the observable photon signal) in addition to an axion-electron interaction (to produce the axions), that work constrained the product of the axion-electron coupling $g_{aee}$ and the axion-photon coupling $g_{a\gamma\gamma}$, rather than just $g_{aee}$ alone. 
\Citet{2022PhRvL.128g1102D} found a bound of $|g_{aee} g_{a\gamma\gamma}| \lesssim 1.3 \times 10^{-25}~\mathrm{GeV}^{-1}$ at 95\% confidence for low axion masses $m_a \ll 10^{-5}~\mathrm{eV}$ based on a search targeting the white dwarf RE J0317-853. A similar analysis was later performed by \citet{2025PhRvD.111j3002N} for searches targeting the two additional white dwarfs PG 0945+246 and WD 1859+148, with the joint limit from all three white dwarfs found to be $|g_{aee} g_{a\gamma\gamma}| \lesssim 1.01 \times 10^{-25}~\mathrm{GeV}^{-1}$ at 95\% confidence for $m_a \lesssim 10^{-6}~\mathrm{eV}$.

More typically, axion searches targeting white dwarfs focus on how the additional energy loss due to axion emission modifies the cooling behaviour of white dwarfs, and in particular the observable consequences of the altered cooling rate on the white dwarf luminosity function of a population of white dwarfs.
Due to the weakness of the interaction between axions and Standard Model particles, axions that are produced in the interior of a white dwarf typically leave the white dwarf without further interaction.
The emission of axions provides an extra mechanism of energy loss for a white dwarf in addition to the standard energy sinks due to the radiation of photons from the white dwarf surface and, for young white dwarfs, the emission of neutrinos from the white dwarf interior.
This causes the white dwarf to cool at a faster rate than predicted by standard cooling processes alone.

Evidence for an additional white dwarf cooling mechanism that is compatible with axion emission has been reported for the empirical white dwarf luminosity functions \citep{Munn:2017,Kilic:2017,Rowell:2011} of white dwarfs in the Galactic thin and thick discs, as well as the halo, and it has been suggested that this hints at the existence of an axion that couples to electrons, with a mass $m_a$ (and angular parameter $\beta$) value of $m_a \sin^2\beta \sim 4 - 10~\mathrm{meV}$ for a DFSZ axion model \citep{2018MNRAS.478.2569I}.
This corresponds to an axion-electron coupling of $g_{aee} \sim \left(1.1 - 2.8\right) \times 10^{-13}$.
The work of \citet{2018MNRAS.478.2569I} expanded upon the earlier work of \citet{2008ApJ...682L.109I} and \citet{2014JCAP...10..069M}, which likewise presented evidence for similar hints of axions based on the shape of some measurements of Galactic white dwarf luminosity functions.
\Citet{2014JCAP...10..069M} also used Galactic white dwarf luminosity function measurements to place a bound of $m_a \sin^2\beta \lesssim 10~\mathrm{meV}$ on the mass of a DFSZ axion at a $3~\sigma$ confidence level, which corresponds to a $3~\sigma$ bound on the axion-electron coupling of $g_{aee} \lesssim 2.8 \times 10^{-13}$.

Recent reviews summarising both hints of axions and bounds on axion parameters from various observations pertaining to stellar evolution are given by \citet{DiLuzio:2020wdo} and \citet{2022JCAP...02..035D}.
These reviews compare bounds reported at a $2~\sigma$ confidence level (which is equivalent to a 95\% confidence level), where the $2~\sigma$ bound from \citet{2014JCAP...10..069M} is given as $g_{aee} \leq 2.1 \times 10^{-13}$.
The bound on $g_{aee}$ has been improved upon since the work of \citet{2018MNRAS.478.2569I} and \citet{2014JCAP...10..069M} using observations of the luminosity of the tip of the red giant branch in globular clusters \citep{2020PhRvD.102h3007C,2020A&A...644A.166S,2025PhyR.1117....1C}.
Axions could be emitted from the electron-degenerate cores of stars on the red giant branch through the process of axion bremsstrahlung from electrons, which would delay the onset of He burning and thus increase the luminosity and temperature of the tip of the red giant branch.
\Citet{2013PhRvL.111w1301V} used the tip of the red giant branch of the Galactic globular cluster M5 (NGC 5904) to derive a rigorous bound of $g_{aee} < 4.3 \times 10^{-13}$  at $2~\sigma$ based on a detailed analysis of statistical and systematic uncertainties.
This analysis was based on a similar earlier study that constrained the neutrino magnetic dipole moment using the tip of the red giant branch in M5 \citep{2013A&A...558A..12V}.
A $2~\sigma$ bound of $g_{aee} \leq 1.6 \times 10^{-13}$ was later found based on an analysis of the tip of the red giant branch of the globular cluster NGC 4258 \citep{2020PhRvD.102h3007C}, and a similar but slightly more stringent bound of $g_{aee} \leq 1.5 \times 10^{-13}$ at $2~\sigma$ was found from a more extensive analysis of the tip of the red giant branch of 22 globular clusters \citep{2020A&A...644A.166S}.
This bound on $g_{aee}$ from the tip of the red giant branch has been updated in the more recent review by \citet{2025PhyR.1117....1C} to $g_{aee} < 0.95 \times 10^{-13}$ (at 95\% confidence level) using updated distances provided by \gaia\ DR3 for 21 of the 22 globular clusters used by \citet{2020A&A...644A.166S} to derive the previous bound.

Globular clusters are good environments for studying stellar evolution because they provide coeval populations of stars with well-controlled values of parameters like distance, reddening, and birthrate.
Globular clusters contain populations of very old stars that were all initially created from the same protostellar material at the same time in a burst of star formation\footnote{Though many globular clusters contain multiple populations, each of these populations formed in an isolated star formation event.}.
The main difference between stars in a globular cluster population is simply the initial mass each star is born with. 
This primarily affects the length of time the star spends on the main sequence burning H in its core, with more massive stars finishing this stage of evolution more quickly. 
All subsequent stages of evolution for stars in globular clusters are so fast that all but the faintest white dwarfs in the population have approximately the same mass \citep{1990PhR...198....1R}.
The more massive white dwarfs are fainter than those studied in this work.
This is in contrast to stars in the solar neighbourhood (i.e. a population of stars in the Galactic disc), for which star formation is a continually ongoing process and the white dwarfs can have a wide range of different masses.

The globular cluster 47 Tucanae (47 Tuc) in particular is a good environment for studying the cooling of white dwarfs, and parameters important for modelling the cooling of white dwarfs in 47 Tuc, such as the typical white dwarf mass and envelope thickness, were studied extensively in \citet{47tuc_deep_acs}.
If axions exist and couple to electrons with the interaction strength hinted at by Galactic white dwarf luminosity functions, there should be a measurable effect on the cooling of white dwarfs in 47 Tuc as well.
The emission of axions affects the white dwarf cooling rate in a manner similar to the emission of neutrinos at early cooling times, and the effect of neutrino emission on the cooling of young white dwarfs in 47 Tuc was studied in detail by \citet{Goldsbury2016}.
Like in the case of neutrino emission, the emission of axions produced in the dense interior of a white dwarf provides an additional mechanism of energy loss that increases the cooling rate compared to what is expected from photon radiation alone, though the effect of axion emission could persist to later cooling ages than neutrino emission due to the particular temperature and density values at which these emission processes are optimised \citep{Nakagawa:1987,Nakagawa:1988}.

\Citet{2015ApJ...809..141H} studied the effect of exotic emission mechanisms in 47 Tuc using some of the same data studied in this work.
These exotic emission mechanisms included axion emission, along with emission due to a potential non-zero neutrino magnetic dipole moment and the emission of Kaluza-Klein modes into extra dimensions.
\Citet{2015ApJ...809..141H} found a bound of $g_{aee} < 8.4 \times 10^{-14}$ at 95\% confidence.
That work constructed empirical bolometric luminosity functions from the data to compare the theoretical predictions with binned data that had been moved to theory space.
Improved statistical techniques for comparing white dwarf cooling predictions to data, in particular the unbinned likelihood method, have since then been used by \citet{Goldsbury2016} to study neutrino cooling in 47 Tuc white dwarfs. 
In the improved method employed by \citet{Goldsbury2016}, the theoretical cooling models are moved to data space using the photometric error distribution function and the unbinned likelihood is then used to compare the models to the data without the need to bin the data.
However, this improved statistical methodology has not yet been applied to the study of axion emission affecting white dwarf cooling.

In this work, a detailed analysis of the cooling of white dwarfs in 47 Tuc is performed in order to look for indirect evidence of axions through the effect of axion emission on the white dwarf cooling rate.
This work follows a procedure similar to the procedure of \citet{Goldsbury2016}, but applied to axions instead of neutrinos. 
We use the same data and many of the same analysis techniques as described in \citet{Goldsbury2016}, but we use new white dwarf cooling models that account for energy loss due to the emission of axions.
The analysis procedure that we use in this work is also similar to that of \citet{47tuc_deep_acs}, where a separate set of data was analysed to study the cooling of older white dwarfs in 47 Tuc.
The white dwarfs studied in \citet{47tuc_deep_acs} were in particular old enough that axion emission would no longer be having an observable effect on their evolution even if it is affecting the evolution of younger white dwarfs, allowing priors to be established using an independent data set for use in the current work.
We make use of the prior knowledge gained from the work of \citet{47tuc_deep_acs} about the typical mass and envelope thickness of white dwarfs in 47 Tuc in our analysis of the effect of axion emission on the cooling of younger white dwarfs in 47 Tuc.
The use of this prior information enables stronger constraints to be placed on the relevant axion parameters.

\section{Data} \label{sec:47tuc_axions_data}

We use archival data from Hubble Space Telescope (HST) observations of 47 Tuc performed as part of the HST Cycle 20 proposal GO-12971 (PI: H. Richer).
The data were collected over 10 orbits, one of which was rejected due to loss of guide stars.
In each orbit, observations were done simultaneously by the Wide Field Camera 3 (WFC3) using the UVIS channel and by the Advanced Camera for Surveys (ACS) using the Wide Field Channel (WFC).
Across all of the orbits, WFC3 observed the inner field at the centre of the cluster while ACS simultaneously observed the outer field in a ring about the WFC3 field of view.
Details of these observations are given in \citet{Goldsbury2016}, along with a detailed description of artificial stars tests that were performed to determine the photometric errors of the data.

For both the WFC3 and ACS observations, the total field of view (after combining the images from all orbits) has a centre that coincides with the centre of 47 Tuc.
The WFC3 field of view extends to a radial distance of $200''$ from the centre (in projection), with full coverage over all azimuthal angles within $160''$ from the centre. The ACS field of view extends from an inner radius of approximately $150''$ to an outer radius of approximately $435''$.
The WFC3 observations were done at wavelengths in the near-ultraviolet, while the ACS observations were done at optical wavelengths.
The WFC3/UVIS exposures were split between the F225W and F336W filters, and the ACS/WFC exposures were split between the F435W and F555W filters.
Note that in the HST naming convention for these filters, the first letter ``F'' simply denotes a filter (as opposed to a grism, which would start with the letter ``G''), the following three-digit number indicates the nominal effective wavelength of the filter's bandpass in units of nm, and the final letter ``W'' denotes a wide bandpass\footnote{Documentation discussing the naming convention for HST filters is available online at e.g. \url{https://hst-docs.stsci.edu/wfc3ihb/chapter-6-uvis-imaging-with-wfc3/6-5-uvis-spectral-elements}.}.
In this work, we furthermore use the standard convention that the name of a filter is used to denote an apparent magnitude measured using that filter.

The data reduction process for the data used in our work followed the same prescription and procedures as described in \citet{2012AJ....143...11K}, just applied to a different HST data set.
The data reduction procedure and corresponding artificial stars tests as applied to the data from \citet{2012AJ....143...11K} are also described in \citet{47tuc_deep_acs}.
Details about the data reduction procedure as applied to the data set studied in our current work are provided in \citet{2015ApJ...809..141H} (see in particular section 2 therein) and more briefly in \citet{Goldsbury2016}.
In summary, all of the WFC3 observations (from each exposure of each visit across the various orbits) were ultimately combined into a single image for each filter.
In the case of the ACS data, there was very little overlap in the field of view for the different orbits, so the observations were combined to produce a separate image for each orbit for each filter.
A catalogue of stars was then generated for each data set (WFC3 and ACS%
\footnote{For the ACS data, the catalogues for the different orbits were combined into a single catalogue after removing any duplicate stars from the limited range of overlap in the fields of view of the different orbits.}%
) from the images for the relevant filters using an iterative point spread function (PSF) fitting technique that measured the photometric and morphological properties of candidate sources.
In this procedure, the reference PSF for the fitting procedure is determined from bright, isolated sources in the image that are highly likely to be stars, then the brightness profiles of candidate sources in the image are fit to this reference profile to determine if they are stars and measure their positions and magnitudes.
The dominant source of observational error comes from this PSF-fitting technique due to contaminant light from neighbouring stars or other sources masking or distorting the brightness profiles of stars, particularly in overcrowded fields.
Artificial stars tests provide a way to quantify this effect.

Artificial stars tests were performed to characterize the photometric errors of both the WFC3/UVIS and ACS/WFC data sets.
In these tests, artificial sources were added to the images and the resulting images were run through the same data reduction procedure that was used to identify sources in the original images.
The artificial sources were generated from the PSF that sources were matched to in the reduction procedure to detect stars and measure their photometry, with the overall brightness of the profile scaled based on the input magnitude assigned to the artificial source being generated.
Artificial sources were generated for a grid of different input magnitudes along the white dwarf cooling sequence, spanning the range of magnitude values relevant for our work.
The artificial sources were each given a unique identifier and tracked to determine both the fraction of stars recovered, which quantifies the completeness, and the difference between the retrieved (output) value of the magnitude and the assigned (input) magnitude of the source for each filter.
The distribution of these magnitude differences given the input magnitudes and radial distance from the cluster centre is the photometric error distribution function, which is used in our analysis in \cref{sec:47tuc_axions_analysis} to compare theoretical cooling models to the HST observations.
In particular, the photometric error distribution function is needed to construct the probability density function predicted by a model for the observables in data space in the unbinned likelihood analysis.

For a given data set that measures magnitudes $m_1$ and $m_2$ in two filters (which are simply labelled ``1'' and ``2'' for generality), the photometric error distribution function will be denoted $E\left(\Delta m_1, \, \Delta m_2 \, ; \, m_{1,\mathrm{in}}, \, m_{2,\mathrm{in}}, \, R\right)$, where $\Delta m_1$ and $\Delta m_2$ are the magnitude errors in each filter, $m_{1,\mathrm{in}}$ and $m_{2,\mathrm{in}}$ are the corresponding input magnitudes, and $R$ is the radial distance from the cluster centre.
The magnitude error in each filter is the difference between the output and input magnitude in that filter found by the artificial stars tests, i.e.
\begin{equation}
    \Delta m_i = m_{i,\mathrm{out}} - m_{i,\mathrm{in}},
\end{equation}
where the subscripts ``in'' and ``out'' denote the input and output values, respectively, of the magnitude in the filter labelled with the index $i \in \left\lbrace 1, \, 2\right\rbrace$.
For the WFC3/UVIS data set, the two magnitudes are $m_1 = \mathrm{F225W}$ and $m_2 = \mathrm{F336W}$, while for the ACS/WFC data set, the two magnitudes are $m_1 = \mathrm{F435W}$ and $m_2 = \mathrm{F555W}$.

The photometric error distribution function is a probability density function, which we normalize to the number of input stars such that integrating over the magnitude errors gives the completeness $C$ as a function of the input magnitudes (as well as the distance from the cluster centre), i.e.
\begin{equation}
    C\left(m_{1,\mathrm{in}}, \, m_{2,\mathrm{in}}, \, R\right) = \int_{-\infty}^{+\infty} \int_{-\infty}^{+\infty} E\left(\Delta m_1, \, \Delta m_2 \, ; \, m_{1,\mathrm{in}}, \, m_{2,\mathrm{in}}, \, R\right) \, \mathrm{d}\Delta m_1 \, \mathrm{d}\Delta m_2.
\end{equation}
The completeness gives the probability of a source being detected and is equivalent to the fraction of artificial sources found by the photometric reduction process in the artificial stars tests.
The value of the completeness can in principle range from zero to unity, though in general the completeness has a value less than unity because not all sources that are physically present will actually be resolved in the observations and recovered by the photometric reduction process.
Overcrowding near the centre of the cluster reduces the completeness of the data, which makes the photometric error distribution function for the WFC3/UVIS data a function of the radial distance from the cluster centre.
The ACS/WFC observations, on the other hand, were taken far enough from the centre of the cluster that the photometric error distribution function for the ACS/WFC data set is not sensitive to the distance from the cluster centre, and thus the $R$ dependence of the functions $E$ and $C$ can be neglected for the ACS/WFC data set.

\section{Models} \label{sec:47tuc_axions_models}

\subsection{MESA Simulations}

Following the procedure of \citet{47tuc_deep_acs}, we performed white dwarf cooling simulations using the stellar evolution software Modules for Experiments in Stellar Astrophysics (MESA; \citep{mesa1,mesa2,mesa3,mesa4,mesa5}) to produce a suite of white dwarf cooling models over a grid of model parameter values.
The simulations are analogous to those described in section 5 of \citet{47tuc_deep_acs} except they additionally implemented energy loss due to the emission of axions, which introduces axion couplings as additional parameters that the models depend on.
The parameters varied in these simulations were the white dwarf mass $M_\mathrm{WD}$, the thickness of the H envelope, and the axion coupling constants.

The standard MESA treatment of diffusion was used for all of the simulations in this work, as was done for \citet{47tuc_deep_acs}.
As the extreme ends of the grid of white dwarf mass values considered in \citet{47tuc_deep_acs} are strongly excluded by the results of that work (with white dwarf masses above $0.5388~M_\odot$ or below $0.5240~M_\odot$ excluded at a $5~\sigma$ level), we restrict our current work to masses spanning the smaller range $0.5240~M_\odot \leq \mwd \leq 0.5388~M_\odot$.
Within this white dwarf mass range, we performed simulations for the same grid of $\mwd$ values considered in \citet{47tuc_deep_acs}, starting from the same initial models as the simulations described in that work. The precise $\mwd$ values considered were $0.5240$, $0.5314$, and $0.5388$~$M_\odot$.

In our MESA simulations of white dwarf cooling, we implement energy loss due to axion emission via the \verb|run_star_extras| module in the \verb|other_neu| subroutine.
We account for energy loss due to both axion (electron) bremsstrahlung, which depends on the axion-electron coupling $g_{aee}$, and the axion Primakoff effect, which depends on the axion-photon coupling $g_{a\gamma\gamma}$.
We implement axion bremsstrahlung emission using the prescription given by \citet{Nakagawa:1987} and \citet{Nakagawa:1988}.
To account for axion Primakoff emission, we use the fitting formula given by \citet{Friedland:2012}, which has previously been implemented (and for which code is publicly available) in the MESA \verb|test_suite| example \verb|axion_cooling|.
We specify the value of each coupling constant, $g_{aee}$ and $g_{a\gamma\gamma}$, through MESA inlist \verb|x_ctrl| parameters.
Note that axion bremsstrahlung is the dominant axion emission mechanism affecting white dwarf cooling. Though axion Primakoff emission is a sub-dominant effect for white dwarf cooling, it was included in the simulations for completeness and to confirm that this effect is negligible enough to neglect any potential dependence of the cooling models on $g_{a\gamma\gamma}$ in our analysis.

We performed simulations for a set of $g_{aee}$ and $g_{a\gamma\gamma}$ values chosen based on the benchmark DFSZ (type I) model.
For this model, the coupling constants $g_{aee}$ and $g_{a\gamma\gamma}$ are related to the axion mass $m_a$ through the expressions%
\footnote{See in particular Table 2.1 of \citet{1990PhR...198....1R} for a summary of the expressions for various axion coupling constants. Note that $\cos\beta$ has been replaced by $\sin\beta$ in the expression for $g_{aee}$. The choice of which definition of $\beta$ to use is simply a matter of convention.} %
\citep{1990PhR...198....1R}
\begin{alignat}{2}
    &g_{a\gamma\gamma} &&= 1.44 \times 10^{-10} \ \left( \frac{m_a}{1~\textrm{eV}} \right) \ \textrm{GeV}^{-1},
    \label{eq:47tuc_axions_gagg} \\
    &g_{aee} &&= 2.83 \times 10^{-11} \ \sin^2\beta \ \left( \frac{m_a}{1~\textrm{eV}} \right),
    \label{eq:47tuc_axions_gaee}
\end{alignat}
where the expression for $g_{aee}$ also depends on the angular parameter $\beta$ with $\tan\beta \in \left[0.25, 170\right]$.
We performed simulations for a grid of $m_a$ values ranging from $0.0$ to $8.0$~meV in increments of $0.5$~meV for the limiting case that $\sin\beta = 1$.
In this case, the cooling models only depend on a single axion parameter, the axion mass $m_a$.
More generally, in parameter regimes where energy loss due to the axion Primakoff effect is negligible, as is the case for the white dwarf cooling regime that we consider in this work, the cooling models only depend on the coupling constant $g_{aee}$ (i.e. the dependence of the cooling models on $g_{a\gamma\gamma}$ is negligible).
The work presented here in terms of $m_a$ (for a DFSZ axion with $\sin\beta = 1$) can thus equivalently be expressed in terms of $g_{aee}$, with the results in terms of $g_{aee}$ applying more generally.
\Cref{tab:47tuc_axions_ma_gaee_conversion} provides the conversion between $m_a$ and the $g_{aee}$ (and $g_{a\gamma\gamma}$) values used for the simulations.

\begin{table}
    \centering
    \begin{tabular}{c c c}
    \toprule
    $m_a \, / \, \left(1~\mathrm{meV}\right)$ & $g_{aee} \, / \, \left(10^{-13}\right)$ & $g_{a\gamma\gamma} \, / \, \left(10^{-12}~\mathrm{GeV}^{-1}\right)$ \\
    \midrule
    0.0 & 0.00 & 0.00 \\
    0.5 & 0.14 & 0.07 \\
    1.0 & 0.28 & 0.14 \\
    1.5 & 0.42 & 0.22 \\
    2.0 & 0.57 & 0.29 \\
    2.5 & 0.71 & 0.36 \\
    3.0 & 0.85 & 0.43 \\
    3.5 & 0.99 & 0.51 \\
    4.0 & 1.13 & 0.58 \\
    4.5 & 1.28 & 0.65 \\
    5.0 & 1.42 & 0.72 \\
    5.5 & 1.56 & 0.79 \\
    6.0 & 1.70 & 0.87 \\
    6.5 & 1.84 & 0.94 \\
    7.0 & 1.98 & 1.01 \\
    7.5 & 2.12 & 1.08 \\
    8.0 & 2.27 & 1.15 \\
    \bottomrule
    \end{tabular}
    \caption{Conversion between axion mass values and the corresponding values used for the coupling constants in the MESA simulations that produced the white dwarf cooling models. These values apply for a DFSZ model with $\sin\beta = 1$.
    For these values of $g_{a\gamma\gamma}$, the effect of axion-photon interactions on white dwarf cooling is negligible; these values are simply reported for completeness.}
    \label{tab:47tuc_axions_ma_gaee_conversion}
\end{table}

\subsection{Cooling Curves}

\begin{figure}
    \centering
    \begin{subfigure}[t]{0.49\textwidth}
        \centering
        \includegraphics[width=\textwidth]{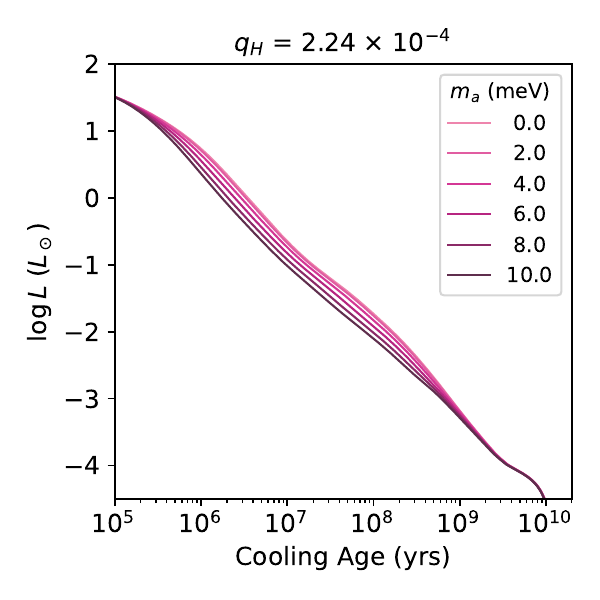}
        \caption{Varying $m_a$ for fixed $q_H$.}
        \label{fig:cooling_curves_maseries}
    \end{subfigure}
    \hfill
    \begin{subfigure}[t]{0.49\textwidth}
        \centering
        \includegraphics[width=\textwidth]{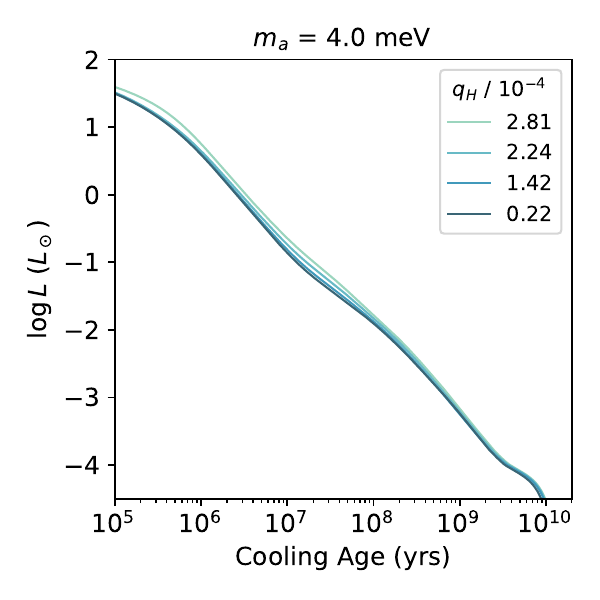}
        \caption{Varying $q_H$ for fixed $m_a$.}
        \label{fig:cooling_curves_qhseries}
    \end{subfigure}
    \caption{Theoretical cooling curves for various axion mass values (left; a) and envelope thickness (right; b) when all other model parameters are fixed.
    In both cases, the white dwarf mass is $M_\mathrm{wd} = 0.5388~M_\odot$.
    For the series of axion mass values shown in the left panel, the envelope thickness always has a fixed value of $q_H = 2.24 \times 10^{-4}$.
    For the series of $q_H$ values shown in the right panel, the axion mass always has a fixed value of $m_a = 4.0~\mathrm{meV}$.}
    \label{fig:cooling_curves_maseries_qhseries}
\end{figure}

\Cref{fig:cooling_curves_maseries_qhseries} shows how the cooling curves vary with axion mass ($m_a$) and H envelope thickness ($q_H$).
The left panel (\cref{fig:cooling_curves_maseries}) shows a series of cooling curves with different $m_a$ values over the range $0 - 10~\mathrm{meV}$ for a fixed envelope thickness of $q_H = 2.24 \times 10^{-4}$, while the right panel (\cref{fig:cooling_curves_qhseries}) shows a series of cooling curves with different $q_H$ values for a fixed axion mass of $m_a = 4.0$. 
For both series of cooling curves, the white dwarf mass is fixed at $M_\mathrm{WD} = 0.5388~M_\odot$.
Note that cooling curves are only shown for a subset of $m_a$ and $q_H$ values in order to facilitate visualisation; the full set of models used in the analysis is more finely spaced in $m_a$ and $q_H$ than what is shown in \cref{fig:cooling_curves_maseries_qhseries}.

For the axion mass range $0 - 10.0~\mathrm{meV}$ considered in our analysis, the emission of axions produced in the core of the white dwarf alters the shape of the cooling curves for cooling ages between $\sim 1 \times 10^{5}~\mathrm{yrs}$ and $3 \times 10^{9}~\mathrm{yrs}$ and over the luminosity range $1.5 \lesssim \log L \lesssim -4$, where $L$ is the luminosity in units of solar luminosity and $\log$ denotes the base-10 logarithm.
This can be seen in \cref{fig:cooling_curves_maseries}, where the cooling curves differ over the luminosity (and cooling age) range where axion emission is an important energy loss mechanism.
Larger values of the axion coupling (corresponding to large values of $m_a$) cause the luminosity of the white dwarf to decrease more quickly with cooling age.
The cooling curves for all of the $m_a$ values re-converge at late cooling times once energy loss due to axion emission becomes negligible compared to energy loss due to photon radiation.

Over the cooling regime where axion emission is an important energy emission mechanism, the morphology of the cooling curve also depends on the H envelope thickness. 
This can be seen in \cref{fig:cooling_curves_qhseries}, where it can be seen that the shape of the cooling curves varies within the relevant luminosity range with varying $q_H$ for a fixed axion mass value.
The cooling curve depends on $q_H$ over this regime because thicker H envelopes result in more residual nuclear burning at the boundary of the envelope, which in turn causes the white dwarf to cool more slowly and remain brighter for a longer period of time early in the cooling process.

\section{Analysis} \label{sec:47tuc_axions_analysis}

Our analysis uses the unbinned likelihood in a procedure very similar to the analysis in \citet{Goldsbury2016}.
Similar methods were also used in \citet{47tuc_deep_acs} and are described in detail therein (see in particular section 7 of \citet{47tuc_deep_acs}).
Unlike in \citet{47tuc_deep_acs}, however, the dependence of the likelihood function on the distance $R$ from the cluster centre cannot be neglected.
The radial dependence is in particular important for analysing the WFC3 data because the WFC3 observations are centred on the cluster centre, where stars belonging to the cluster are most densely concentrated.
Overcrowding in the observations near the cluster centre due to this dense concentration results in reduced completeness at small values of $R$.
The completeness of the WFC3 observations is also reduced to a lesser extent at large values of $R$ due to the geometry of the observations, with less overlap in the fields at larger values of $R$ resulting in fewer fields that an object may be detected in.
Furthermore, the radial density distribution (in projection) of a globular cluster in general, and 47 Tuc in particular, varies most rapidly with $R$ at small values of $R$ where the cluster is most concentrated; the distribution is less sensitive to $R$ in the tails where the density varies more slowly with $R$.
The ACS observations were taken far enough from the centre that radial dependence can be neglected for the ACS data.

For a particular data set, the number density distribution function accounting for radial dependence is given by the expression \citep{Goldsbury2016}
\begin{align}
\begin{split}
    f(m_1, m_2, R; \, \theta) 
    &= \dot{N} \, \frho\left(R; \, r_0, r_t\right)
    \int_{-\infty}^{\infty} \int_{-\infty}^{\infty}
    f_M\left(m_1^\prime, m_2^\prime; \, \theta_M\right)\\
    &\quad \times E\left(m_1-m_1^\prime, m_2-m_2^\prime; \, m_1^\prime, m_2^\prime, R\right)
    \, \mathrm{d}m_1^\prime \, \mathrm{d}m_2^\prime,
\end{split}
\label{eq:47tuc_axions_dist_func}
\end{align}
where the photometric error distribution function $E$ is parametrized by $R$ (in addition to the input magnitudes $m_1^\prime$ and $m_2^\prime$),
$\frho$ is the projected radial (surface) density distribution after integrating over the azimuth angle, 
$f_M$ is a function that quantifies the cooling rate, 
$\theta_M$ denotes the set of parameters that the cooling models depends on (after being moved from theory space to data space),
and $\theta = \left\lbrace \theta_M, \, \dot{N}, \, r_0, \, r_t \right\rbrace$ denotes the set of all parameters that the full model depends on, which includes the white dwarf birthrate $\dot{N}$ and any parameters that parametrize the radial density distribution.
In the case that radial dependence can be neglected, \cref{eq:47tuc_axions_dist_func} reduces to the expression used for the distribution function in the analysis of \citet{47tuc_deep_acs}, where the completeness reduction factor discussed in that work is here implicitly equal to unity over the entire magnitude range considered.

The radial density distribution $\frho$ is, stated properly, a probability density function that we take to be normalised over the range of $R$ values defining the relevant data space.
We use the same radial density distribution\footnote{Note that the function $\frho(R)$ in our work is called $\rho(R)$ in \citet{Goldsbury2016}. We use the notation $\frho$ to make the nature of this quantity as a two-dimensional probability density function more clear.} as \citet{Goldsbury2016}, which is a King-Michie model \citep{1963MNRAS.126..499M,1966AJ.....71...64K} with a King radius of $r_0 = 32''$ and tidal radius of $r_t = 3800''$.
The parametric dependence of $\frho$ is written explicitly in \cref{eq:47tuc_axions_dist_func} to make the model dependence on these parameters in general clear; however, these parameters are held fixed in the analysis performed in this work.
We follow the notation of \citet{Goldsbury2016} and \citet{Goldsbury2013} whereby a lowercase $r$ denotes a three-dimensional radial distance from the cluster centre and an uppercase $R$ denotes the corresponding two-dimensional radius in projection.
The two-dimensional nature of $\frho$ is made more transparent by expressing it as
\begin{equation}
    \frho\left(R\right) = \frac{2 \pi \, R \, \Sigma\left(R\right)}{\int 2 \pi \, R \, \Sigma\left(R\right) \, \mathrm{d}R},
\end{equation}
where $\Sigma$ is the projected surface density given by an Abel transform of the three-dimensional density distribution, as described in \citet{Goldsbury2013},
and the integration to normalise $f_R$ is performed over the $R$ limits defining the data space.
The procedure for evaluating $\Sigma$ as a function of $R$ is described in \citet{Goldsbury2013}, which entails numerically solving the system of equations describing the King-Michie model and then performing an Abel transform on the result.

The magnitudes represented by $m_1$ and $m_2$ (and their primed equivalents) are $m_1 = \text{F225W}$ and $m_2 = \text{F336W}$ in the case of the WFC3 data, while they are $m_1 = \text{F435W}$ and $m_2 = \text{F555W}$ in the case of the ACS data.
The function $f_M$ quantifies the theoretical cooling rate predicted by the cooling model as a function of both magnitudes before accounting for photometric errors. Note that the quantity $\dot{N} \, f_M$ is the theoretical number density distribution of the magnitude values before accounting for photometric errors.
The expression for $f_M$ given in \citet{47tuc_deep_acs} and related discussion also applies here,
\begin{equation}
    f_M\left(m_1^\prime, m_2^\prime; \, \theta_M\right) = \frac{\mathrm{d}t}{\mathrm{d}m_1^\prime} \, \delta\left[m_2^\prime - m_{2,\mathrm{mod}}\left(m_1^\prime; \, \theta_M \right)\right],
    \label{eq:47tuc_axions_fM_repeated}
\end{equation}
though some of the parameters differ in the set of model parameters represented by $\theta_M$.
The theoretical cooling models before moving to data space are parametrized by the white dwarf mass $M_\mathrm{WD}$, the H envelope thickness parameter $\lqh$, and the axion mass $m_a$.
Moving the theoretical models to data space depends on the distance modulus $\mu$ and colour excess $\ebv$, so the function $f_M$ is parametrized by the parameter set 
\begin{equation}
    \theta_M = \left\lbrace m_a, \, \mwd, \, \lqh, \, \mu, \, \ebv \right\rbrace.
\end{equation}
Since converting the cooling rate to a number density depends on the white dwarf birthrate $\dot{N}$ and the radial density distribution depends on the additional set of parameters $\theta_R = \left\lbrace r_0, \, r_t \right\rbrace$, the full set of parameters that the likelihood depends on is 
\begin{equation}
    \theta = \left\lbrace m_a, \, \mwd, \, \lqh, \, \dot{N}, \, \mu, \, \ebv, \, r_0, \, r_t \right\rbrace.
\end{equation}

As was done in \citet{47tuc_deep_acs}, the values of $\mu$ and $\ebv$ are held fixed, with a value of $\mu = 13.24$ used for the distance modulus \citep{Chen2018} and a value of $\ebv = 0.04$ used for the colour excess \citep{1996AJ....112.1487H}.
For this value of the colour excess, the corresponding value of the total V-band extinction is $A_V = 0.124$, where the relative visibility has been taken to be $R_V = 3.1$, the typical value for the Milky Way.
The values of all of the parameters that are held fixed are summarised in \cref{tab:47tuc_axions_fixed_params}.
With $\mu$ and $\ebv$ held fixed in addition to $r_0$ and $r_t$, as described above, the set of parameters $\theta_M$ that the cooling rate depends on in the analysis is reduced to $\theta_M = \left\lbrace m_a, \, \mwd, \, \lqh, \right\rbrace$, and the full model (and thus the likelihood) depends only on the reduced set of parameters $\theta = \left\lbrace m_a, \, \mwd, \, \lqh, \, \dot{N} \right\rbrace$.

\begin{table}
    \centering
    \begin{tabular}{l l}
    \toprule 
        Parameter & Value\\
    \midrule 
        $r_0$ & $32''$\\
        $r_t$ & $3800''$\\
        $\mu$ & $13.24$\\
        $\ebv$ & $0.04$\\
    \bottomrule 
    \end{tabular}
    \caption{Summary of model parameter values held fixed in analysis.}
    \label{tab:47tuc_axions_fixed_params}
\end{table}

The input magnitudes before accounting for photometric errors, denoted by a prime symbol in \cref{eq:47tuc_axions_dist_func,eq:47tuc_axions_fM_repeated}, are calculated from the relevant theory-space cooling model variables using bolometric corrections via the procedure described in section 7 of \citet{47tuc_deep_acs}, but with extinction values appropriate for the magnitude filters of the data considered in this work.
The relevant extinctions for the WFC3/UVIS filters are given by
\begin{align}
    A_\mathrm{F225W} &= 2.62940 \, A_V,\\
    A_\mathrm{F336W} &= 1.67536 \, A_V,
\end{align}
and the relevant extinctions for the ACS/WFC filters are given by
\begin{align}
    A_\mathrm{F435W} &= 1.33879 \, A_V,\\
    A_\mathrm{F555W} &= 1.03065 \, A_V.
\end{align}
Note that these extinctions were determined using the same extinction law\footnote{The extinction values were retrieved using \url{http://stev.oapd.inaf.it/cgi-bin/cmd_3.7}, which uses this extinction law.}, that of \citet{1989ApJ...345..245C} and \citet{1994ApJ...422..158O}, as was used in \citet{47tuc_deep_acs}.

The natural logarithm of the unbinned likelihood is calculated from the distribution function $f$ given by \cref{eq:47tuc_axions_dist_func} using the expression
\begin{equation}
    \ln \mathcal{L}\left(\theta\right) = \sum_i \ln f_i - N_\mathrm{pred},
    \label{eq:47tuc_axions_likelihood}
\end{equation}
where  $f_i$ is the distribution function evaluated at the data coordinate $d_i = \left(m_{1i}, m_{2i}, R_i\right)$ observed for the $i$th data point,
\begin{equation}
    f_i = f\left(m_{1i}, m_{2i}, R_i; \theta\right),
\end{equation}
and $N_\mathrm{pred}$ is the model prediction for the total number of white dwarfs in the data space,
\begin{equation}
    N_\mathrm{pred} = \iiint\limits_{\text{data space}} f\left(m_1, m_2, R; \theta\right) \, \mathrm{d}R \, \mathrm{d}m_1 \, \mathrm{d}m_2.
\end{equation}

\begin{figure}
    \centering
    \begin{subfigure}[t]{0.495\textwidth}
        \centering
        \includegraphics{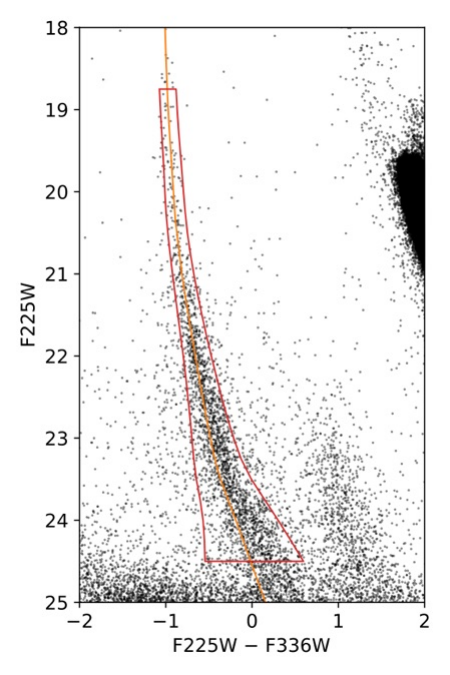}
        \caption{WFC3/UVIS data set}
        \label{fig:47tuc_axions_dataspace_wfc3}
    \end{subfigure}
    \hfill
    \begin{subfigure}[t]{0.495\textwidth}
        \centering
        \includegraphics{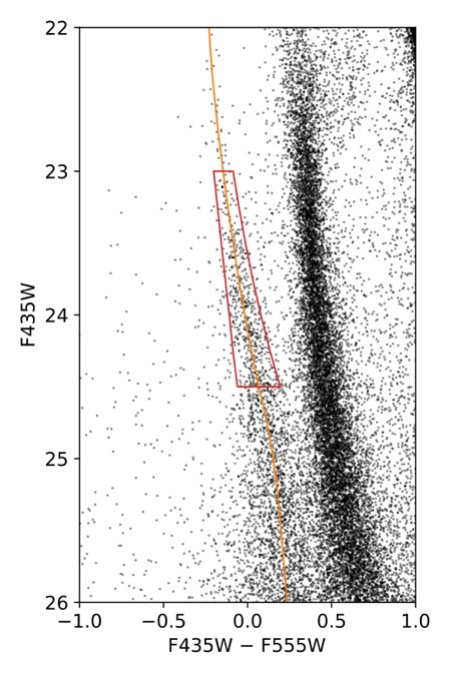}
        \caption{ACS/WFC data set}
        \label{fig:47tuc_axions_dataspace_acs}
    \end{subfigure}
    \caption{CMDs showing the data space selections used in the unbinned likelihood analysis for the WFC3/UVIS data (left; a) and ACS/WFC data (right; b).
    The plots are focused on the white dwarf cooling sequence, with the data shown as black points.
    The boundaries of the white dwarf data space selection for each data set are indicated by the red curves.
    The evolutionary track predicted by the model in each case before accounting for photometric errors is shown as an orange curve.}
    \label{fig:47tuc_axions_dataspace}
\end{figure}

The data space boundaries as they appear in a colour-magnitude diagram (CMD) for each data set considered in our analysis are shown in \cref{fig:47tuc_axions_dataspace}.
Note that the analysis is actually performed in magnitude-magnitude space. The data space is simply shown in colour-magnitude space for better visualisation, and it is straightforward to move between colour-magnitude and magnitude-magnitude space.
The WFC3/UVIS data and corresponding data space selection are shown in \cref{fig:47tuc_axions_dataspace_wfc3}, while the ACS/WFC data and corresponding data space selection are shown in \cref{fig:47tuc_axions_dataspace_acs}.
These plots are focused on the white dwarf cooling sequence of 47 Tuc, with the data space boundaries shown as red curves enclosing a large part of the cooling sequence.
For each data set, the evolutionary track predicted by the model for the relevant magnitudes before accounting for photometric errors is shown by the orange curve that lies along the 47 Tuc cooling sequence and passes through the data space.

The particular reference model shown in \cref{fig:47tuc_axions_dataspace} corresponds to parameter values of $M_\mathrm{WD} = 0.5388~M_\odot$, $m_a = 0~\mathrm{meV}$, and $\lqh = -3.60$, though it should be noted that the evolutionary track in the CMD shows very little variation over the range of parameter values considered in our analysis.
It can be seen visually from \cref{fig:47tuc_axions_dataspace} that the theoretical cooling tracks align reasonably well with the observed data in the CMD.
For the WFC3 data, the model lies left of the centre line of the empirical cooling sequence because the photometric errors are not distributed equally in the two filters and the error distribution has not yet been applied to the model shown in \cref{fig:47tuc_axions_dataspace}.
For the ACS data, the photometric error distribution is more symmetric in the two filters, and it can correspondingly be seen in \cref{fig:47tuc_axions_dataspace} that the model lies more closely along the centre line of the empirical white dwarf sequence of the ACS data.

The WFC3 data space boundaries are chosen to be approximately $3 \, \sigma$ in colour from the model cooling track, while the ACS data space boundaries are chosen to be approximately $2 \, \sigma$ from the model track.
In each case, the error in colour was determined as a function of magnitude $m_1$ (the \textit{y}-axis magnitude in the relevant sub-figure of \cref{fig:47tuc_axions_dataspace}) from the photometric error distribution given by artificial stars tests.
These data space boundaries are chosen such that they enclose the 47 Tuc white dwarf sequence tightly enough in colour to avoid contamination from the Small Magellanic Cloud (SMC).
In \cref{fig:47tuc_axions_dataspace}, the SMC corresponds to the sequence of stars located to the right of the 47 Tuc white dwarf sequence and running approximately parallel to it.
The ACS data space is shorter than the WFC3 data space because the SMC sequence begins to intersect with the 47 Tuc white dwarf sequence at brighter magnitudes (corresponding to earlier cooling times) in the ACS data than in the WFC3 data, as can be seen in \cref{fig:47tuc_axions_dataspace}.

In the case of the WFC3 data (for which the variable $R$ is included in the analysis), the data space definition also includes a cut in the radial distance from the cluster centre of $R \leq 4,000~\mathrm{pixels}$\footnote{This cut was also used in the analysis of \citet{Goldsbury2016}, though it is not explicitly mentioned in that paper.}.
The WFC3/UVIS pixel scale is 0.04 arcseconds per pixel, so this corresponds to a cut of $R \leq 160''$ in physical units.
This cut removes objects located at the outermost radii of the total WFC3 field of view where there is not full coverage about the entire circumference of a circle of that radius; thus, this cut circumvents any potential concerns about a reduction in completeness due to the field geometry at these outermost radii.

The birthrate of white dwarfs for a particular data set is specific to the field of view for those observations, so the WFC3 and ACS data are expected to have different birthrates.
A combined analysis of both of these data sets thus depends on two birthrate parameters, which will be denoted as $\Ndotwfc$ for WFC3 and $\Ndotacs$ for ACS.
This is accounted for in a combined analysis of the WFC3 and ACS data sets as follows.

The theoretical cooling models predict separate distribution functions for each set of data: $f_\mathrm{WFC3}\left(\mathrm{F225W}, \mathrm{F336W}, R; \, \theta\right)$ with $\theta = \left\lbrace \Ndotwfc, \, \theta_M \right\rbrace$ for the WFC3 data and $f_\mathrm{ACS}\left(\mathrm{F435W}, \mathrm{F555W}; \, \theta\right)$ with $\theta = \left\lbrace \Ndotacs, \, \theta_M \right\rbrace$ for the ACS data.
Likelihoods can thus be calculated for each data set using \cref{eq:47tuc_axions_likelihood} for the appropriate distribution function.
Letting $D_\mathrm{WFC3}$ and $D_\mathrm{ACS}$ denote the set of WFC3 data points and ACS data points, respectively, in the relevant data space for each field, i.e. 
\begin{align}
    D_\mathrm{WFC3} &\equiv \left\lbrace \mathrm{F225W}_i, \mathrm{F336W}_i, R_i \right\rbrace_{i \, \in \, \mathrm{WFC3~data~points}},\\
    D_\mathrm{ACS} &\equiv \left\lbrace \mathrm{F435W}_i, \mathrm{F555W}_i \right\rbrace_{i \, \in \, \mathrm{ACS~data~points}},
\end{align}
the corresponding likelihoods $\mathcal{L}_\mathrm{WFC3}$ and $\mathcal{L}_\mathrm{ACS}$ for each of these data sets separately are defined as
\begin{align}
    \mathcal{L}_\mathrm{WFC3}\left(\Ndotwfc, \, \theta_M\right) &\equiv p\left(D_\mathrm{WFC3} \, | \, \Ndotwfc, \, \theta_M \right),\\
    \mathcal{L}_\mathrm{ACS}\left(\Ndotacs, \, \theta_M\right) &\equiv p\left(D_\mathrm{ACS} \, | \, \Ndotacs, \, \theta_M \right),
\end{align}
where $p$ denotes a probability density function.
The likelihood $\mathcal{L}_\mathrm{comb}$ of both data sets combined is defined as
\begin{equation}
    \mathcal{L}_\mathrm{comb}\left(\theta\right)
    \equiv p\left(D_\mathrm{WFC3}, \, D_\mathrm{ACS} \, | \, \Ndotwfc, \, \Ndotacs, \, \theta_M \right),
\end{equation}
where $\theta = \left\lbrace \Ndotwfc, \, \Ndotacs, \, \theta_M \right\rbrace$ is the set of all parameters considered in the combined analysis.

Since each data point is statistically independent, it follows that
\begin{align}
\begin{split}
    &p\left(D_\mathrm{WFC3}, \, D_\mathrm{ACS} \, | \, \Ndotwfc, \, \Ndotacs, \, \theta_M \right)\\
    &\quad = p\left(D_\mathrm{WFC3} \, | \, \Ndotwfc, \, \theta_M \right) \, p\left(D_\mathrm{ACS} \, | \, \Ndotacs, \, \theta_M \right).
\end{split}
\end{align}
The log-likelihood of the combined data is thus simply given by the sum of the log-likelihoods of the separate data sets, i.e.
\begin{equation}
    \ln \mathcal{L}_\mathrm{comb}\left(\theta\right)
    = \ln \mathcal{L}_\mathrm{WFC3}\left(\Ndotwfc, \, \theta_M\right) + \ln \mathcal{L}_\mathrm{ACS}\left(\Ndotacs, \, \theta_M\right).
\end{equation}
The joint posterior for the combined analysis is then given by
\begin{equation}
    p\left(\theta \, | \, D_\mathrm{WFC3}, \, D_\mathrm{ACS}\right)
    = \frac{p\left(\theta\right) \, \mathcal{L}_\mathrm{comb}\left(\theta\right)}{\int p\left(\theta\right) \, \mathcal{L}_\mathrm{comb}\left(\theta\right) \, \mathrm{d}\theta},
    \label{eq:47tuc_axions_posterior}
\end{equation}
where the joint prior distribution $p\left(\theta\right)$ for all of the parameters can be expressed as
\begin{equation}
    p\left(\theta\right)
    = p\left(\Ndotwfc\right) \, p\left(\Ndotacs\right) \, p\left(\mwd, \, \lqh\right) \, p\left(m_a\right)
\end{equation}
and the integral in the denominator of \cref{eq:47tuc_axions_posterior} is performed over the entire parameter space.

For the main results reported in \cref{sec:47tuc_axions_results} below, uniform priors were used for both of the birthrate parameters.
The option of using Gaussian birthrate priors with the same values as used by \citet{Goldsbury2016} is considered in \cref{sec:appendix_additional_analyses,sec:appendix_47tuc_axions_comparison}.
However, it is found in \cref{sec:appendix_47tuc_axions_wfc3_only,sec:appendix_47tuc_axions_acs_only} (for each data set individually) and shown below in \cref{sec:47tuc_axions_results} (for the combined analysis) that the birthrate priors of \citet{Goldsbury2016} both overestimate $\dot{N}_\mathrm{WFC3}$ and underestimate $\dot{N}_\mathrm{ACS}$ by a similar amount.
This suggests that there may be a relevant effect that was not accounted for in determining (or applying) those birthrate priors, such as the effect of cluster relaxation \citep{2015ApJ...804...53H,Heyl2017} causing stars to leave the WFC3 field and enter the ACS field over time.
Though using Gaussian priors for the birthrates could potentially give tighter bounds on the parameters, uniform priors are used for the birthrates to avoid the risk of imposing an incorrect prior value.

The results of \citet{47tuc_deep_acs} provided priors for $\mwd$ and $\lqh$.
Our main analysis, for which the results are presented below in \cref{sec:47tuc_axions_results}, used a joint prior $p(\mwd, \, \lqh)$ for $\mwd$ and $\lqh$ given by the joint posterior distribution from \citet{47tuc_deep_acs} after marginalising over the birthrate.
This corresponds to the distribution plotted in Fig. 11 in \citet{47tuc_deep_acs}.
Using uniform priors for $\mwd$ and $\lqh$ instead yields similar best-fitting parameter values (when uniform birthrate priors are used) but a more extended posterior distribution (with some degeneracy between $\mwd$ and $\lqh$ for the WFC3 data in particular), so using the \citet{47tuc_deep_acs} joint prior for $\mwd$ and $\lqh$ provides tighter parameter constraints.
This is shown in detail in \cref{sec:appendix_additional_analyses,sec:appendix_47tuc_axions_comparison}.
In all cases considered, in both \cref{sec:47tuc_axions_results} and \cref{sec:appendix_additional_analyses,sec:appendix_47tuc_axions_comparison}, the prior for $m_a$ was taken to be uniform.

\section{Results} \label{sec:47tuc_axions_results}

The results for the combined fit of the cooling models to both the WFC3/{\allowbreak}UVIS and ACS/{\allowbreak}WFC data are presented in this section.
The combined fit to both of these data sets gives stronger constraints than fitting either one of these data sets separately.
Nevertheless, separate fits of the models to each of these data sets should give similar results to the combined fit if each of the data sets are reasonably well-fitted by the models, which is verified in \cref{sec:appendix_additional_analyses,sec:appendix_47tuc_axions_comparison}.
In addition to the combined analysis presented in this section, which gives the main results of our work, we also independently fitted the cooling models to each of the two data sets and present those results in \cref{sec:appendix_47tuc_axions_wfc3_only,sec:appendix_47tuc_axions_acs_only}.
The results for the WFC3 data alone are given in \cref{sec:appendix_47tuc_axions_wfc3_only}, while the results for the ACS data alone are given in \cref{sec:appendix_47tuc_axions_acs_only}.
The key results of the separate WFC3 and ACS analyses are compared with the results of the combined analysis in \cref{sec:appendix_47tuc_axions_comparison}.

\subsection{Posterior Distributions} \label{sec:47tuc_axions_postdists}

\begin{figure}
    \centering
    \includegraphics[width=0.625\textwidth]{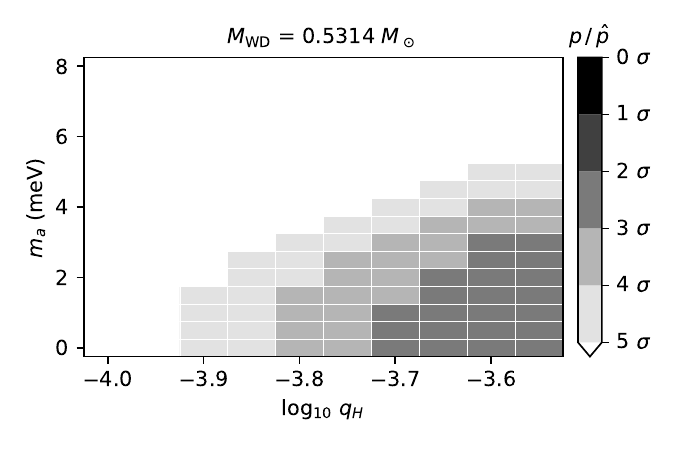}\\
    \includegraphics[width=0.625\textwidth]{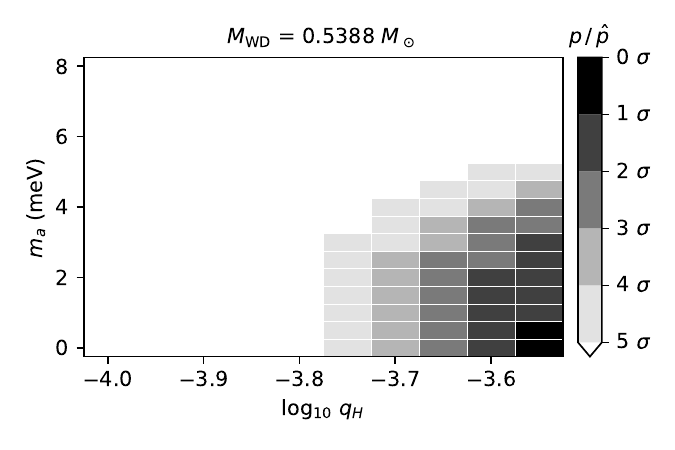}\\
    \caption{Joint posterior probability density distribution after marginalising over the birthrates. 
    Slices of the distribution as a function of axion mass ($m_a$) and envelope thickness ($q_{H}$) are shown for fixed values of white dwarf mass ($M_\mathrm{WD}$).
    The probability density ($p$) has been scaled by its maximum value ($\hat{p}$) so that the plotted quantity is $p / \hat{p}$. The filled contours are drawn at level values corresponding to the $\sigma$ levels indicated on the legend for a two-dimensional normal distribution.
    The lowest mass case ($\mwd = 0.5240~M_\odot$) is not shown because it is excluded at $5~\sigma$.}
    \label{fig:47tuc_axions_joint_density}
\end{figure}

The joint posterior density distribution after marginalising over both of the birthrates is shown in \cref{fig:47tuc_axions_joint_density} as filled contour plots. 
Each plot shows a slice of this distribution as a function of $m_a$ and $\log_{10} q_H$  for a particular value of $M_\mathrm{WD}$.
In these plots, the posterior density $p$ (marginalised over $\Ndotwfc$ and $\Ndotacs$) has been scaled by its maximum value $\hat{p}$ attained on the parameter grid, and the contour levels are drawn at $p/\hat{p}$ values of -0.5, -2.0, -4.5, -8.0, and -12.5.
These level values correspond to the analogously scaled probability density values of a normal distribution evaluated at 1, 2, 3, 4, and 5~$\sigma$, with $p/\hat{p} = \exp\left(-0.5 \, n_\sigma^2\right)$ for a level value corresponding to $n_\sigma \, \sigma$.
Plots are only shown for the $M_\mathrm{WD}$ values from the parameter grid that are most significant for the posterior distribution: 0.5314 and 0.5388~$M_\odot$. For the case of $\mwd = 0.5240~M_\odot$, which has been omitted from \cref{fig:47tuc_axions_joint_density}, the value of $p/\hat{p}$ is less than the smallest displayed level, i.e. $p/\hat{p} < 5~\sigma$, for every point on the grid of $m_a$ and $\lqh$ values.

\begin{figure}
    \centering
    \includegraphics[width=0.75\textwidth]{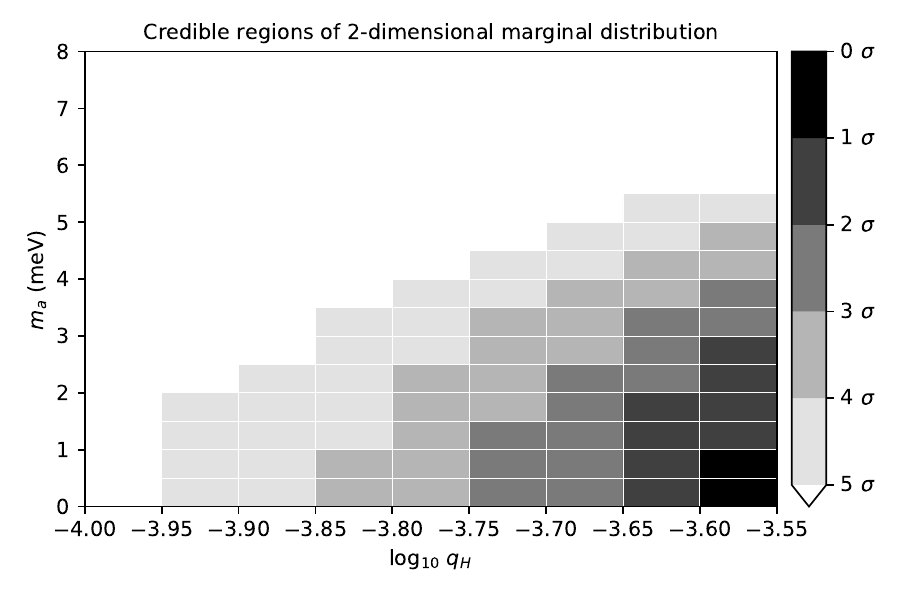}
    \caption{Two-dimensional joint credible regions of axion mass ($m_a$) and envelope thickness ($q_H$) after marginalising over the other parameters. 
    The filled contours show regions of enclosed probability containing the same probability as enclosed by a two-dimensional normal distribution for the $\sigma$ levels indicated by the legend.}
    \label{fig:47tuc_axions_2d_CRs}
\end{figure}

It should be emphasised that the filled contours in \cref{fig:47tuc_axions_joint_density} indicate ranges of probability density values, not regions of enclosed probabilities, so they are not credible regions.
The two-dimensional credible regions in the joint parameter space of $m_a$ and $\log_{10} q_H$ are shown in \cref{fig:47tuc_axions_2d_CRs}.
The credible regions shown in \cref{fig:47tuc_axions_2d_CRs} are more specifically the highest posterior density credible regions calculated from the joint posterior distribution shown in \cref{fig:47tuc_axions_joint_density} after marginalising over $M_\mathrm{WD}$ as well as $\Ndotwfc$ and $\Ndotacs$.
The filled contours in \cref{fig:47tuc_axions_2d_CRs} demarcate regions of enclosed probability, and the total probability enclosed by each contour (including the regions coloured darker than that contour level) is indicated on the colour bar in terms of the number of standard deviations of a (spherically symmetric) two-dimensional normal distribution that encloses the same probability.
The contour levels of 1, 2, 3, 4, and 5~$\sigma$ respectively correspond to total enclosed probabilities of 39.35, 86.47, 98.89, 99.97, and 99.9996\%.

\Cref{fig:47tuc_axions_joint_density} and \cref{fig:47tuc_axions_2d_CRs} both show that smaller $m_a$ and larger $\lqh$ values are favoured, with the peak of the joint posterior distribution occurring for $m_a = 0~\mathrm{meV}$ and $\lqh = -3.55$ on the parameter grid.
\Cref{fig:47tuc_axions_joint_density} and \cref{fig:47tuc_axions_2d_CRs} also highlight the degeneracy between $m_a$ and $\lqh$:
larger $m_a$ values are more probable when the value of $\lqh$ is also large, though small $m_a$ values are still favoured even at the largest $\lqh$ values. 
As can be seen in \cref{fig:47tuc_axions_joint_density}, the morphology of the marginal posterior distribution as a function of $m_a$ and $\lqh$ is similar regardless of $\mwd$, with the distribution simply becoming more tightly concentrated on the highest $\lqh$ values for larger $\mwd$.
The peak in the posterior distribution occurs for $\mwd = 0.5388~M_\odot$.
This is the largest value of $\mwd$ considered in the current analysis, but larger values on the extended parameter grid considered in \citet{47tuc_deep_acs} are strongly disfavoured by the results of that work (which were used as the prior on $\mwd$ and $\lqh$ in this analysis).

The maximum of the posterior distribution also occurs at the (upper) edge of the $\lqh$ grid, but this grid cannot be extended to larger $\lqh$ values (with the current grid spacing) because this is a physical limit. 
In creating the suite of white dwarf cooling simulations used in this work, white dwarfs that started with thicker H envelopes simply burned away the extra H in the envelope through residual nuclear burning at very early cooling times and ended up with the same envelope thickness at the reference cooling age of $10~\mathrm{Myr}$ at which $\lqh$ is defined in this work.
Furthermore, the cooling models with the same $\lqh$ value (at this reference time) produced by these simulations were the same over the magnitude range of interest for this work, regardless of the initial thickness at very early times.

\begin{figure}
    \centering
    \begin{subfigure}[t]{0.49\textwidth}
        \centering
        \includegraphics[width=\textwidth]{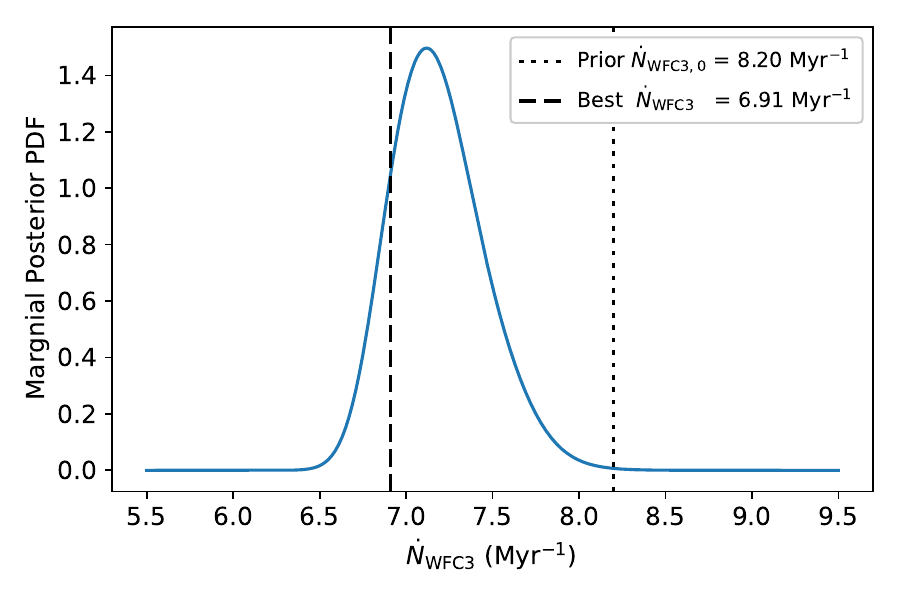}
        \caption{WFC3 birthrate}
        \label{fig:47tuc_axions_1d_marg_dist_Ndot_wfc3}
    \end{subfigure}
    \hfill
    \begin{subfigure}[t]{0.49\textwidth}
        \centering
        \includegraphics[width=\textwidth]{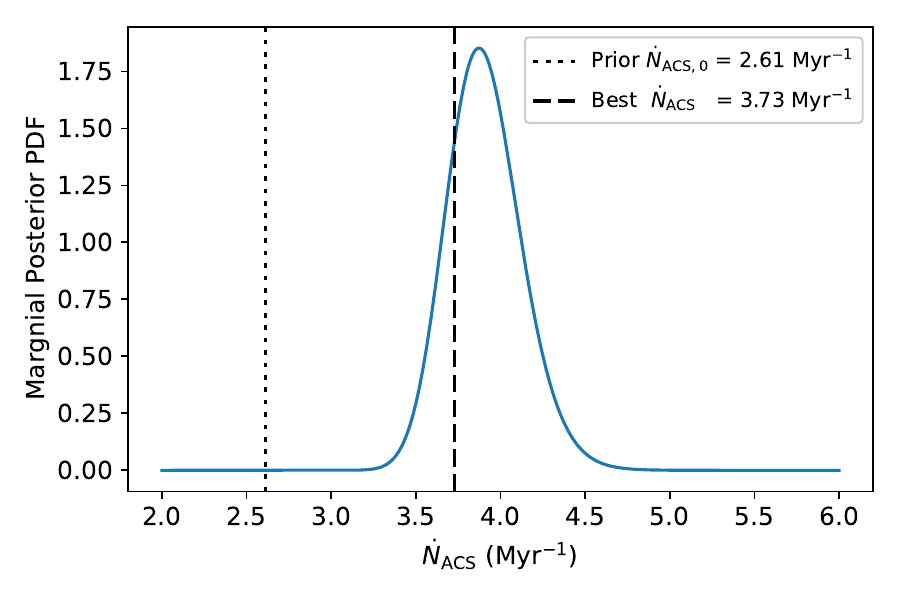}
        \caption{ACS birthrate}
        \label{fig:47tuc_axions_1d_marg_dist_Ndot_acs}
    \end{subfigure}
    \\[1.5ex]
    \begin{subfigure}[t]{0.49\textwidth}
        \centering
        \includegraphics[width=\textwidth]{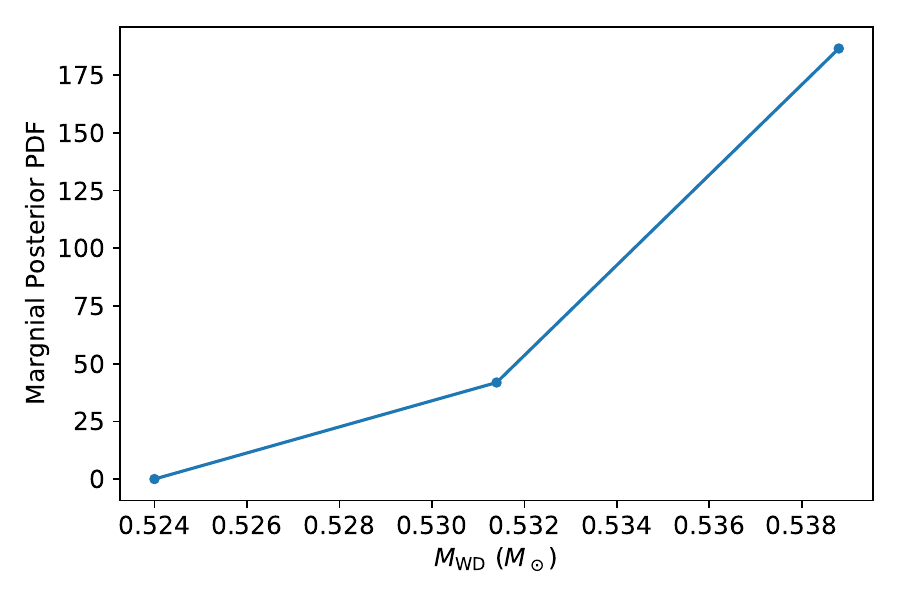}
        \caption{White dwarf mass}
        \label{fig:47tuc_axions_1d_marg_dist_MWD}
    \end{subfigure}
    \hfill
    \begin{subfigure}[t]{0.49\textwidth}
        \centering
        \includegraphics[width=\textwidth]{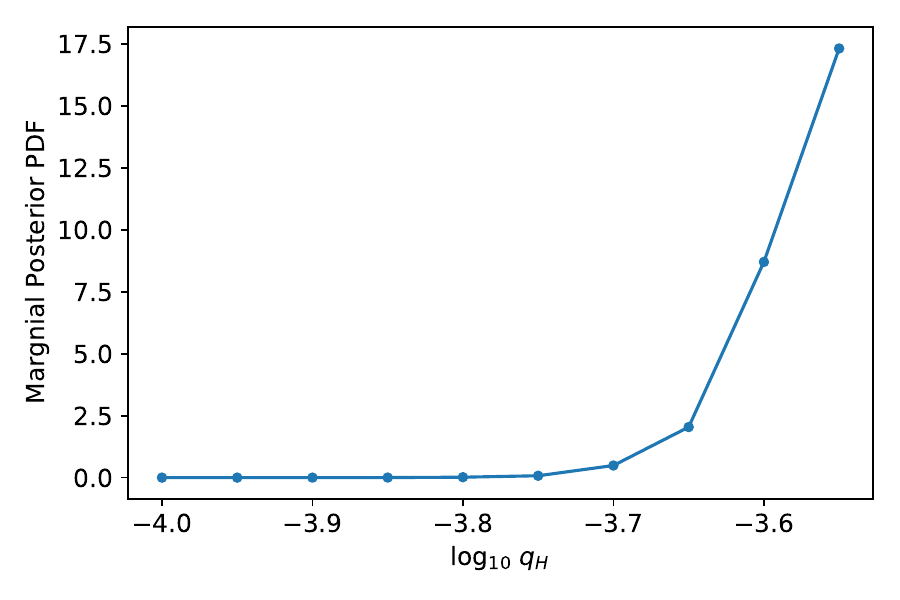}
        \caption{Envelope thickness}
        \label{fig:47tuc_axions_1d_marg_dist_lqh}
    \end{subfigure}
    \\[1.5ex]
    \begin{subfigure}[t]{0.49\textwidth}
        \centering
        \includegraphics[width=\textwidth]{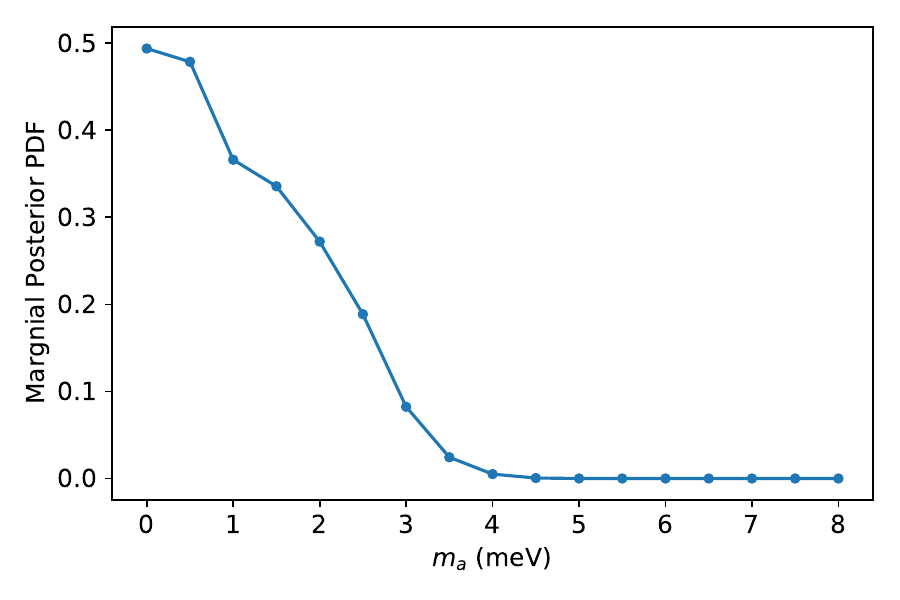}
        \caption{Axion mass}
        \label{fig:47tuc_axions_1d_marg_dist_ma}
    \end{subfigure}
    \\[1.5ex]
    \caption{One-dimensional posterior probability density distributions for each parameter after marginalising over all other model parameters.}
    \label{fig:47tuc_axions_1d_marginal_distributions}
\end{figure}

The one-dimensional posterior probability density distributions of each model parameter after marginalising over all other parameters are shown in \cref{fig:47tuc_axions_1d_marginal_distributions}.
The posterior distribution for the WFC3 birthrate is shown in \cref{fig:47tuc_axions_1d_marg_dist_Ndot_wfc3}, while the posterior distribution for the ACS birthrate is shown in \cref{fig:47tuc_axions_1d_marg_dist_Ndot_acs}.
For each birthrate distribution, the value of $\dot{N}$ that maximises the posterior distribution is indicted by a dashed vertical line and the prior value $\dot{N}_0$ used by \citet{Goldsbury2016} is indicated by a dotted vertical line for comparison.
It can be seen in these plots that the posterior for $\Ndotwfc$ is notably smaller than the prior value of \citet{Goldsbury2016}, while the posterior for $\Ndotacs$ is larger than the prior value of \citet{Goldsbury2016}.
This may be an indication that an appreciable number of white dwarfs are leaving the WFC3 field of view (i.e. the inner field) and entering the ACS field of view (i.e. the outer field).

\Cref{fig:47tuc_axions_1d_marg_dist_MWD}, \cref{fig:47tuc_axions_1d_marg_dist_lqh}, and \cref{fig:47tuc_axions_1d_marg_dist_ma} show the one-dimensional posterior distributions for $\mwd$, $\lqh$, and $m_a$, respectively.
Some of the trends noted above for these parameters based on \cref{fig:47tuc_axions_joint_density} and \cref{fig:47tuc_axions_2d_CRs} can be seen clearly in \cref{fig:47tuc_axions_1d_marg_dist_MWD,fig:47tuc_axions_1d_marg_dist_lqh,fig:47tuc_axions_1d_marg_dist_ma}.
\Cref{fig:47tuc_axions_1d_marg_dist_MWD} highlights the overall increased probability of larger $\mwd$ values over the limited range of values considered in this analysis, which was noted above based on \cref{fig:47tuc_axions_joint_density}.
\Cref{fig:47tuc_axions_1d_marg_dist_lqh} and \cref{fig:47tuc_axions_1d_marg_dist_ma} show that large $\lqh$ and small $m_a$ values are favoured, as is also clearly seen in the plot of the two-dimensional credible regions for these parameters, \cref{fig:47tuc_axions_2d_CRs}.
Using the one-dimensional marginal posterior distributions to calculate the individual 95\% confidence levels of each of these parameters, it is found that $-3.67 \leq \lqh \leq -3.55$ and $m_a \leq 2.85~\mathrm{meV}$ at 95\% confidence.
This limit for the DFSZ axion mass corresponds to a limit for the coupling constant of $g_{aee} \leq 0.81 \times 10^{-13}$ for any axion or axion-like particle model with an axion-electron interaction.

\subsection{Best-Fitting Model}  \label{sec:47tuc_axions_cumdists}

The combination of parameter values for which the full joint posterior distribution is maximised on the discrete parameter grid for the combined analysis is summarised in \cref{tab:47tuc_axions_best_model_params}.
These parameter values correspond to the optimal model on the parameter grid for the combined analysis after accounting for all of the priors and before marginalising over any parameters.
The 95\% credible regions calculated from the one-dimensional marginal posterior distributions shown in \cref{fig:47tuc_axions_1d_marginal_distributions} are given in \cref{tab:47tuc_axions_best_model_params} as errors on the parameter values of the best-fitting model.
Note that these parameter values are not necessarily the values that optimise the marginal posteriors from which the credible regions were calculated,
though it can be seen from \cref{fig:47tuc_axions_1d_marginal_distributions} that these values are similar (and indeed exactly the same for $\mwd$, $\lqh$, and $m_a$).
Also note that although the values of $\mwd$, $\lqh$, and $m_a$ (but not $\Ndotwfc$ or $\Ndotacs$) for the best-fitting model are restricted to the parameter grid of the cooling models, the limits of the credible regions are not restricted to the grid points because linear interpolation was used when calculating the credible regions.
For easy reference, the $g_{aee}$ values corresponding to the best-fitting value and credible region limits of $m_a$ have also been given in \cref{tab:47tuc_axions_best_model_params}.

\begin{table}
    \centering
    \begin{tabular}{l D{,}{.}{2.10}}
        \toprule
        Parameter & \multicolumn{1}{l}{~Value} \\
        \midrule
        $\Ndotwfc~(\mathrm{Myr}^{-1})$ & 6,91_{-0.23}^{+0.82} \\[1ex]
        $\Ndotacs~(\mathrm{Myr}^{-1})$ & 3,73_{-0.24}^{+0.62} \\[1ex]
        $\mwd~(M_\odot)$ & 0,5388_{-0.0106}^{+0.0000} \\[1ex]
        $\lqh$ & -3,55_{-0.12}^{+0.00} \\[1ex]
        $m_a~(\mathrm{meV})$ & 0,00_{-0.00}^{+2.85} \\
        \addlinespace[1.5ex] \midrule \addlinespace[1.5ex]
        $g_{aee} \, / \, 10^{-13}$ & 0,00_{-0.00}^{+0.81} \\
        \bottomrule
    \end{tabular}
    \caption{Parameter values of optimal model for combined analysis. The values are reported for the combination of parameter values that maximises the joint posterior distribution on the full parameter grid. The errors indicate the 95\% credible region calculated from the one-dimensional marginal posterior distribution for each parameter.}
    \label{tab:47tuc_axions_best_model_params}
\end{table}

\begin{figure}
    \centering
    \includegraphics[width=0.75\textwidth]{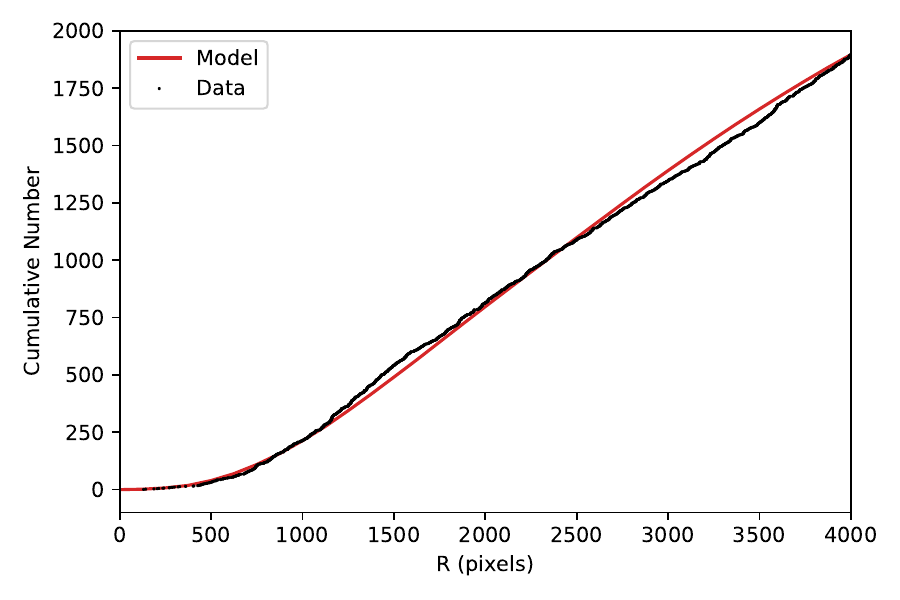}
    \caption{Cumulative number distribution of radial distance ($R$) from cluster centre for optimal model (red curve) and WFC3/UVIS data (black points).}
    \label{fig:47tuc_axions_rad_cumdist}
\end{figure}

\begin{figure}
    \centering
    \begin{subfigure}[t]{0.49\textwidth}
        \centering
        \includegraphics[width=\textwidth]{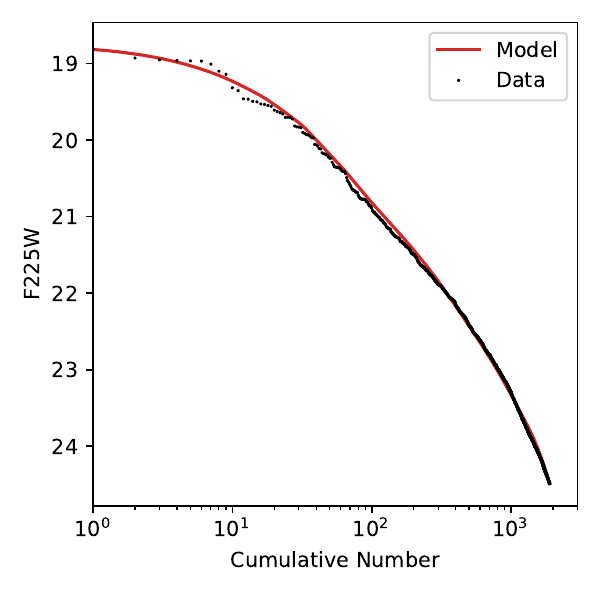}
        \caption{WFC3/UVIS, F225W magnitude}
        \label{fig:47tuc_axions_invLFs_wfc3_f225w}
    \end{subfigure}
    \hfill
    \begin{subfigure}[t]{0.49\textwidth}
        \centering
        \includegraphics[width=\textwidth]{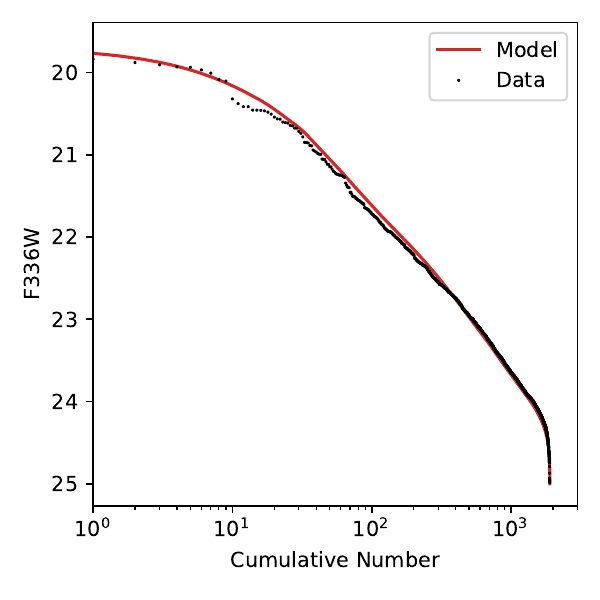}
        \caption{WFC3/UVIS, F336W magnitude}
        \label{fig:47tuc_axions_invLFs_wfc3_f336w}
    \end{subfigure}
    \\[1.5ex]
    \begin{subfigure}[t]{0.49\textwidth}
        \centering
        \includegraphics[width=\textwidth]{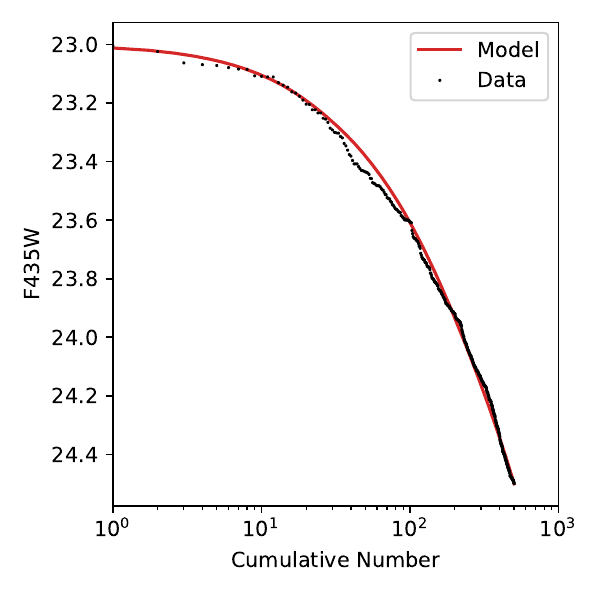}
        \caption{ACS/WFC, F435W magnitude}
        \label{fig:47tuc_axions_invLFs_acs_f435w}
    \end{subfigure}
    \hfill
    \begin{subfigure}[t]{0.49\textwidth}
        \centering
        \includegraphics[width=\textwidth]{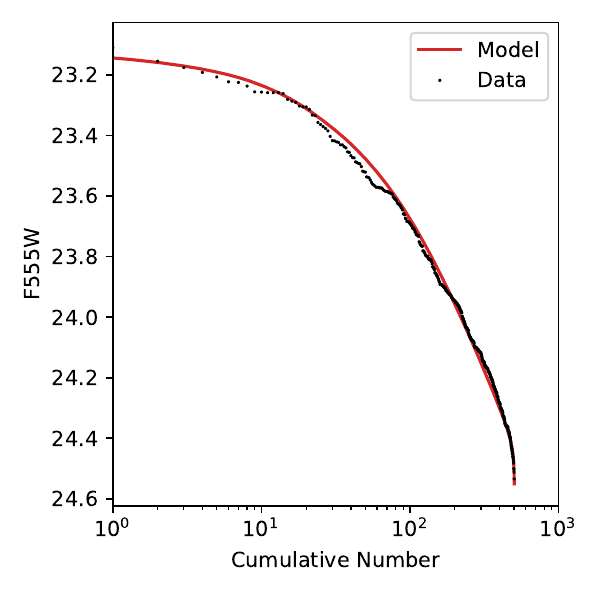}
        \caption{ACS/WFC, F555W magnitude}
        \label{fig:47tuc_axions_invLFs_acs_f555w}
    \end{subfigure}
    \\[1.5ex]
    \caption{Inverse cumulative luminosity function of optimal model from combined fit (red curve) compared to the two sets of HST data used in the combined fit (black points). The different sub-figures show the empirical distributions for the WFC3/UVIS data (top row) with respect to F225W magnitude (left column; a) and F336W magnitude (right column; b) and the distributions for the ACS/WFC data (bottom row) with respect to F435W magnitude (left column; c) and F555W magnitude (right column; d).}
    \label{fig:47tuc_axions_invLFs}
\end{figure}

To visually verify the goodness of fit of the optimal model, we plot the one-dimensional marginal cumulative number distributions predicted by the optimal model over the corresponding empirical distributions for each data component of each of the two data sets.
These are shown in \cref{fig:47tuc_axions_rad_cumdist,fig:47tuc_axions_invLFs}, with the model prediction shown as the red curve and the data shown as black points.
In all of these cases, the incompleteness of the empirical observations is accounted for in the model instead of applying any completeness correction to the data.

\Cref{fig:47tuc_axions_rad_cumdist} shows the cumulative number distribution with respect to the radial distance ($R$) from the cluster centre for the WFC3/UVIS data, with $R$ given in units of WFC3/UVIS pixels.
Note that the unit conversion between WFC3/UVIS pixels and arcseconds is $0.04''$ per pixel \citep{WFC3DataHandbook,WFC3InstrumentHandbook}.
The one-dimensional marginal number density distributions with respect to $R$ predicted by the model for the WFC3/UVIS data is calculated by integrating the full three-dimensional number density distribution $f_\mathrm{WFC3}\left(\mathrm{F225W},\mathrm{F336W},R\right)$ with respect to the magnitudes F225W and F336W. The corresponding marginal cumulative number distribution with respect to $R$ (plotted as the red curve in \cref{fig:47tuc_axions_rad_cumdist}) is then calculated from this marginal number density distribution by integrating up to each value of $R$ that is plotted.
Note that since our models for the ACS/WFC data do not depend on $R$, there is no plot analogous to \cref{fig:47tuc_axions_rad_cumdist} to be made for the ACS/WFC data.

\Cref{fig:47tuc_axions_invLFs} shows the inverse cumulative luminosity functions for each magnitude of each data set.
Note that the cumulative luminosity function with respect to a particular magnitude is simply the cumulative number distribution for that magnitude.
For a particular magnitude, the model prediction is given by the cumulative number density distribution after marginalising over $R$ (if applicable) and the other magnitude that $f$ depends on.
The top row of \cref{fig:47tuc_axions_invLFs} shows the (inverse) distributions for the WFC3 data with respect to the two different WFC3 magnitudes, $m_1 = \mathrm{F225W}$ in \cref{fig:47tuc_axions_invLFs_wfc3_f225w} and $m_2 = \mathrm{F336W}$ in \cref{fig:47tuc_axions_invLFs_wfc3_f336w}.
The theoretical distributions (predicted by the optimal model) that are shown in \cref{fig:47tuc_axions_invLFs_wfc3_f225w,fig:47tuc_axions_invLFs_wfc3_f336w} are both calculated from the same distribution function $f_\mathrm{WFC3}\left(\mathrm{F225W},\mathrm{F336W},R\right)$, but with a different magnitude marginalised over in each case (and $R$ marginalised over in both cases).
Likewise, the bottom row of \cref{fig:47tuc_axions_invLFs} shows the (inverse) distributions for the ACS data with respect to the two different ACS magnitudes, $m_1 = \mathrm{F435W}$ in \cref{fig:47tuc_axions_invLFs_acs_f435w} and $m_2 = \mathrm{F555W}$ in \cref{fig:47tuc_axions_invLFs_acs_f555w},
and the theoretical distributions that are shown in \cref{fig:47tuc_axions_invLFs_acs_f435w,fig:47tuc_axions_invLFs_acs_f555w} are both calculated from the same distribution function $f_\mathrm{ACS}\left(\mathrm{F435W},\mathrm{F555W}\right)$, but with a different magnitude marginalised over in each case.

It can be seen from \cref{fig:47tuc_axions_rad_cumdist,fig:47tuc_axions_invLFs} that the distribution predicted by the optimal model for each data variable well reproduces the corresponding empirical distribution for that variable, indicating good agreement between the optimal model and the data.
\Cref{fig:47tuc_axions_rad_cumdist} shows that the form of the radial density distribution $f_R$ used for the model was a reasonable choice. The small differences between the model distribution and the empirical distribution for the cumulative number with respect to $R$ seen in \cref{fig:47tuc_axions_rad_cumdist} are well balanced across the range of $R$ values, so allowing the King radius (or tidal radius) to vary as another parameter of the model is not likely to provide much improvement to the fit.
\Cref{fig:47tuc_axions_invLFs} shows how well the optimal cooling model fits the data.
Note that the cumulative number is shown on a linear scale in \cref{fig:47tuc_axions_rad_cumdist}, whereas a logarithmic scale is used for the cumulative number in each of the plots shown in \cref{fig:47tuc_axions_invLFs}.
The minor differences between the model and the empirical distributions with respect to magnitude seen in \cref{fig:47tuc_axions_invLFs} occur at smaller magnitudes where there are fewer white dwarfs. 
These early (small magnitude) regions of the distributions in \cref{fig:47tuc_axions_invLFs} are emphasised in the plots due the logarithmic scale used for the cumulative number, but they are actually less important to the fit than the later (larger magnitude) regions where there are much more white dwarfs and where the model distributions very closely match the empirical distributions.

As a quick and simple quantitative test of the goodness of fit of the optimal model, we perform a set of one-sample Kolmogorov-Smirnov (KS) tests for each of the one-dimensional marginal distribution functions predicted by the optimal model for each of the two data sets used in the combined fit.
The results of the KS tests for each of the cases considered are summarised in \cref{tab:47tuc_axions_KS_test_results}.
The pairs of model and empirical distributions compared by these KS tests correspond to the distributions shown in \cref{fig:47tuc_axions_rad_cumdist,fig:47tuc_axions_invLFs} after normalisation.
Note that the KS test actually compares the cumulative probability density distributions (measured by the cumulative fraction), rather than the (inverse) cumulative number distributions shown in \cref{fig:47tuc_axions_rad_cumdist,fig:47tuc_axions_invLFs}.
The \textit{p}-value returned by the (one-sample) KS test is the probability that, for a sample drawn from the model distribution, the resultant sample distribution differs from the model distribution by at least as much as the observed empirical distribution differs from the model distribution.
The large \textit{p}-values found for the KS tests comparing the optimal model to the data (see \cref{tab:47tuc_axions_KS_test_results}) are indicative of a good fit.
These \textit{p}-values are all well above a reasonable threshold value of $10^{-4}$.
The KS test results are particularly good for the ACS data, and are also quite reasonable for the WFC3 data.

\begin{table}
    \centering
    \begin{tabular}{l @{\hspace*{3\tabcolsep}} l @{\hspace*{3\tabcolsep}} d{2.4}}
    \toprule
        Data Set & Data Variable & \heading{\textit{p}-value} \\
    \midrule
        WFC3/UVIS & $R$ & 0.0156 \\
        ~ & F225W & 0.0362 \\
        ~ & F336W & 0.0132 \\
    \addlinespace
        ACS/WFC & F435W & 0.3527 \\
        ~ & F555W & 0.2739 \\
    \bottomrule
    \end{tabular}
    \caption{Results of KS tests for combined analysis comparing the one-dimensional marginal cumulative probability distribution functions predicted by the optimal model to the corresponding empirical (cumulative fraction) distribution.}
    \label{tab:47tuc_axions_KS_test_results}
\end{table}

\section{Discussion}  \label{sec:47tuc_axions_discussion}

The constraints found in this work could potentially be improved upon by applying non-uniform priors for the white dwarf birthrates in the unbinned likelihood analysis, but determining  appropriate prior values for these birthrates requires further analysis that is beyond the scope of this work.
The birthrate priors used by \citet{Goldsbury2016}, which were calculated from red giant branch stars that have just left the main sequence, seem to simultaneously overestimate the birthrate for the WFC3 field and underestimate it for the ACS field.
\Citet{Goldsbury2016} used a WFC3 birthrate prior of $\Ndotwfc = 8.2 \pm 0.3~\mathrm{Myr}$ and an ACS birthrate prior of $\Ndotacs = 2.61 \pm 0.07~\mathrm{Myr}$.
In contrast, the optimal model found in our work has a WFC3 birthrate of $\Ndotwfc = 6.91_{-0.23}^{+0.82}~\mathrm{Myr}$, which is smaller than the corresponding \citet{Goldsbury2016} prior value, and an ACS birthrate of $\Ndotacs = 3.73_{-0.24}^{+0.62}~\mathrm{Myr}$, which is larger than the corresponding \citet{Goldsbury2016} prior value.
A similar difference between the posterior and prior birthrate values can also be seen in the results of \citet{Goldsbury2016}, though it was not discussed in that work.
From the one-dimensional posterior distributions plotted in \citet{Goldsbury2016}, it can be seen that the posterior distribution for the WFC3 birthrate is concentrated about smaller values than the corresponding prior value while the distribution for the ACS birthrate is concentrated about larger values than the corresponding prior value, with the average posterior value in each case differing by more than 3 $\sigma$ from the prior.

This may indicate that cluster relaxation has a measurable effect that needs to be accounted for when determining the white dwarf birthrate.
Stars move away from the centre of a star cluster over time due to diffusion, a phenomenon that was observed for 47 Tuc by \citet{2015ApJ...804...53H} and \citet{Heyl2017}.
Stars leaving the inner WFC3 field over time should cause $\Ndotwfc$ to decrease over time.
The stars that leave the WFC3 field then enter the ACS field, which would cause $\Ndotacs$ to increase (as long as there are more stars entering than leaving the field).
This effect could potentially be accounted for by modelling the change in $\Ndotwfc$ and $\Ndotacs$ over time as part of the analysis, in which case these parameters would really be an ``effective'' birthrate equal to the net rate of white dwarf formation minus the rate of white dwarfs leaving (or plus the rate of white dwarfs entering) the field of view. However, this is an additional complication that requires careful consideration, so it is left for future work.

Using horizontal branch stars instead of red giant branch stars could also potentially give a better estimate of the white dwarf birthrate, since horizontal branch stars are closer to reaching their white dwarf stage.
However, since axions affect the length of time that stars spend on the horizontal branch, this effect would need to be accounted for in the stellar evolution models used to calculate the prior.
While axions can also have an effect on stellar evolution along the red giant branch, this effect is less important at the start of the red giant branch (while being more important for the tip of the red giant branch).

Stronger constraints could also potentially be achieved if the ACS data space could be extended to fainter magnitudes, but this would require carefully accounting for the SMC.
In this work, the issue of SMC contamination has been avoided by simply ending the ACS data space before the SMC begins to intersect the 47 Tuc white dwarf sequence.
The ACS data space could be extended by modelling the SMC and incorporating the model for the SMC contaminants into the full model used for the analysis.
However, as the ACS data alone favours similar cooling model parameters to the WFC3 data alone (see \cref{sec:appendix_additional_analyses,sec:appendix_47tuc_axions_comparison}), with both data sets favouring large $\lqh$ and small $m_a$ values, extending the ACS data space to fainter magnitudes would likely only provide a minor improvement in the final credible regions.

It is quite natural to compare our results with those of \citet{2015ApJ...809..141H} which found a slightly weaker constraint of $g_{aee}<0.84\times 10^{-13}$ at 95\% confidence using only a fraction of the data from the HST observations (the ACS data).
On the other hand, like \citet{Goldsbury2016}, we use the entire HST dataset which has more than twice the white-dwarf stars considered in \citet{2015ApJ...809..141H}.
A comparison of the 95\%-confidence intervals on the neutrino emission rate between the two papers reveals the greater statistical power of the complete dataset: $0.83<f_\nu<1.22$ for \citet{Goldsbury2016} versus $0.6<f_\nu<1.7$ for \citet{2015ApJ...809..141H} --- a constraint tighter by nearly a factor of three.
Furthermore, by using the results of \citet{Chen2018} and \citet{47tuc_deep_acs} we narrow the priors on the distance to the cluster and the thickness of the hydrogen layer.
Despite this, the constraint on the axion coupling is only marginally better than \citet{2015ApJ...809..141H}.
In the decade since \citet{2015ApJ...809..141H} and \citet{Goldsbury2016}, the theoretical treatment of white dwarfs within the MESA software (which all these works draw upon) has matured.
In particular, the treatments of the equation of state of degenerate matter, the conductive and radiative opacities, the nuclear reaction rates (for residual nuclear burning), and compressional heating within the white dwarf have improved in accuracy.
Furthermore, neither of the previous works included element diffusion, which provides energy and determines the structure of the outer layers of the white dwarfs.
Although only the equation of state will directly affect the axion and neutrino emission, the rest will determine the photon emission from the surface, which is what we actually measure to constrain the evolution of the white dwarfs.
Consequently, our constraint of $g_{aee} < 0.81 \times 10^{-13}$ at 95\% confidence is the state of the art from a theoretical and observational point of view.

Furthermore, the lack of a dramatic shift in the bound despite stronger priors and the addition of many more stars in the neutrino and axion cooling regime may indicate that these results define the floor of what can be achieved from measurements of white dwarf luminosity functions.
The extra stars have reduced the statistical uncertainty, but the recent models and the data from fainter white dwarfs in \citet{47tuc_deep_acs} point toward appreciably different white dwarf envelope structures (especially much thicker H layers), so the axion limit is now being set largely by the physical sensitivity of the cooling models rather than by sample size.
In that regime, one should not expect the upper bound on $g^2_{aee}$ to improve nearly as fast as $\sqrt{N}$, even with many more white dwarfs.
Thicker hydrogen layers may have made the cooling sequence much more sensitive to $q_H$ through diffusion and residual hydrogen burning, increasing the covariance between $q_H$ and axion cooling.
Consequently, the larger white dwarf sample improves the determination of $q_H$ far more than it improves the marginalized upper limit on $g_{aee}$.
The stars within the core of 47 Tuc studied in this paper exhibit these larger values for mass of the hydrogen layer (see \cref{sec:appendix_additional_analyses}) even without input from the older white dwarfs studied in \citet{47tuc_deep_acs}.
Although one would not expect all the white dwarfs to have precisely the same envelope mass, hydrogen layers much thicker than observed are limited by the instability of the layer to hydrogen burning, so in a physical sense, these thicker layers do not rely on a fine tuning of the preceding stellar evolution to yield a layer much thinner than the physical maximum.
As the white dwarfs lack companions and all have very similar compositions and progenitor masses, we expect their evolution to be similar as well, resulting in thicker hydrogen layers and layers of similar thicknesses.

It is also important to understand what systematic uncertainties may remain.
Salaris {\em et al.}\ \citep{2009ApJ...692.1013S} outline how the cooling ages of white dwarfs of different effective temperatures and surface gravities are sensitive to assumptions on the underlying physics.
Our sample of white dwarfs have effective temperatures greater than 12,000~K and surface gravities very close to $10^8$~cm~s$^{-2}$.
In fact, Fig.~8 of \citet{2025MNRAS.541.1390S} shows that the white dwarfs in our sample, which are brighter than F150W2 of 24, all are expected to have nearly the same mass of about 0.55~M$_\odot$ and form from stars with a mass very close to 0.9~M$_\odot$.
\Citet{2009ApJ...692.1013S} argued that the strongest sensitivity is on neutrino cooling.
Variation of the neutrino emissivity by a factor of two above and below that expected from the Standard Model can result in a nearly 40\% deviation in the cooling timescale.
The key point is that \citet{2009ApJ...692.1013S} explicitly states that the uncertainties in the neutrino emissivity within the Standard Model are at most a few percent, a factor of one hundred smaller than the variation performed.
Instead of highlighting a potential remaining systematic uncertainty, this result highlights how our sample is exquisitely sensitive to physics beyond the Standard Model such as an axion.
Using the results of \citet{2015ApJ...809..141H}, we can estimate that our constraint on the axion corresponds to about a 60\% constraint on the neutrino emissivity or about a 10\% constraint on the cooling timescale using the sensitivity analysis of \citet{2009ApJ...692.1013S}.
The next largest uncertainty that they highlight is the software used to perform the evolution.
The two algorithms (from 2000-2008) do not share the same underlying physics inputs so it is difficult to quantify exactly where this sensitivity originates.
However, in the context of white dwarf evolution simulations two decades later, the agreement between different algorithms with the same physics input is better than a few percent for young white dwarfs of about 0.6~M$_\odot$ \citep{2013A&A...555A..96S,2026ApJS..283...41B} that we consider.

Salaris {\em et al.}\ \citep{2009ApJ...692.1013S} argue that the third most important uncertainty originates from the treatment of electron conduction.
In fact, Blouin {\em et al.}\ \citep{2020ApJ...899...46B} argued for a substantial revision of electron conductivity in the weakly degenerate layers of the white dwarf envelope.
Later Cassisi {\em et al.}\ \citep{2021A&A...654A.149C} compared white dwarf cooling times for the conductivity used in this work with those proposed in \citet{2020ApJ...899...46B}.
For the crucial regime of $10^{-1}>L/\textrm{L}_\odot > 10^{-3}$ where the white dwarfs lie in our sample, Cassisi {\em et al.}\ \citep{2021A&A...654A.149C} found differences up to about 10\% in the cooling timescale depending on the underlying assumptions from \citet{2020ApJ...899...46B}.
In fact for strongly damped corrections, the difference in cooling time is at most 4\% in this regime. This systematic uncertainty lies at a level similar to our axion constraint.
Finally, Salaris {\em et al.}\ \citep{2009ApJ...692.1013S} argue that the composition of the core of the white dwarf (they vary the oxygen abundance at the centre from 30\% to 90\%) could affect the cooling timescale at the 5\% level.
As our white dwarf evolution begins with an internal structure from a full stellar evolution of the progenitor, we argue that the uncertainty in the interior model is much smaller than considered by Salaris {\em et al.}\ \citep{2009ApJ...692.1013S}.
In particular, these stellar models have been benchmarked with observations of the post-main sequence stars of 47 Tuc (that are nearly identical to the white dwarf progenitors), and refs.~\citep{2015ApJ...810..127H,2016ApJ...826...88P,2016ApJ...830..139P} find detailed agreement that would make such a large discrepancy between the core composition within our model and the stars of 47 Tuc difficult to achieve.

Even with these provisos and without any of these potential improvements to the analysis, the bounds derived for the axion-electron coupling (and DFSZ axion mass) are an improvement upon previous bounds reported in the literature.
For comparison, a recent summary of bounds (and hints) for various axion couplings (including $g_{aee}$ and $g_{a\gamma\gamma}$) from stellar evolution is given by \citet{2022JCAP...02..035D} (see especially Table 1 of \citet{2022JCAP...02..035D}).
The new bound on $g_{aee}$ found in our work improves upon the leading bound from white dwarf cooling (derived using the Galactic white dwarf luminosity function) \citep{2014JCAP...10..069M}, as well as the bound recently found by \citet{2025PhyR.1117....1C} from the tip of the red giant branch of 21 globular clusters.
Of particular interest is that our newly derived (DFSZ axion) bound of $m_a \leq 2.85~\mathrm{meV}$ (at 95\% confidence) excludes the favoured range of $m_a \sim 4 - 10~\mathrm{meV}$ hinted at by the cooling anomaly reported for the white dwarf luminosity functions of the Galactic disc and halo \citep{2008ApJ...682L.109I,2018MNRAS.478.2569I}.
As the bound derived in this work is based on the analysis of only a single globular cluster, applying the analysis procedure used in our work to white dwarf cooling data from multiple globular clusters (like has recently been done for the tip of the red giant branch) could potentially probe even smaller values of the axion-electron coupling in the future.

Contemporaneous with the work of \citet{2025PhyR.1117....1C}, a separate analysis by \citet{2025JETPL.121..159T} of the tip of the red giant branch of seven globular clusters found a somewhat stronger bound, stated to be $g_{aee} < 0.52 \times 10^{-13}$ at 95\% confidence level.
\Citet{2025JETPL.121..159T} followed the same analysis procedure and used distances from the same data set as \citet{2025PhyR.1117....1C} did, with both \citet{2025PhyR.1117....1C} and \citet{2025JETPL.121..159T} following the procedure of \citet{2020A&A...644A.166S} and using the \gaia\ DR3 distances determined by \citet{2021MNRAS.505.5957B}.
It is thus strange that \citet{2025JETPL.121..159T}, which used a smaller number of globular clusters in the analysis, found a stronger bound than \citet{2025PhyR.1117....1C}.
To check the bound reported by \citet{2025JETPL.121..159T}, we performed the analysis described in that work ourselves and found that for two of the clusters, NGC 288 and NGC 6362, the observed bolometric magnitude of the tip of the red giant branch differed significantly (by more than $5~\sigma$) from the theoretical prediction even for $g_{aee} = 0$, which was the best-fitting value for both the combined analysis of all the clusters and the analysis of each of these two clusters individually.
This indicates that the model does not give a good fit for these two anomalous clusters.
We note that the tip bolometric magnitude value of $-2.770 \pm 0.065$ listed for NGC 288 in Table 1 of \citet{2025JETPL.121..159T} is a particularly notable outlier in that table and also differs significantly from the value of $-3.77 \pm 0.24$ found by \citet{2020A&A...644A.166S} for the same cluster.
Using a value of $-3.770$ instead of $-2.770$ for the tip bolometric magnitude of NGC 288 and excluding the cluster NGC 6362 from our reanalysis, we find a bound of $g_{aee} < 1.2 \times 10^{-13}$ at 95\% confidence level.
This is also the bound that we find if we exclude both NGC 288 and NGC 6362 from our reanalysis.

This bound is weaker than the bound of $g_{aee} < 0.95 \times 10^{-13}$ (at 95\% confidence level) found by \citet{2025PhyR.1117....1C} using 21 globular clusters with \gaia\ DR3 distances, though it is stronger than the older bound of $g_{aee} < 1.48 \times 10^{-13}$ previously found by \citet{2020A&A...644A.166S} using 22 globular clusters with distances based on the zero-age horizontal branch.
We find the revised bound from our reanalysis of the \citet{2025JETPL.121..159T} data to be a more sensible result than the bound reported by \citet{2025JETPL.121..159T} because, while improved distance measurements can give a stronger bound on the coupling even if a smaller number of globular clusters are included in the sample, an analysis of a much smaller number of globular clusters with distances from the same set of measurements should not give a notably stronger bound.
We thus do not consider the stated bound of $g_{aee} < 0.52 \times 10^{-13}$ from \citet{2025JETPL.121..159T} to be credible. We instead consider the bound of $g_{aee} < 0.95 \times 10^{-13}$ from \citet{2025PhyR.1117....1C} to be the leading bound from observations of the tip of the red giant branch and the overall previous leading bound on the axion-electron coupling prior to our current work.
Excluding the bound given by \citet{2025JETPL.121..159T}, our newly derived bound of $g_{aee} \leq 0.81 \times 10^{-13}$ from white dwarf cooling in 47 Tuc is thus the current leading bound on the axion-electron coupling.
Furthermore, it is subject to a different and smaller set of systematics than the red-giant bounds.

\section{Conclusions}  \label{sec:47tuc_axions_conclusions}

If axions exist and couple to electrons, then axions can be produced in the electron-degenerate core of a white dwarf via axion bremsstrahlung from electrons scattering off ions.
The emission of these axions from the white dwarf would provide an extra energy loss mechanism that would affect the rate of white dwarf cooling.
For a sufficiently large axion-electron coupling strength, this would have an observable effect on the shape of the white dwarf luminosity function, and hints of axions affecting the rate of white dwarf cooling have been suggested based on a cooling anomaly observed in white dwarf luminosity functions of the Galactic disc and halo \citep{2008ApJ...682L.109I,2018MNRAS.478.2569I}.

In this work, we analysed the cooling of white dwarfs in the globular cluster 47 Tuc to search for indirect evidence of axions produced via axion bremsstrahlung from electrons in the white dwarf interiors.
White dwarf cooling models were created by using the stellar evolution software MESA to perform simulations of white dwarf cooling that accounted for the energy loss associated with the emission of axions produced in the white dwarf. 
A suite of such models was created for a grid of parameter values that varied the axion-electron coupling strength, the white dwarf mass, and the thickness of the H envelope of the white dwarf.
These cooling models were compared to HST observations of young white dwarfs, for which axion emission is expected to be the dominant cooling mechanism, by performing an unbinned likelihood analysis similar to that of \citet{47tuc_deep_acs} and \citet{Goldsbury2016}.

The white dwarf mass and envelope thickness are nuisance parameters in the analysis which were constrained using prior information from \citet{47tuc_deep_acs}.
The results of that work, which analysed the cooling of older white dwarfs in 47 Tuc using a separate set of HST data that included white dwarfs old enough for their envelopes to have become convectively coupled to their cores, are particularly important for constraining the envelope thickness.
There is some degeneracy between the axion-electron coupling constant and the envelope thickness, so including this prior information for the envelope thickness in particular is important for constraining the value of the coupling constant.
From the analysis performed in this work, we find that models with thick envelopes and no axion emission are favoured.

We find that the axion-electron coupling constant is constrained to be $g_{aee} \leq 0.81 \times 10^{-13}$ at 95\% confidence.
This bound applies for any axion or axion-like particle model that includes a coupling to electrons. For a DFSZ model, this corresponds to a constraint on the axion mass (and angular parameter) of $m_a \sin^2\beta \leq 2.85~\mathrm{meV}$.
As this constraint was derived using a Bayesian likelihood analysis, the constraint is conditional on the adopted priors and white dwarf cooling models, and there are some uncertainties in the conductive opacities, core composition, and related microphysics in the modelling of white dwarf cooling that have not been fully propagated through the Bayesian analysis.
However, despite these caveats, the analysis presented in this work and the derived constraint are nevertheless an improvement upon earlier work in the literature.
The bound on $g_{aee}$ found in this work is more stringent than previous astrophysical bounds from both the Galactic white dwarf luminosity function \citep{2014JCAP...10..069M} and the tip of the red giant branch of globular clusters \citep{2025PhyR.1117....1C}.
We do not see evidence of any cooling anomaly in the cooling of 47 Tuc white dwarfs, and the updated bound on $g_{aee}$ excludes the values favoured for the hint of axions suggested by the white dwarf luminosity functions of the Galactic disc and halo.


\appendix
\clearpage
\section{Analyses of Individual Data Sets} \label[appendix]{sec:appendix_additional_analyses}

\subsection{Overview} \label[subappendix]{sec:appendix_47tuc_axions_overview}

In \cref{sec:appendix_47tuc_axions_wfc3_only} and \cref{sec:appendix_47tuc_axions_acs_only}, results are presented for analyses analogous to the analysis in the main text but performed separately for each of the data sets used in the combined analysis.
The results of fitting only the WFC3/UVIS data are presented in \cref{sec:appendix_47tuc_axions_wfc3_only}, while the results of fitting only the ACS/WFC data are presented in \cref{sec:appendix_47tuc_axions_acs_only}.
For each of these data sets, the analysis of that data in isolation is performed for three different cases of priors, which are summarised in \cref{tab:app_47tuc_axions_priors_cases}.
In the first case, all priors are taken to be uniform.
Case 2 (in comparison to case 1) tests the effect of using the Gaussian prior from \citet{Goldsbury2016} for the birthrate, with the other parameters taken to have the same uniform priors as in case 1.
These birthrate priors are $\Ndotwfc = 8.2 \pm 0.3~\mathrm{Myr}$ for the WFC3 data and $\Ndotacs = 2.61 \pm 0.07~\mathrm{Myr}$ for the ACS data.
Case 3 (in comparison to case 1) tests the effect of using the results of \citet{47tuc_deep_acs} for the priors on $\mwd$ and $\lqh$, with the birthrate taken to have the same uniform prior as in case 1.
In all three cases, a uniform prior is used for $m_a$.
Note that case 3 is the combination of priors that was used in the main text for the combined (WFC3 and ACS) analysis.
In \cref{sec:appendix_47tuc_axions_comparison}, the final results of the WFC3 only and ACS only analyses for case 3 of the priors are compared to the results of the combined analysis given in \cref{sec:47tuc_axions_results} of the main text, where the same set of priors was used.

\begin{figure}[!h]
    \centering
    \setlength\figwidth{0.625\textwidth}
    \includegraphics[width=\figwidth]{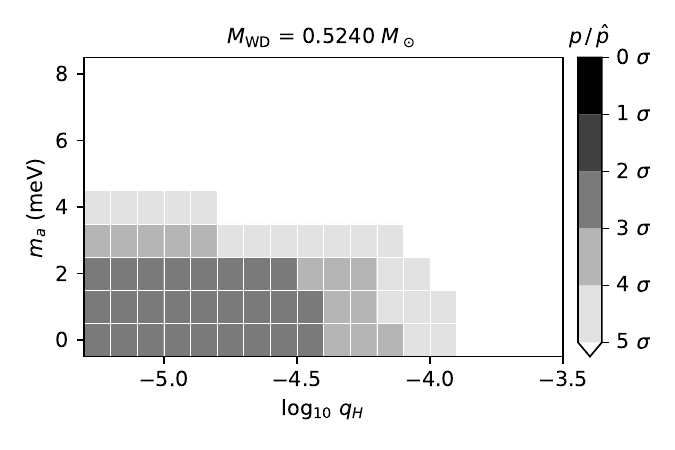}\\
    \includegraphics[width=\figwidth]{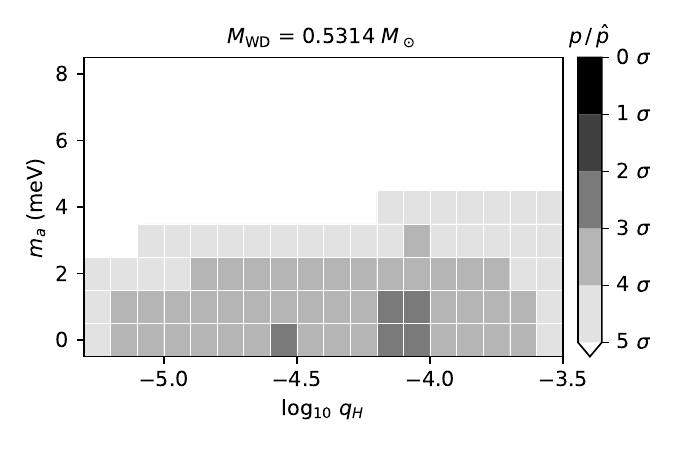}\\
    \includegraphics[width=\figwidth]{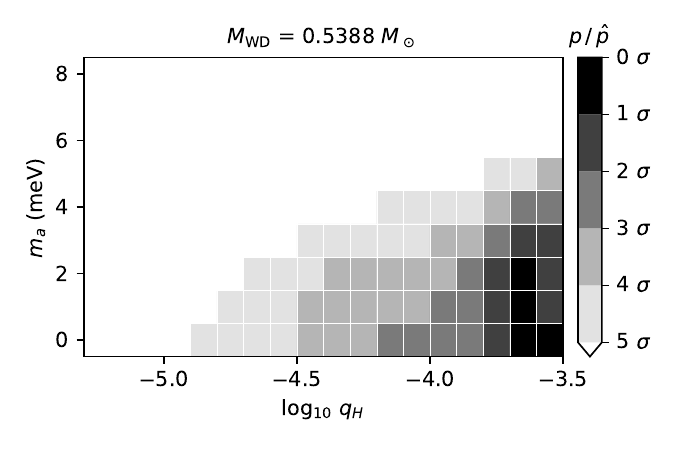}\\
    \caption{WFC3, case 1: joint posterior probability density distribution after marginalising over the birthrate.}
    \label{fig:app_47tuc_axions_joint_density_wfc3_case1}
\end{figure}

For the WFC3 only analyses in \cref{sec:appendix_47tuc_axions_wfc3_only}, plots of the posterior distributions and credible regions for each case of priors are shown in \cref{sec:appendix_47tuc_axions_postdists_wfc3}.
Three figures are shown for each case in \cref{sec:appendix_47tuc_axions_postdists_wfc3}.
The first figure shows slices of the three-dimensional joint posterior distribution after marginalising over the birthrate, the second figure shows the two-dimensional joint credible regions for $\lqh$ and $m_a$, and the third figure shows the one-dimensional marginal posterior distributions of all the parameters.
These figures are analogous to \cref{fig:47tuc_axions_joint_density}, \cref{fig:47tuc_axions_2d_CRs}, and \cref{fig:47tuc_axions_1d_marginal_distributions}, respectively, from \cref{sec:47tuc_axions_results} of the main text.
The corresponding figures for the WFC3 only analyses are:
i) \cref{fig:app_47tuc_axions_joint_density_wfc3_case1}, \cref{fig:app_47tuc_axions_2d_CRs_wfc3_case1}, and \cref{fig:app_47tuc_axions_1d_marginal_distributions_wfc3_case1} for WFC3 case 1,
ii) \cref{fig:app_47tuc_axions_joint_density_wfc3_case2}, \cref{fig:app_47tuc_axions_2d_CRs_wfc3_case2}, and \cref{fig:app_47tuc_axions_1d_marginal_distributions_wfc3_case2} for WFC3 case 2,
and iii) \cref{fig:app_47tuc_axions_joint_density_wfc3_case3}, \cref{fig:app_47tuc_axions_2d_CRs_wfc3_case3}, and \cref{fig:app_47tuc_axions_1d_marginal_distributions_wfc3_case3} for WFC3 case 3.
Note that for \cref{fig:app_47tuc_axions_joint_density_wfc3_case1,fig:app_47tuc_axions_joint_density_wfc3_case2,fig:app_47tuc_axions_joint_density_wfc3_case3} showing the joint posterior distribution (analogous to \cref{fig:47tuc_axions_joint_density}), all three $\mwd$ slices are shown instead of just the two slices shown in \cref{fig:47tuc_axions_joint_density}.
In \cref{sec:appendix_47tuc_axions_cumdists_wfc3_cases}, plots of the (inverse) cumulative luminosity functions and cumulative radial distributions for the best-fitting model of each case are shown.
The cumulative luminosity functions for the best-fitting models of all three cases are compared in \cref{fig:app_47tuc_axions_invLFs_wfc3_cases}, which shows a set of plots analogous to (the top row of) \cref{fig:47tuc_axions_invLFs}. Each row of \cref{fig:app_47tuc_axions_invLFs_wfc3_cases} corresponds to a separate case.
Likewise, \cref{fig:app_47tuc_axions_rad_cumdist_wfc3_cases} shows a set of plots analogous to \cref{fig:47tuc_axions_rad_cumdist}, with each row showing the cumulative radial distribution for the best-fitting model of a particular case.
Tables summarising the best-fitting model results for each WFC3 case, \cref{tab:app_47tuc_axions_best_model_params_wfc3_cases} and \cref{tab:app_47tuc_axions_KS_test_results_wfc3_cases}, are also given in \cref{sec:appendix_47tuc_axions_cumdists_wfc3_cases}.
\Cref{tab:app_47tuc_axions_best_model_params_wfc3_cases} gives the best-fitting model parameters for each case.
\Cref{tab:app_47tuc_axions_KS_test_results_wfc3_cases} gives the \textit{p}-values from KS tests comparing the distributions predicted by the best-fitting model of each case to the corresponding empirical distribution.
Finally, the results of the WFC3 only analyses are also discussed in \cref{sec:appendix_47tuc_axions_cumdists_wfc3_cases}.

\begin{table}
    \centering
    \begin{tabular}{c c c c}
    \toprule
        ~ & $\dot{N}$ & $M_\mathrm{WD}$ & $\log_{10} q_H$ \\
    \midrule
        Case 1 & Uniform & Uniform & Uniform \\
        Case 2 & Gaussian & Uniform & Uniform \\
        Case 3 & Uniform & \Citet{47tuc_deep_acs} & \Citet{47tuc_deep_acs} \\
    \bottomrule
    \end{tabular}
    \caption{Summary of priors used in each case for the analysis of only the WFC3 data and only the ACS data.
    The Gaussian prior for the birthrate ($\dot{N}$) is the same prior used for the corresponding field in \citet{Goldsbury2016}.
    The \citet{47tuc_deep_acs} priors for the white dwarf mass ($M_\mathrm{WD}$) and envelope thickness parameter ($\log_{10} q_H$) refer to the joint posterior distribution from \citet{47tuc_deep_acs} for the standard diffusion scenario after marginalising over the birthrate.}
    \label{tab:app_47tuc_axions_priors_cases}
\end{table}

The structure of \cref{sec:appendix_47tuc_axions_acs_only} for the ACS only analyses is the same as \cref{sec:appendix_47tuc_axions_wfc3_only}.
Plots of the posterior distributions and credible regions are given in \cref{sec:appendix_47tuc_axions_postdists_acs} for each case.
Like for the WFC3 only analysis, the three figures shown for each case of the ACS only analysis in \cref{sec:appendix_47tuc_axions_postdists_acs} are analogous to \cref{fig:47tuc_axions_joint_density}, \cref{fig:47tuc_axions_2d_CRs}, and \cref{fig:47tuc_axions_1d_marginal_distributions} from \cref{sec:47tuc_axions_results} of the main text.
The equivalent figures for the ACS only analyses are:
i) \cref{fig:app_47tuc_axions_joint_density_acs_case1}, \cref{fig:app_47tuc_axions_2d_CRs_acs_case1}, and \cref{fig:app_47tuc_axions_1d_marginal_distributions_acs_case1} for ACS case 1,
ii) \cref{fig:app_47tuc_axions_joint_density_acs_case2}, \cref{fig:app_47tuc_axions_2d_CRs_acs_case2}, and \cref{fig:app_47tuc_axions_1d_marginal_distributions_acs_case2} for ACS case 2,
and iii) \cref{fig:app_47tuc_axions_joint_density_acs_case3}, \cref{fig:app_47tuc_axions_2d_CRs_acs_case3}, and \cref{fig:app_47tuc_axions_1d_marginal_distributions_acs_case3} for ACS case 3.
Results for the best-fitting models of all of these cases are given in \cref{sec:appendix_47tuc_axions_cumdists_acs_cases}, including plots of the cumulative luminosity functions and tables summarising the results.
The cumulative luminosity functions for the best-fitting models of the three ACS cases are compared in \cref{fig:app_47tuc_axions_invLFs_acs_cases}.
This plot is analogous to \cref{fig:app_47tuc_axions_invLFs_wfc3_cases} of the WFC3 only analysis and contains a set of plots analogous to (the bottom row of) \cref{fig:47tuc_axions_invLFs}, with each row of \cref{fig:app_47tuc_axions_invLFs_acs_cases} corresponding to a different case.
Note that for the ACS only analysis, there is no radial distribution figure (like \cref{fig:app_47tuc_axions_rad_cumdist_wfc3_cases} or \cref{fig:47tuc_axions_rad_cumdist}) because the ACS analysis does not include $R$ dependence.
\Cref{tab:app_47tuc_axions_best_model_params_acs_cases} gives the best-fitting model parameters for the ACS cases, and \cref{tab:app_47tuc_axions_KS_test_results_acs_cases} gives the \text{p}-values of KS tests comparing the best-fitting model distributions to the corresponding empirical distributions.
\Cref{tab:app_47tuc_axions_best_model_params_acs_cases} and \cref{tab:app_47tuc_axions_KS_test_results_acs_cases} are both given in \cref{sec:appendix_47tuc_axions_cumdists_acs_cases}, along with a discussion of the results.

\clearpage
\subsection{WFC3/UVIS Data Only} \label[subappendix]{sec:appendix_47tuc_axions_wfc3_only}

\subsubsection{Posterior Distributions} \label[subsubappendix]{sec:appendix_47tuc_axions_postdists_wfc3}


\clearpage

\begin{figure}
    \centering
    \includegraphics[width=0.75\textwidth]{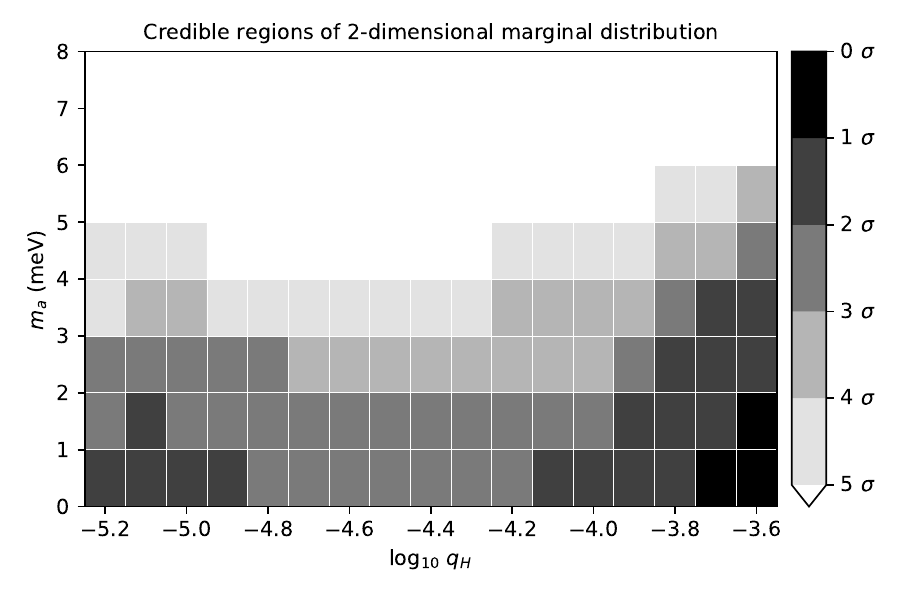}
    \caption{WFC3, case 1: two-dimensional joint credible regions of axion mass ($m_a$) and envelope thickness ($q_H$) after marginalising over the other parameters.}
    \label{fig:app_47tuc_axions_2d_CRs_wfc3_case1}
\end{figure}

\begin{figure}
    \centering
    \begin{subfigure}[t]{0.49\textwidth}
        \includegraphics[width=\textwidth]{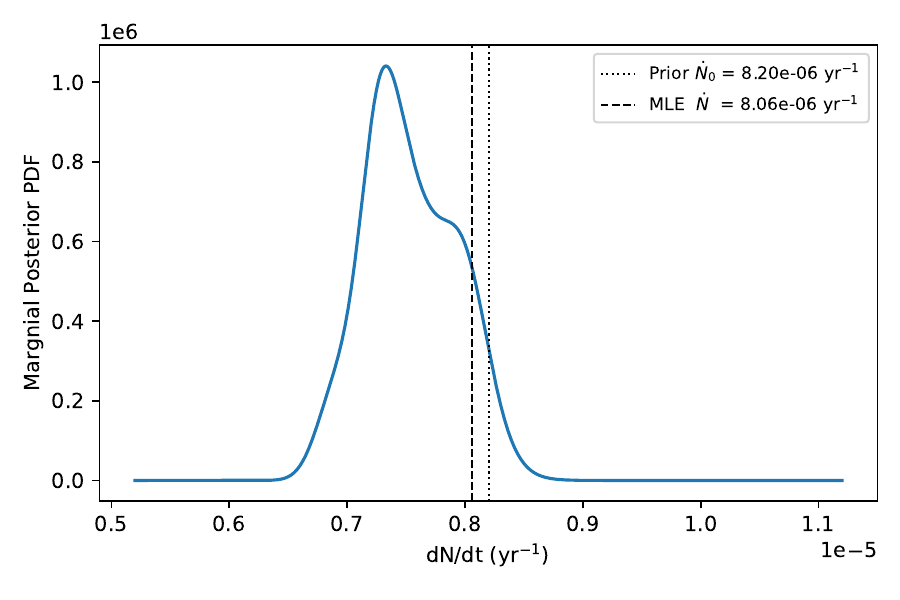}
        \caption{Birthrate}
        \label{fig:app_47tuc_axions_1d_marg_dist_Ndot_wfc3_case1}
    \end{subfigure}
    \hfill
    \begin{subfigure}[t]{0.49\textwidth}
        \includegraphics[width=\textwidth]{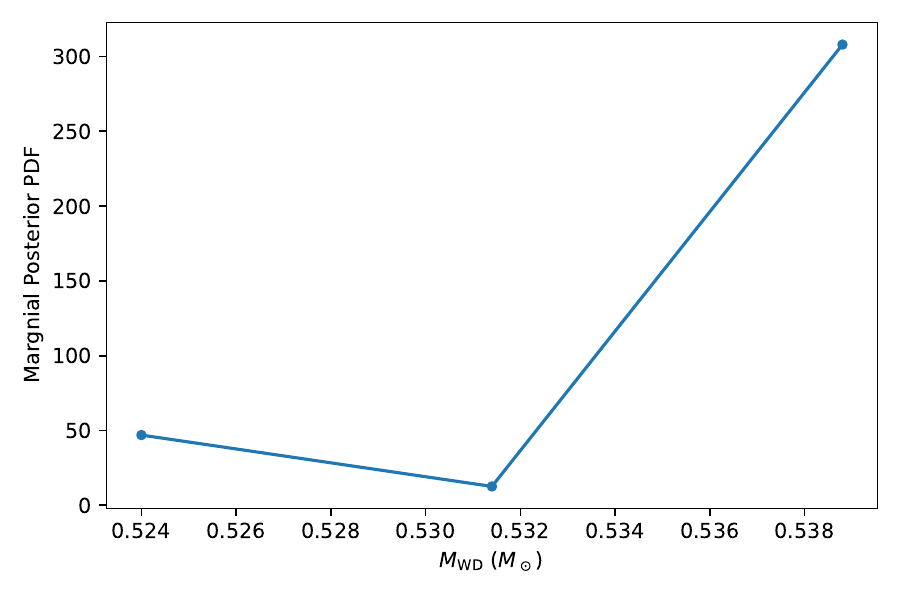}
        \caption{White dwarf mass}
        \label{fig:app_47tuc_axions_1d_marg_dist_MWD_wfc3_case1}
    \end{subfigure}
    \\[1.5ex]
    \begin{subfigure}[t]{0.49\textwidth}
        \includegraphics[width=\textwidth]{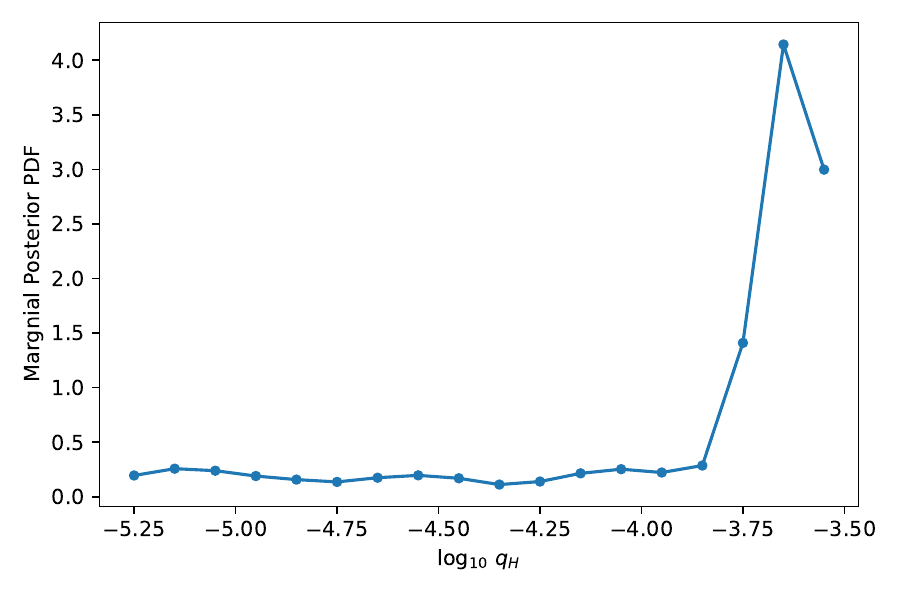}
        \caption{Envelope thickness}
        \label{fig:app_47tuc_axions_1d_marg_dist_lqh_wfc3_case1}
    \end{subfigure}
    \hfill
    \begin{subfigure}[t]{0.49\textwidth}
        \includegraphics[width=\textwidth]{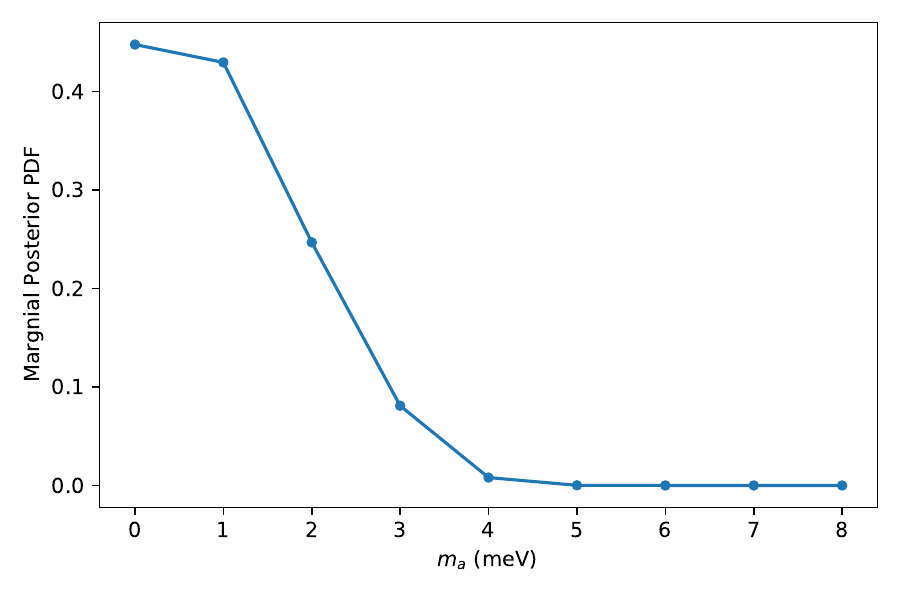}
        \caption{Axion mass}
        \label{fig:app_47tuc_axions_1d_marg_dist_ma_wfc3_case1}
    \end{subfigure}
    \\[1.5ex]
    \caption{WFC3, case 1: one-dimensional posterior density distributions for each parameter after marginalising over all other model parameters.}
    \label{fig:app_47tuc_axions_1d_marginal_distributions_wfc3_case1}
\end{figure}
\clearpage


\begin{figure}[!h]
    \centering
    \setlength\figwidth{0.625\textwidth}
    \includegraphics[width=\figwidth]{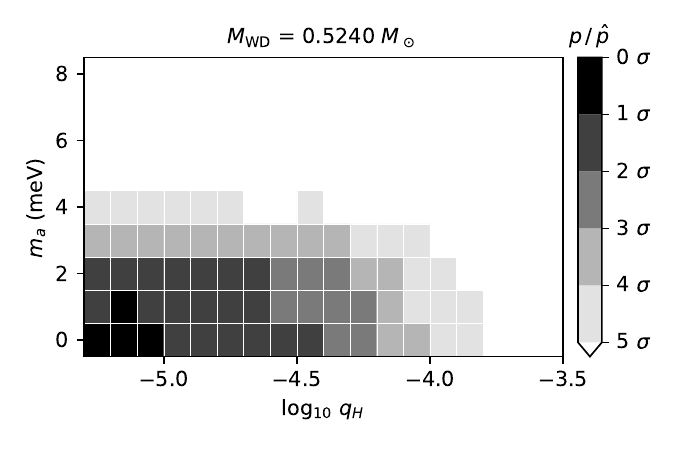}\\
    \includegraphics[width=\figwidth]{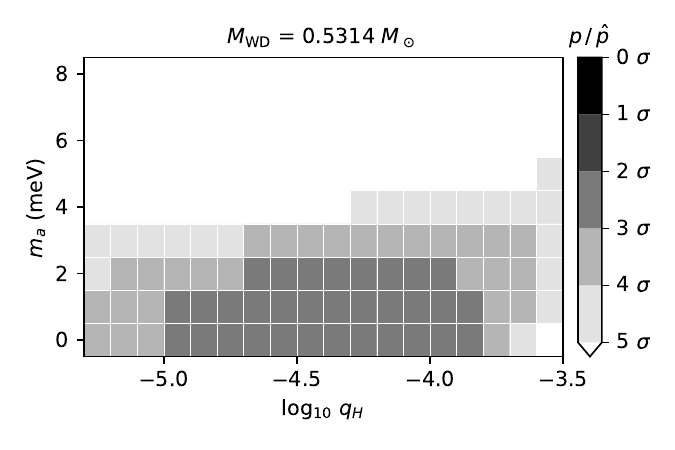}\\
    \includegraphics[width=\figwidth]{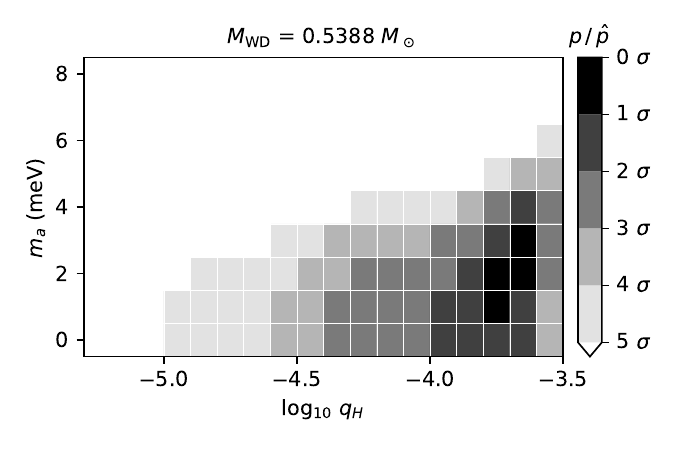}\\
    \caption{WFC3, case 2: joint posterior probability density distribution after marginalising over the birthrate.}
    \label{fig:app_47tuc_axions_joint_density_wfc3_case2}
\end{figure}
\clearpage

\begin{figure}
    \centering
    \includegraphics[width=0.75\textwidth]{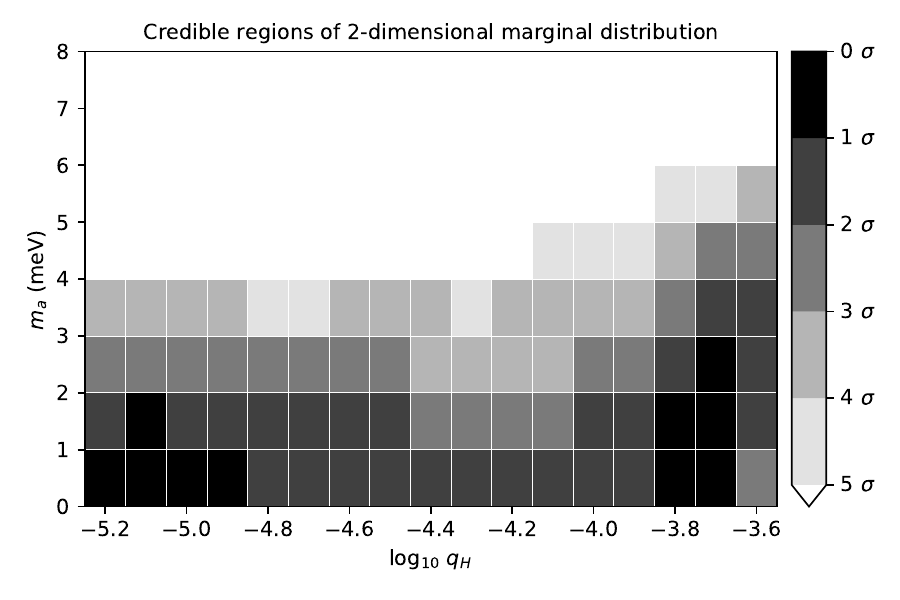}
    \caption{WFC3, case 2: two-dimensional joint credible regions of axion mass ($m_a$) and envelope thickness ($q_H$) after marginalising over the other parameters.}
    \label{fig:app_47tuc_axions_2d_CRs_wfc3_case2}
\end{figure}

\begin{figure}
    \centering
    \begin{subfigure}[t]{0.49\textwidth}
        \includegraphics[width=\textwidth]{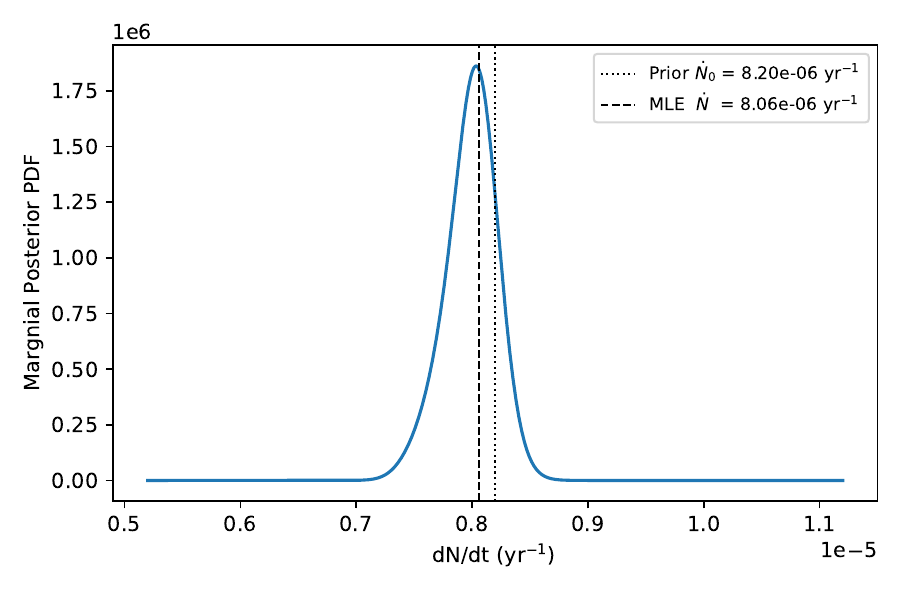}
        \caption{Birthrate}
        \label{fig:app_47tuc_axions_1d_marg_dist_Ndot_wfc3_case2}
    \end{subfigure}
    \hfill
    \begin{subfigure}[t]{0.49\textwidth}
        \includegraphics[width=\textwidth]{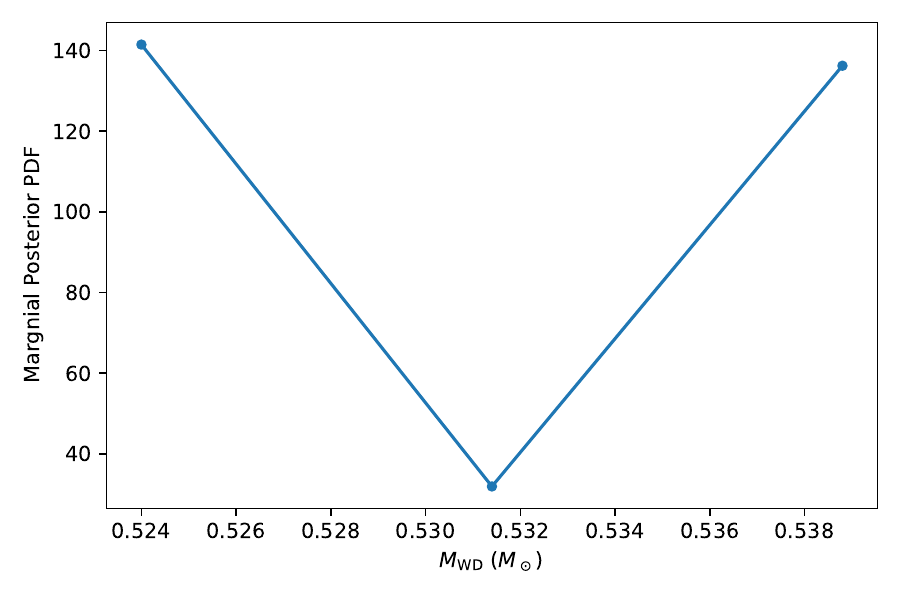}
        \caption{White dwarf mass}
        \label{fig:app_47tuc_axions_1d_marg_dist_MWD_wfc3_case2}
    \end{subfigure}
    \\[1.5ex]
    \begin{subfigure}[t]{0.49\textwidth}
        \includegraphics[width=\textwidth]{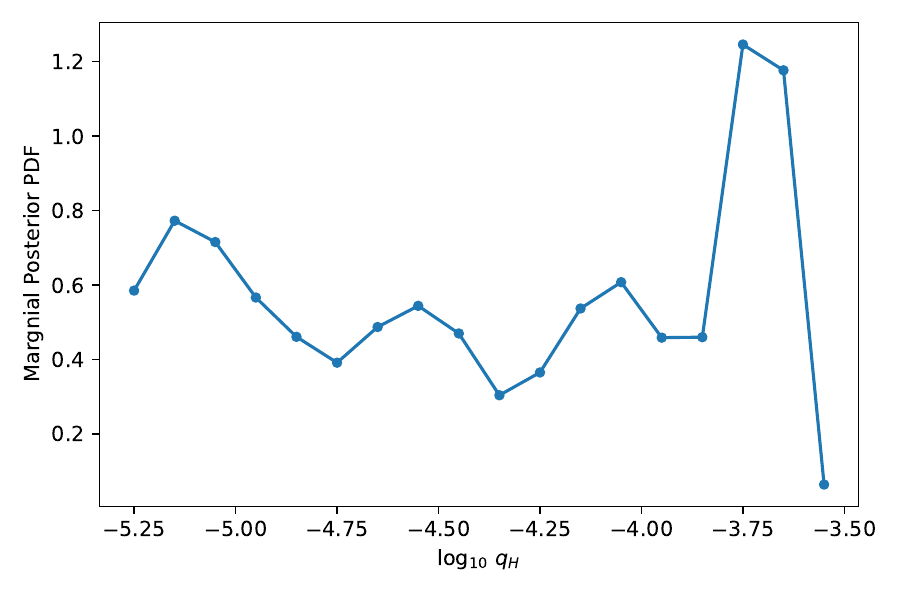}
        \caption{Envelope thickness}
        \label{fig:app_47tuc_axions_1d_marg_dist_lqh_wfc3_case2}
    \end{subfigure}
    \hfill
    \begin{subfigure}[t]{0.49\textwidth}
        \includegraphics[width=\textwidth]{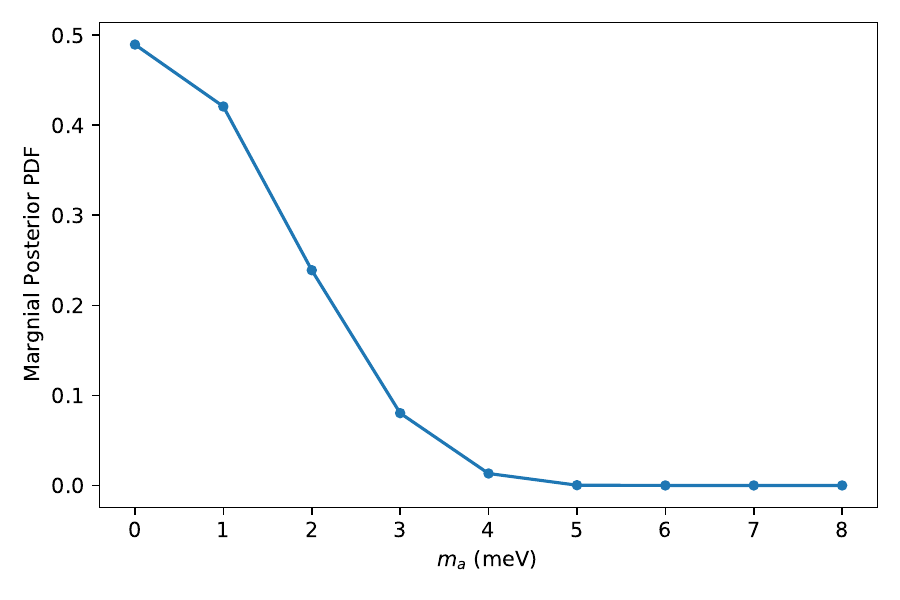}
        \caption{Axion mass}
        \label{fig:app_47tuc_axions_1d_marg_dist_ma_wfc3_case2}
    \end{subfigure}
    \\[1.5ex]
    \caption{WFC3, case 2: one-dimensional posterior density distributions for each parameter after marginalising over all other model parameters.}
    \label{fig:app_47tuc_axions_1d_marginal_distributions_wfc3_case2}
\end{figure}
\clearpage


\begin{figure}[!h]
    \centering
    \setlength\figwidth{0.625\textwidth}
    \includegraphics[width=\figwidth]{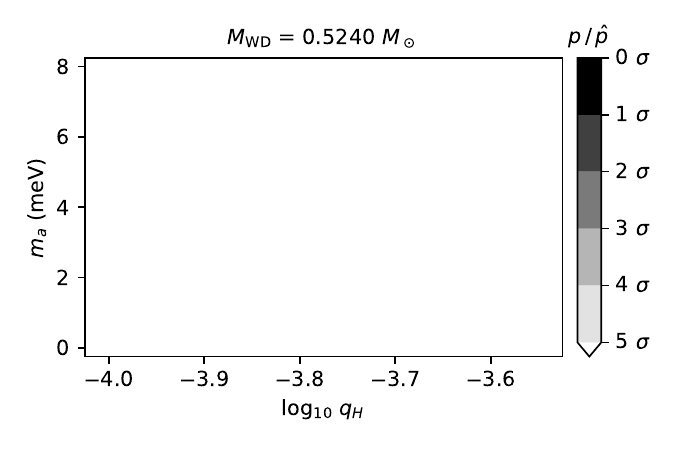}\\
    \includegraphics[width=\figwidth]{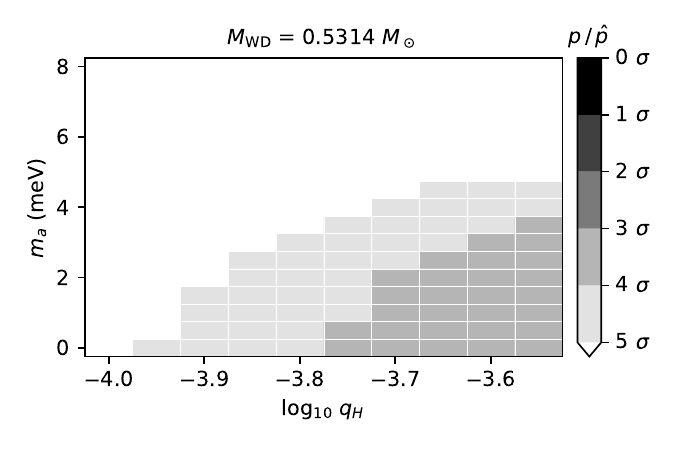}\\
    \includegraphics[width=\figwidth]{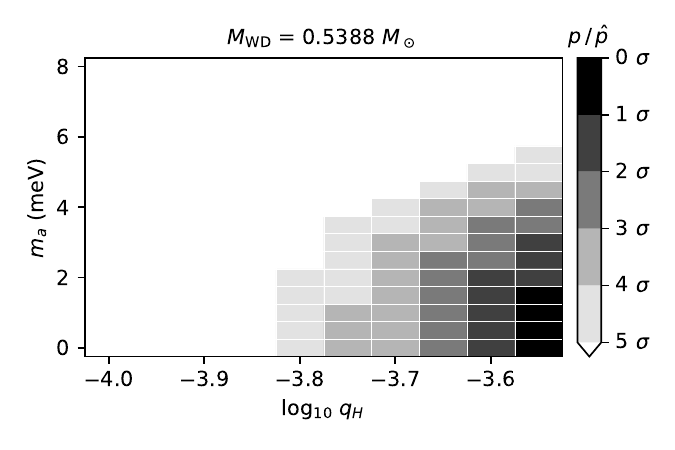}\\
    \caption{WFC3, case 3: joint posterior probability density distribution after marginalising over the birthrate.}
    \label{fig:app_47tuc_axions_joint_density_wfc3_case3}
\end{figure}
\clearpage

\begin{figure}
    \centering
    \includegraphics[width=0.75\textwidth]{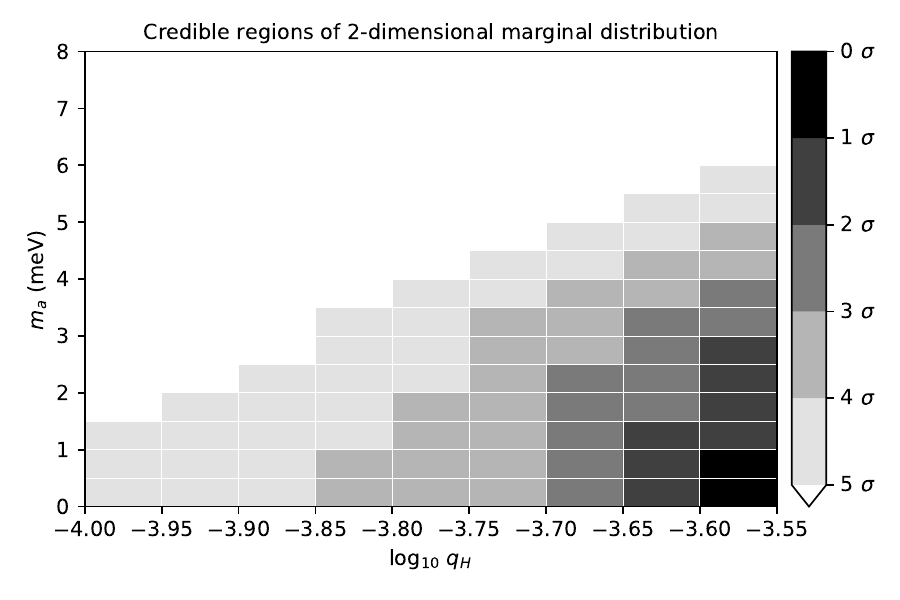}
    \caption{WFC3, case 3: two-dimensional joint credible regions of axion mass ($m_a$) and envelope thickness ($q_H$) after marginalising over the other parameters.}
    \label{fig:app_47tuc_axions_2d_CRs_wfc3_case3}
\end{figure}

\begin{figure}
    \centering
    \begin{subfigure}[t]{0.49\textwidth}
        \includegraphics[width=\textwidth]{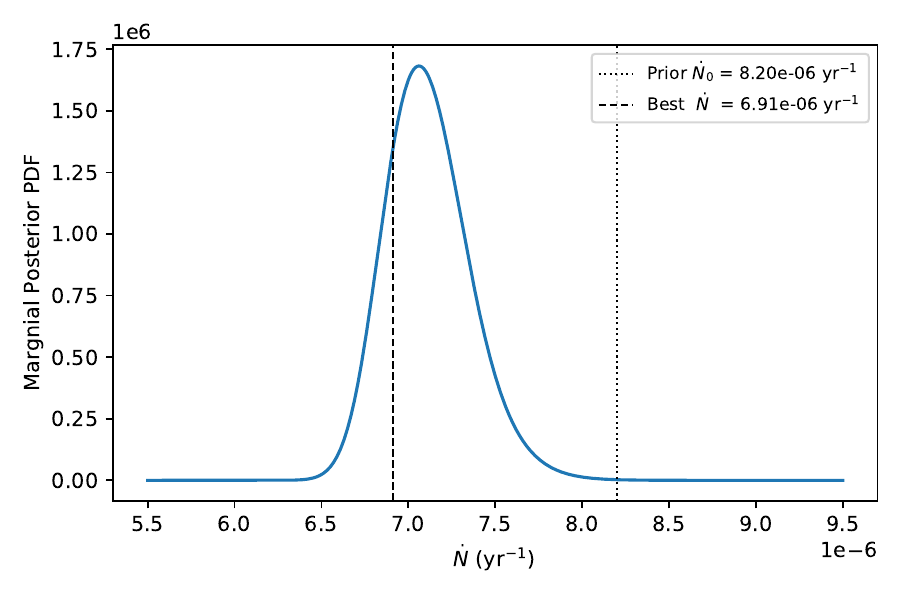}
        \caption{Birthrate}
        \label{fig:app_47tuc_axions_1d_marg_dist_Ndot_wfc3_case3}
    \end{subfigure}
    \hfill
    \begin{subfigure}[t]{0.49\textwidth}
        \includegraphics[width=\textwidth]{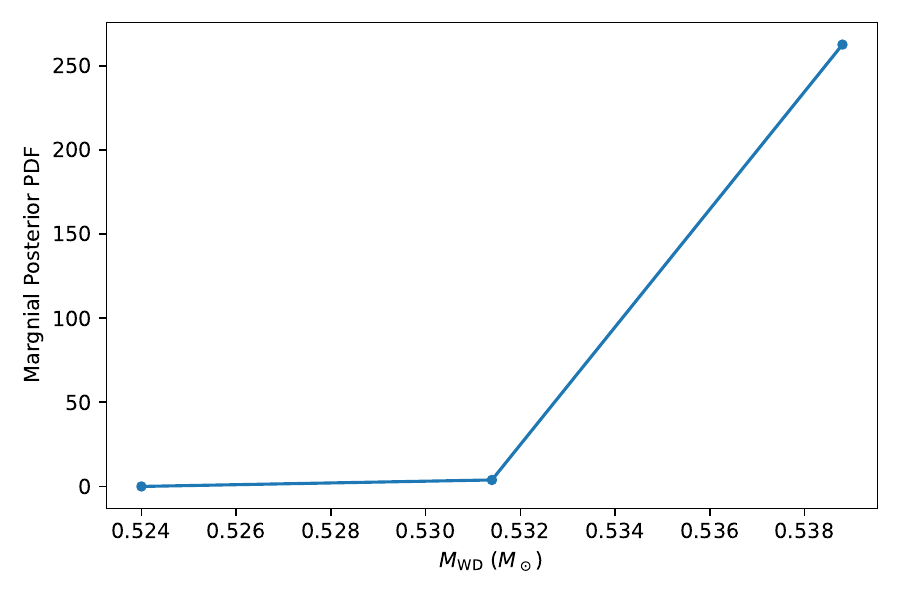}
        \caption{White dwarf mass}
        \label{fig:app_47tuc_axions_1d_marg_dist_MWD_wfc3_case3}
    \end{subfigure}
    \\[1.5ex]
    \begin{subfigure}[t]{0.49\textwidth}
        \includegraphics[width=\textwidth]{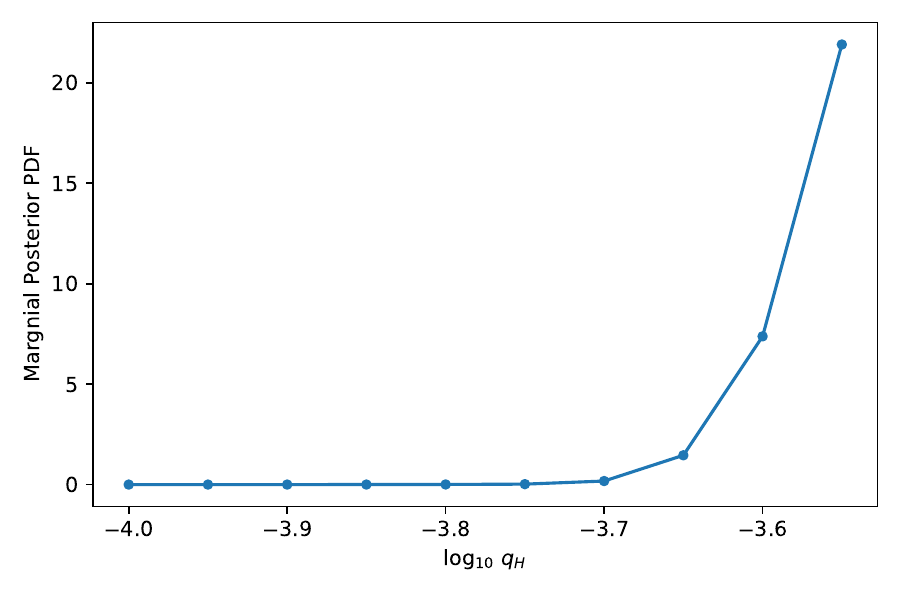}
        \caption{Envelope thickness}
        \label{fig:app_47tuc_axions_1d_marg_dist_lqh_wfc3_case3}
    \end{subfigure}
    \hfill
    \begin{subfigure}[t]{0.49\textwidth}
        \includegraphics[width=\textwidth]{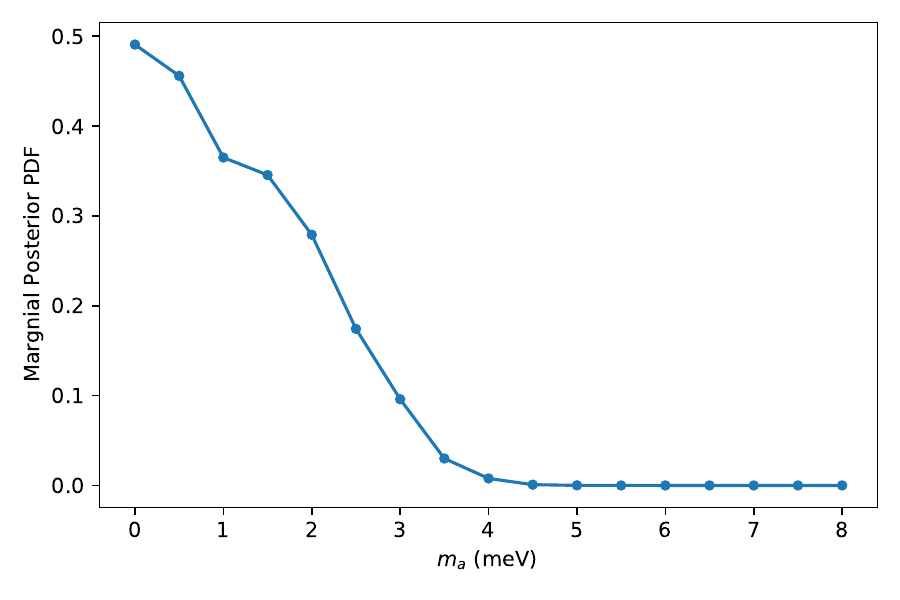}
        \caption{Axion mass}
        \label{fig:app_47tuc_axions_1d_marg_dist_ma_wfc3_case3}
    \end{subfigure}
    \\[1.5ex]
    \caption{WFC3, case 3: one-dimensional posterior density distributions for each parameter after marginalising over all other model parameters.}
    \label{fig:app_47tuc_axions_1d_marginal_distributions_wfc3_case3}
\end{figure}
\clearpage

\subsubsection{Best-Fitting Models Comparison} \label[subsubappendix]{sec:appendix_47tuc_axions_cumdists_wfc3_cases}

\begin{figure}[!h]
    \centering
    \setlength{\figwidth}{0.35\textwidth} 
    \begin{subfigure}[t]{0.49\textwidth}
        \centering
        \includegraphics[width=\figwidth]{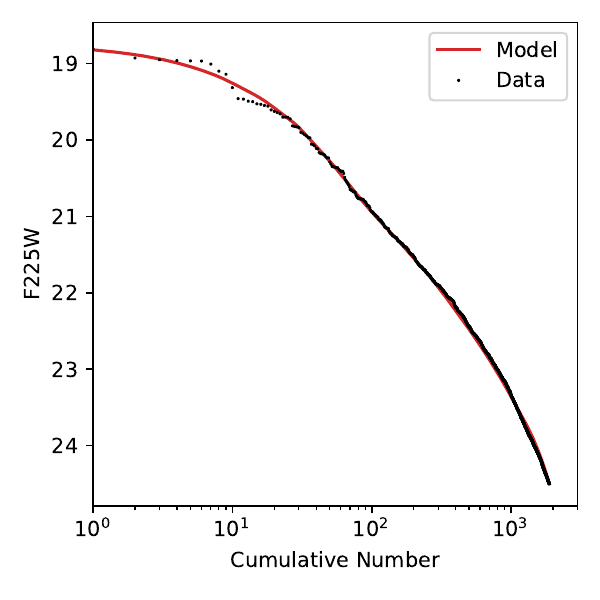}
        \caption{Case 1, F225W}
        \label{fig:app_47tuc_axions_invLFs_wfc3_f225w_case1}
    \end{subfigure}
    \hfill
    \begin{subfigure}[t]{0.49\textwidth}
        \centering
        \includegraphics[width=\figwidth]{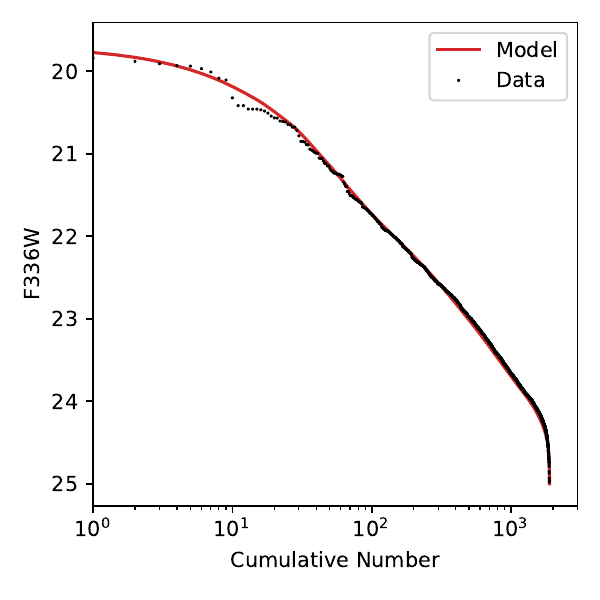}
        \caption{Case 1, F336W}
        \label{fig:app_47tuc_axions_invLFs_wfc3_f336w_case1}
    \end{subfigure}\\[1.5ex]
    \begin{subfigure}[t]{0.49\textwidth}
        \centering
        \includegraphics[width=\figwidth]{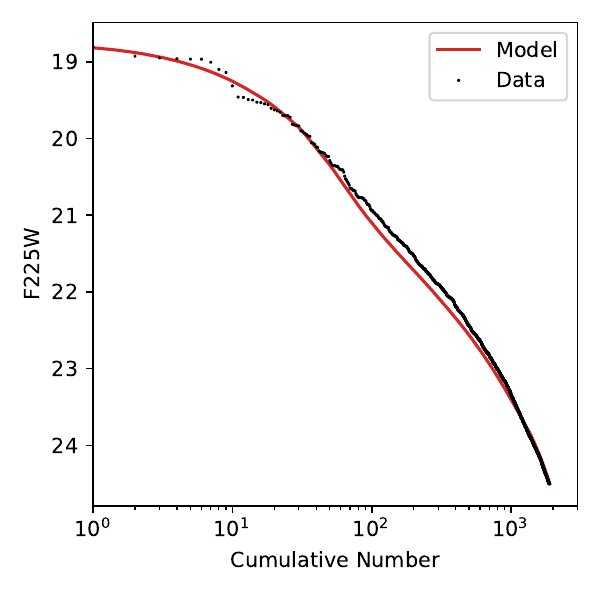}
        \caption{Case 2, F225W}
        \label{fig:app_47tuc_axions_invLFs_wfc3_f225w_case2}
    \end{subfigure}
    \hfill
    \begin{subfigure}[t]{0.49\textwidth}
        \centering
        \includegraphics[width=\figwidth]{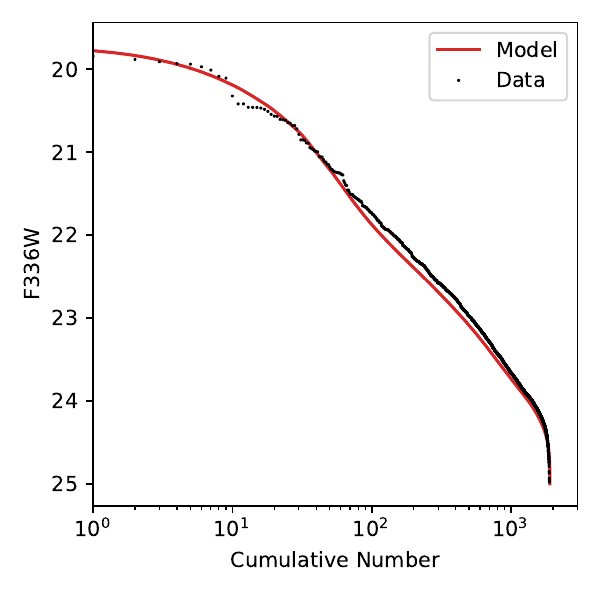}
        \caption{Case 2, F336W}
        \label{fig:app_47tuc_axions_invLFs_wfc3_f336w_case2}
    \end{subfigure}\\[1.5ex]
    \begin{subfigure}[t]{0.49\textwidth}
        \centering
        \includegraphics[width=\figwidth]{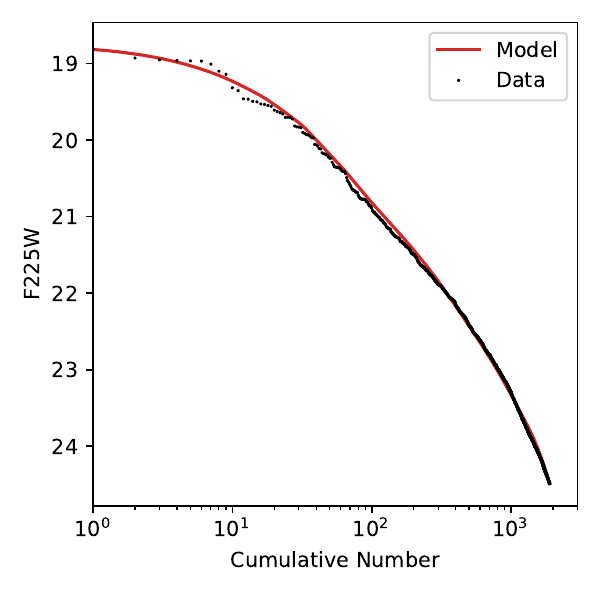}
        \caption{Case 3, F225W}
        \label{fig:app_47tuc_axions_invLFs_wfc3_f225w_case3}
    \end{subfigure}
    \hfill
    \begin{subfigure}[t]{0.49\textwidth}
        \centering
        \includegraphics[width=\figwidth]{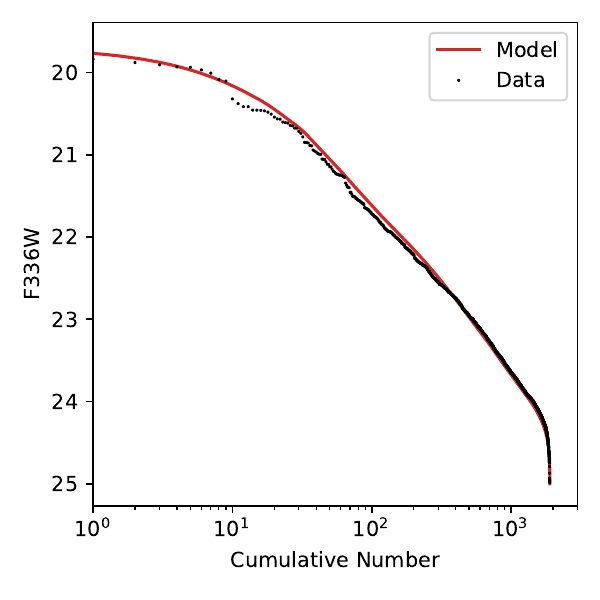}
        \caption{Case 3, F336W}
        \label{fig:app_47tuc_axions_invLFs_wfc3_f336w_case3}
    \end{subfigure}
    \\[1.5ex]
    \caption{WFC3, all cases: inverse cumulative luminosity function of optimal model (red curve) compared to the data (black points) for both F225W (left column) and F336W (right column).
    Each row shows a different case: case 1 (top), case 2 (middle), and case 3 (bottom).}
    \label{fig:app_47tuc_axions_invLFs_wfc3_cases}
\end{figure}
\clearpage

\begin{figure}
    \centering
    \setlength{\figwidth}{0.44\textwidth}
    \begin{subfigure}[t]{\figwidth}
        \includegraphics[width=\textwidth]{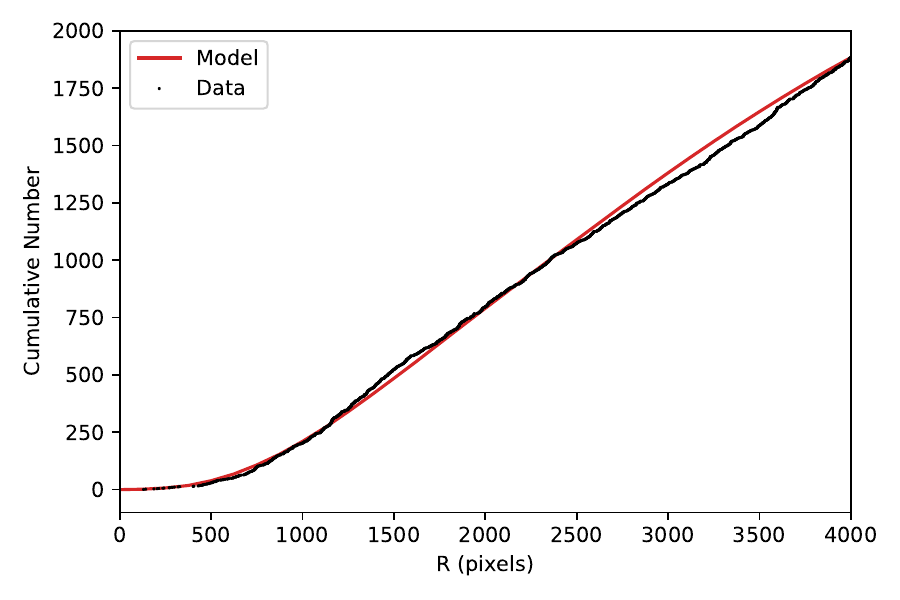}
        \caption{Case 1}
        \label{fig:app_47tuc_axions_rad_cumdist_wfc3_case1}
    \end{subfigure}\\[1.5ex]
    \begin{subfigure}[t]{\figwidth}
        \includegraphics[width=\textwidth]{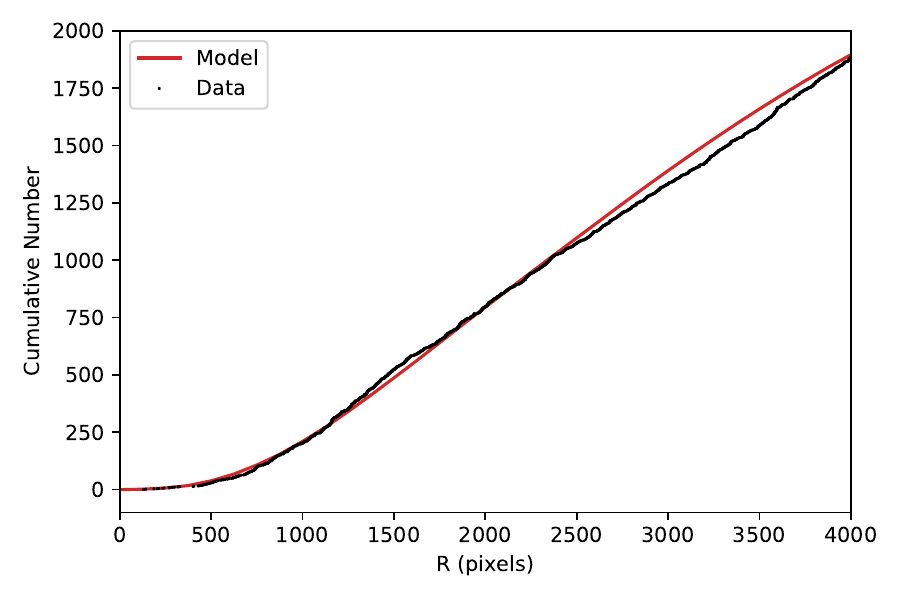}
        \caption{Case 2}
        \label{fig:app_47tuc_axions_rad_cumdist_wfc3_case2}
    \end{subfigure}\\[1.5ex]
    \begin{subfigure}[t]{\figwidth}
        \includegraphics[width=\textwidth]{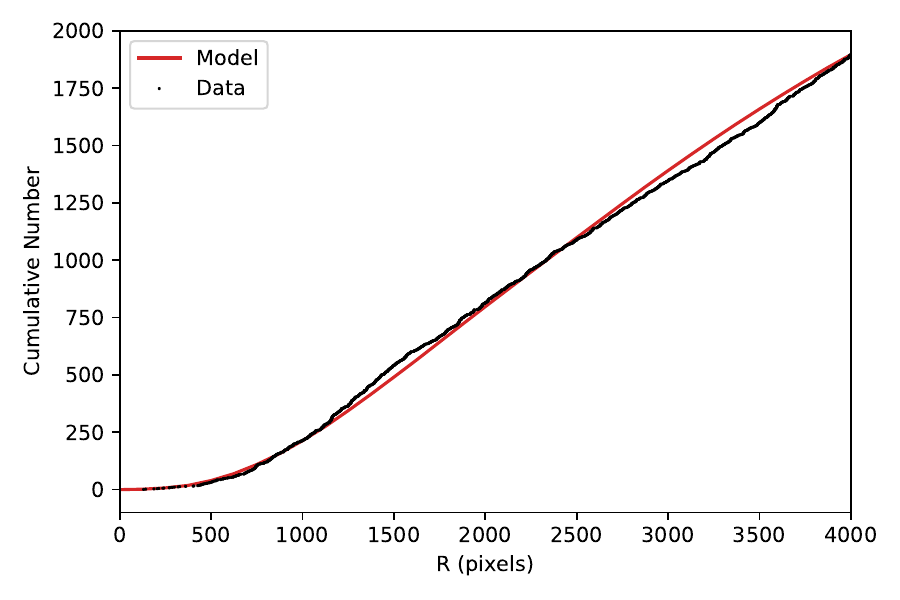}
        \caption{Case 3}
        \label{fig:app_47tuc_axions_rad_cumdist_wfc3_case3}
    \end{subfigure}\\
    \caption{WFC3, all cases: cumulative number distribution of radial distance ($R$) from cluster centre for optimal model (red curve) and WFC3/UVIS data (black points). Each row shows a different case: case 1 (top), case 2 (middle), and case 3 (bottom).}
    \label{fig:app_47tuc_axions_rad_cumdist_wfc3_cases}
\end{figure}

\begin{table}
    \centering
    \begin{tabular}{c c c c c}
        \toprule
        ~ & $\dot{N}~(\mathrm{Myr}^{-1})$ & $M_\mathrm{WD}~(M_\odot)$ & $\log_{10} q_H$ & $m_a~(\mathrm{meV})$ \\
        \midrule
        Case 1 & $7.28$ & $0.5388$ & $-3.65$ & $1.0$ \\
        Case 2 & $8.06$ & $0.5240$ & $-5.15$ & $0.0$ \\
        Case 3 & $6.91$ & $0.5388$ & $-3.55$ & $0.0$ \\
        \bottomrule
    \end{tabular}
    \caption{Parameter values of the optimal model for the different WFC3 cases. This is the combination of values that maximises the joint posterior distribution on the parameter grid.}
    \label{tab:app_47tuc_axions_best_model_params_wfc3_cases}
\end{table}
\clearpage

\begin{table}
    \centering
    \begin{tabular}{c c c c}
    \toprule
        ~ & F225W & F336W & $R$ \\
    \midrule
        Case 1 & 0.0343 & 0.0124 & 0.0097 \\
        Case 2 & 0.0057 & $4 \times 10^{-6}$ & 0.0087 \\
        Case 3 & 0.0362 & 0.0132 & 0.0156 \\
    \bottomrule
    \end{tabular}
    \caption{Results of KS tests for the different WFC3 cases. The \textit{p}-values are reported for KS tests comparing the one-dimensional marginal cumulative probability distribution functions predicted by the optimal model for each case to the corresponding empirical distribution.}
    \label{tab:app_47tuc_axions_KS_test_results_wfc3_cases}
\end{table}

\Cref{tab:app_47tuc_axions_best_model_params_wfc3_cases} summarises the parameter values of the optimal model for each of the cases considered in the analysis of the WFC3 data alone. 
The optimal models for cases 1 and 3 (which both use uniform birthrate priors) correspond to very similar parameter values, both notably favouring large $\lqh$ values.
This is in contrast to case 2 (which used a Gaussian birthrate prior), which favours a much smaller $\lqh$ value and larger $\dot{N}$. The case 2 result is pushed to larger $\dot{N}$ values by the Gaussian birthrate prior used in this case, which seems to overestimate the birthrate. Forcing a birthrate value that is too large drives the fit to smaller $\lqh$ values (for the data space considered in this work).
Note that regardless of which priors are used, all of the cases favour $m_a$ values near zero.

\Cref{tab:app_47tuc_axions_KS_test_results_wfc3_cases} summarises the results of the one-sample KS tests for the WFC3 cases.
Cases 1 and 3 have very similar \textit{p}-values (when comparing the same data variable), and these \textit{p}-values are reasonable for all of the data variables (i.e. F225W, F336W, and $R$). This is particularly the case for the magnitudes (F225W and F336W); the \textit{p}-value for $R$ is somewhat better for case 3 than case 1, though both are reasonable.
The KS test \textit{p}-values are not a direct measure of goodness-of-fit, they simply provide a check as to whether it is reasonably likely that the data are drawn from the model distribution (and can be used to reject this null hypothesis in the case of a poor fit if the \textit{p}-values are sufficiently small).
The similarity of the KS test results for cases 1 and 3 is expected given the similarity of the best-fitting parameter values summarised in \cref{tab:app_47tuc_axions_best_model_params_wfc3_cases}.
The smaller \textit{p}-values for case 2 (which used a Gaussian instead of uniform birthrate prior) compared to case 1 (and case 3) are indicative of a worse fit.
This is particularly notable in the \textit{p}-values for the F336W magnitude (and to a lesser extent the F225W magnitude), with the case 2 value for F336W so small ($< 10^{-4}$) that we can reject the null hypothesis that the empirical F336W values are drawn from the corresponding distribution predicted by the best-fitting model of case 2.
This suggests that the Gaussian WFC3 birthrate prior from \citet{Goldsbury2016} is not an appropriate choice, at least for the data space used in this work.
This may be due to the different definitions of the data space between this work and \citet{Goldsbury2016}; axion emission can affect white dwarf cooling until later cooling times than the neutrino emission that was the focus of \citet{Goldsbury2016}, so we use a data space that extends to larger magnitudes (corresponding to older white dwarfs) than did \citet{Goldsbury2016}.
The older white dwarfs included in our data space could have a different birthrate than the younger white dwarfs.

\clearpage 
\subsection{ACS/WFC Data Only} \label[subappendix]{sec:appendix_47tuc_axions_acs_only}

\subsubsection{Posterior Distributions} \label[subsubappendix]{sec:appendix_47tuc_axions_postdists_acs}


\begin{figure}[!h]
    \centering
    \setlength{\figwidth}{0.625\textwidth}
    \includegraphics[width=\figwidth]{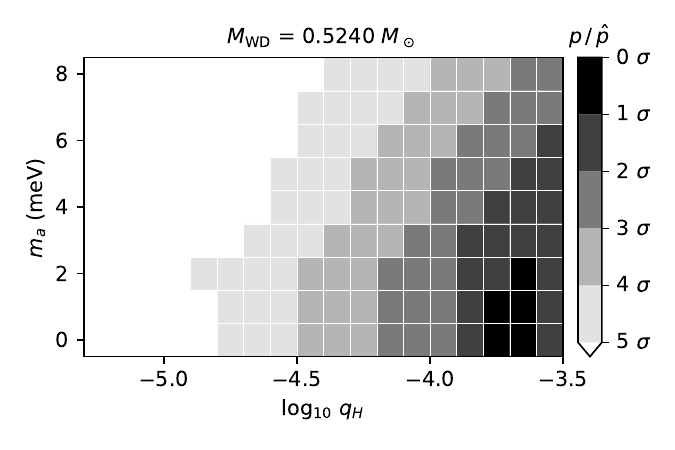}\\
    \includegraphics[width=\figwidth]{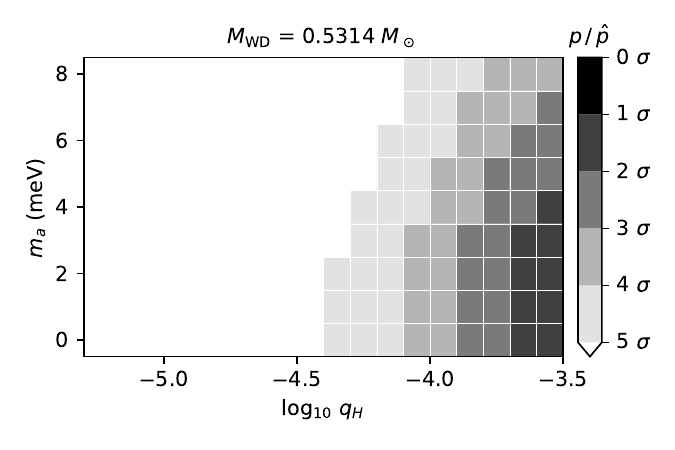}\\
    \includegraphics[width=\figwidth]{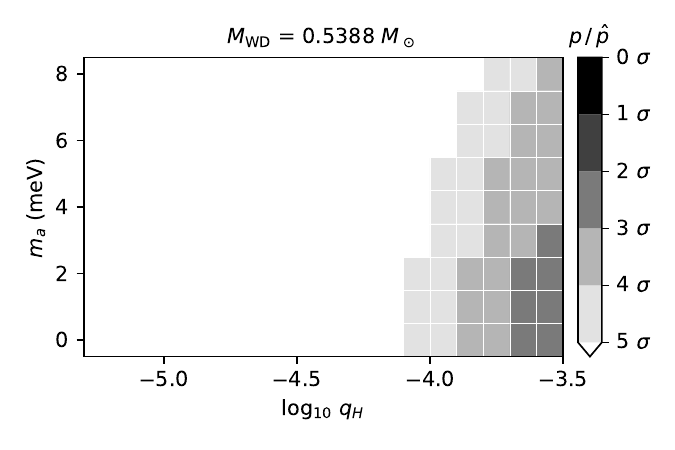}\\
    \caption{ACS, case 1: joint posterior probability density distribution after marginalising over the birthrate.}
    \label{fig:app_47tuc_axions_joint_density_acs_case1}
\end{figure}
\clearpage

\begin{figure}
    \centering
    \includegraphics[width=0.75\textwidth]{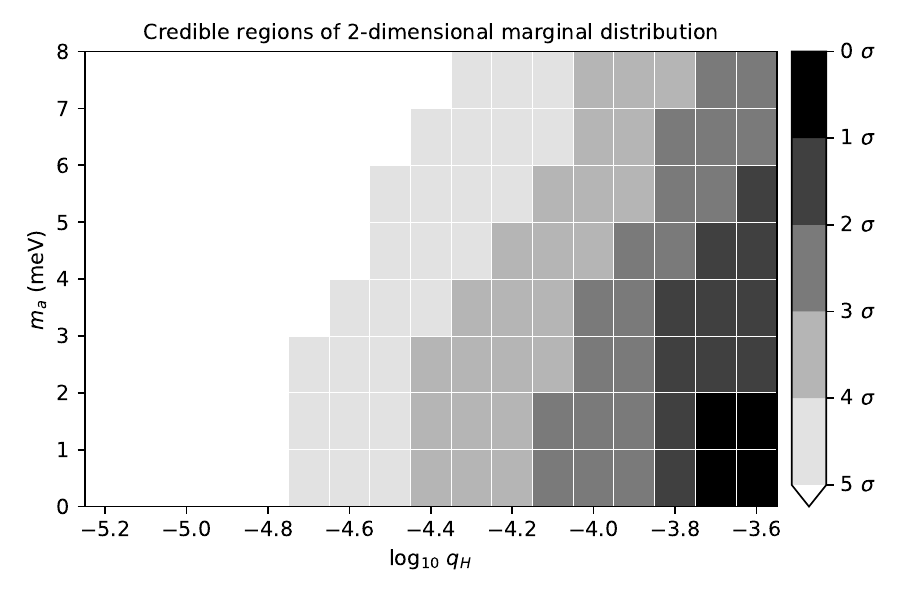}
    \caption{ACS, case 1: two-dimensional joint credible regions of axion mass ($m_a$) and envelope thickness ($q_H$) after marginalising over the other parameters.}
    \label{fig:app_47tuc_axions_2d_CRs_acs_case1}
\end{figure}

\begin{figure}
    \centering
    \begin{subfigure}[t]{0.49\textwidth}
        \includegraphics[width=\textwidth]{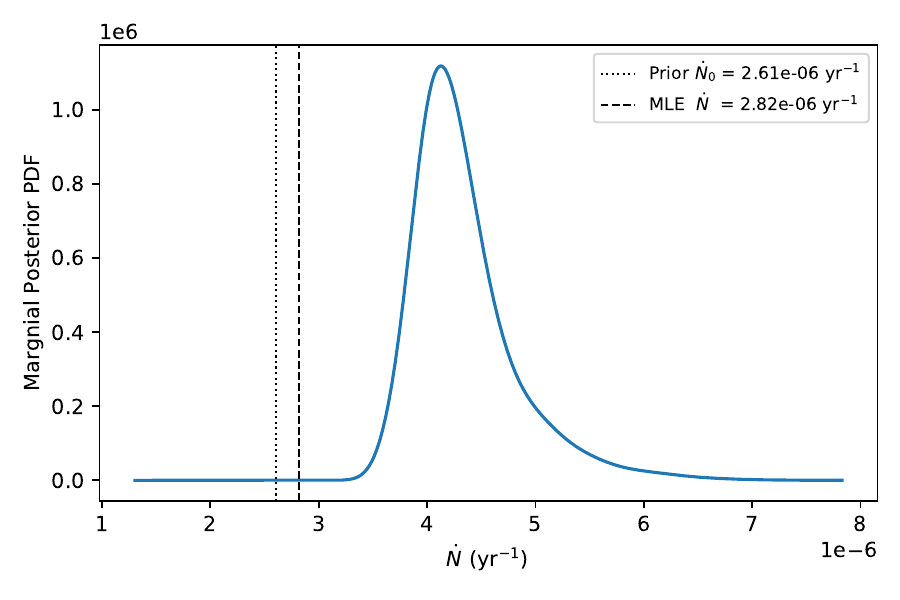}
        \caption{Birthrate}
        \label{fig:app_47tuc_axions_1d_marg_dist_Ndot_acs_case1}
    \end{subfigure}
    \hfill
    \begin{subfigure}[t]{0.49\textwidth}
        \includegraphics[width=\textwidth]{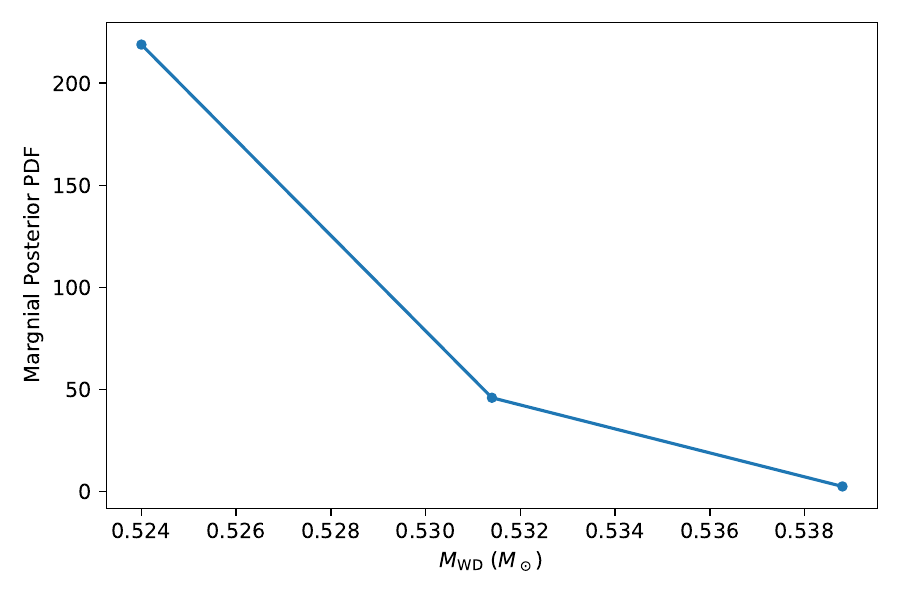}
        \caption{White dwarf mass}
        \label{fig:app_47tuc_axions_1d_marg_dist_MWD_acs_case1}
    \end{subfigure}
    \\[1.5ex]
    \begin{subfigure}[t]{0.49\textwidth}
        \includegraphics[width=\textwidth]{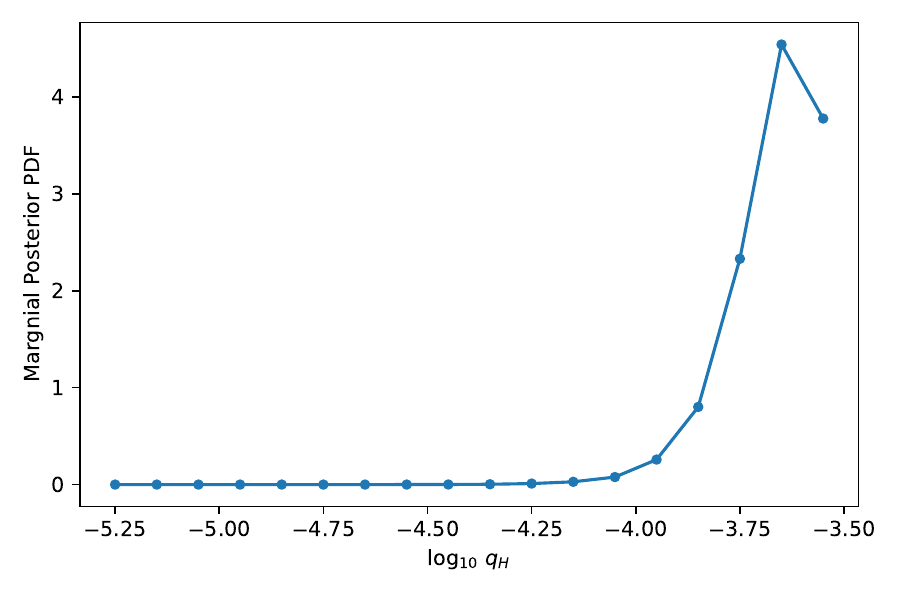}
        \caption{Envelope thickness}
        \label{fig:app_47tuc_axions_1d_marg_dist_lqh_acs_case1}
    \end{subfigure}
    \hfill
    \begin{subfigure}[t]{0.49\textwidth}
        \includegraphics[width=\textwidth]{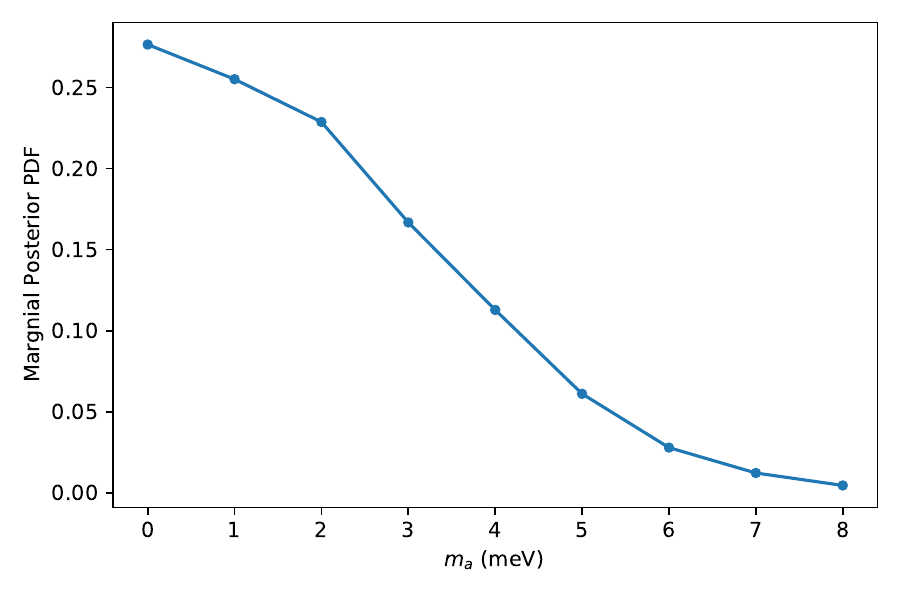}
        \caption{Axion mass}
        \label{fig:app_47tuc_axions_1d_marg_dist_ma_acs_case1}
    \end{subfigure}
    \\[1.5ex]
    \caption{ACS, case 1: one-dimensional posterior density distributions for each parameter after marginalising over all other model parameters.}
    \label{fig:app_47tuc_axions_1d_marginal_distributions_acs_case1}
\end{figure}
\clearpage


\begin{figure}[!h]
    \centering
    \setlength{\figwidth}{0.625\textwidth}
    \includegraphics[width=\figwidth]{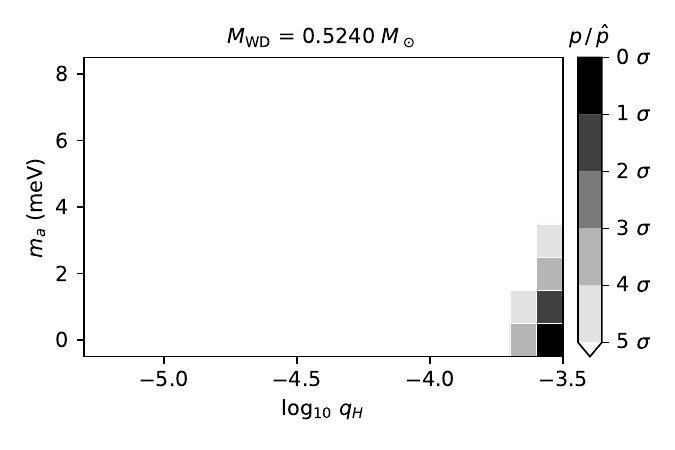}\\
    \includegraphics[width=\figwidth]{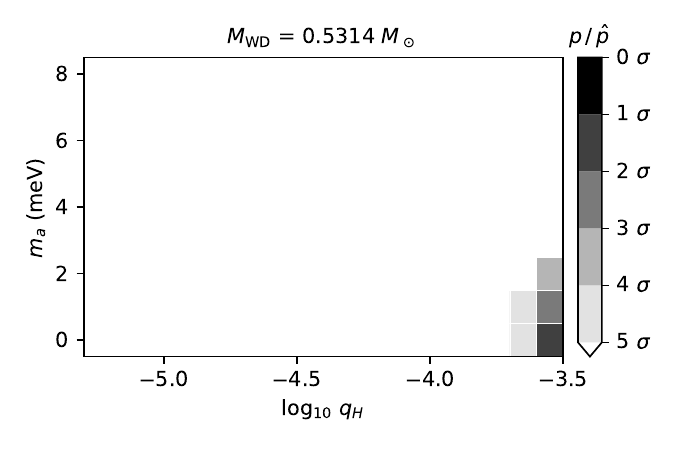}\\
    \includegraphics[width=\figwidth]{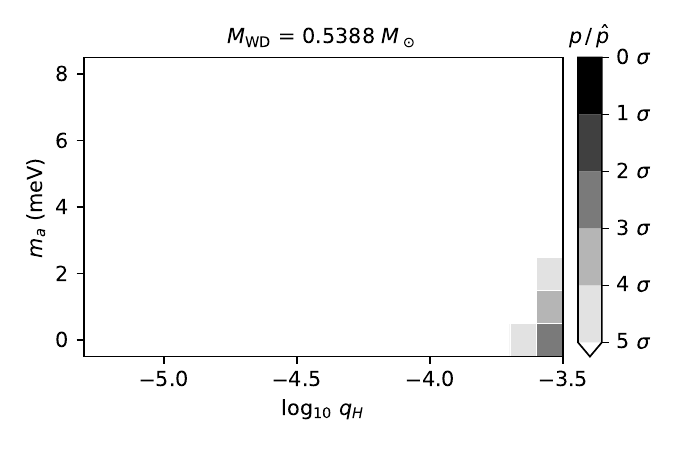}\\
    \caption{ACS, case 2: joint posterior probability density distribution after marginalising over the birthrate.}
    \label{fig:app_47tuc_axions_joint_density_acs_case2}
\end{figure}
\clearpage

\begin{figure}
    \centering
    \includegraphics[width=0.75\textwidth]{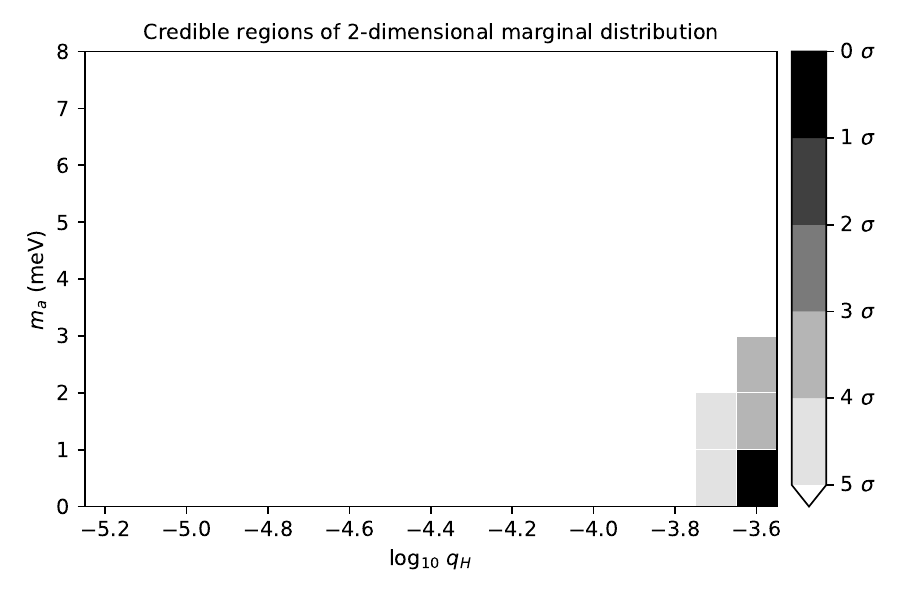}
    \caption{ACS, case 2: two-dimensional joint credible regions of axion mass ($m_a$) and envelope thickness ($q_H$) after marginalising over the other parameters.}
    \label{fig:app_47tuc_axions_2d_CRs_acs_case2}
\end{figure}

\begin{figure}
    \centering
    \begin{subfigure}[t]{0.49\textwidth}
        \includegraphics[width=\textwidth]{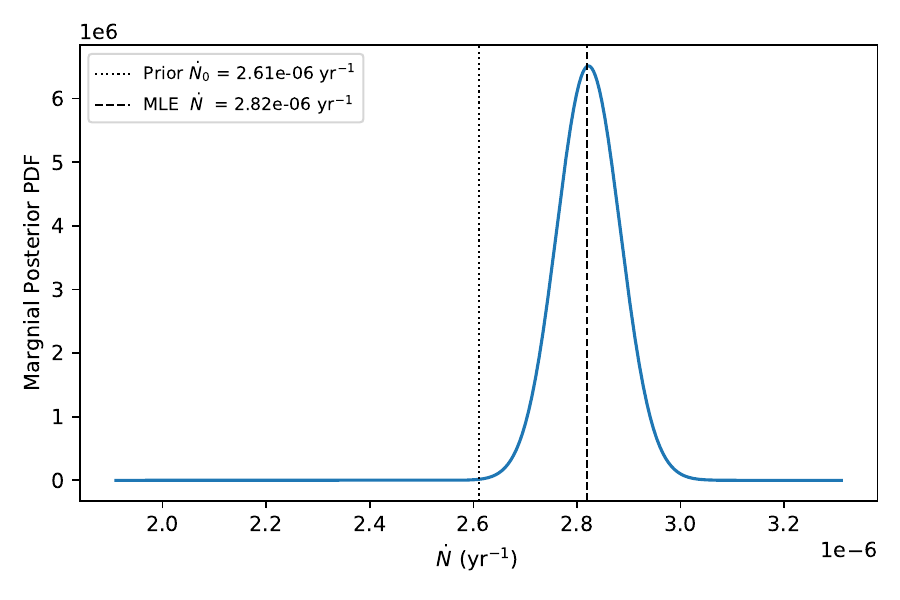}
        \caption{Birthrate}
        \label{fig:app_47tuc_axions_1d_marg_dist_Ndot_acs_case2}
    \end{subfigure}
    \hfill
    \begin{subfigure}[t]{0.49\textwidth}
        \includegraphics[width=\textwidth]{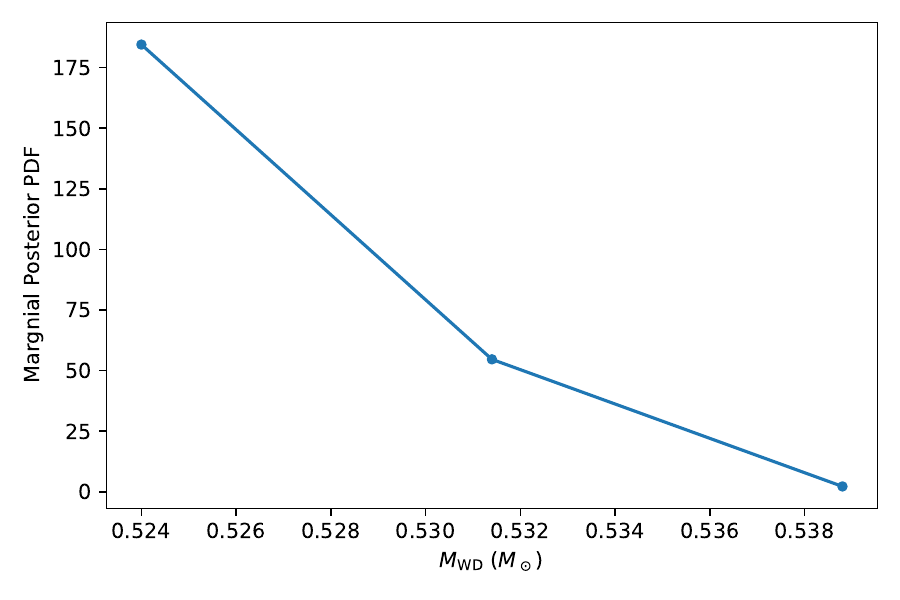}
        \caption{White dwarf mass}
        \label{fig:app_47tuc_axions_1d_marg_dist_MWD_acs_case2}
    \end{subfigure}
    \\[1.5ex]
    \begin{subfigure}[t]{0.49\textwidth}
        \includegraphics[width=\textwidth]{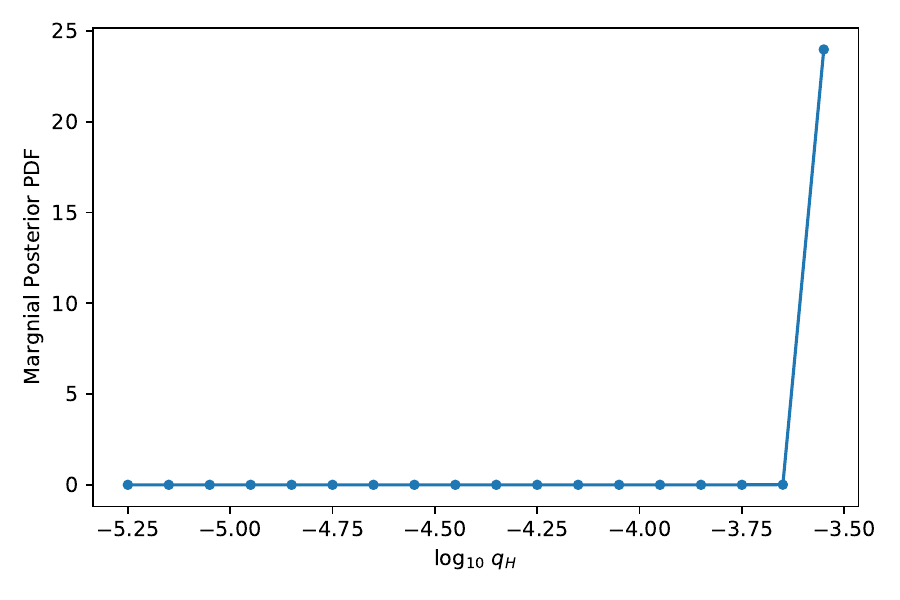}
        \caption{Envelope thickness}
        \label{fig:app_47tuc_axions_1d_marg_dist_lqh_acs_case2}
    \end{subfigure}
    \hfill
    \begin{subfigure}[t]{0.49\textwidth}
        \includegraphics[width=\textwidth]{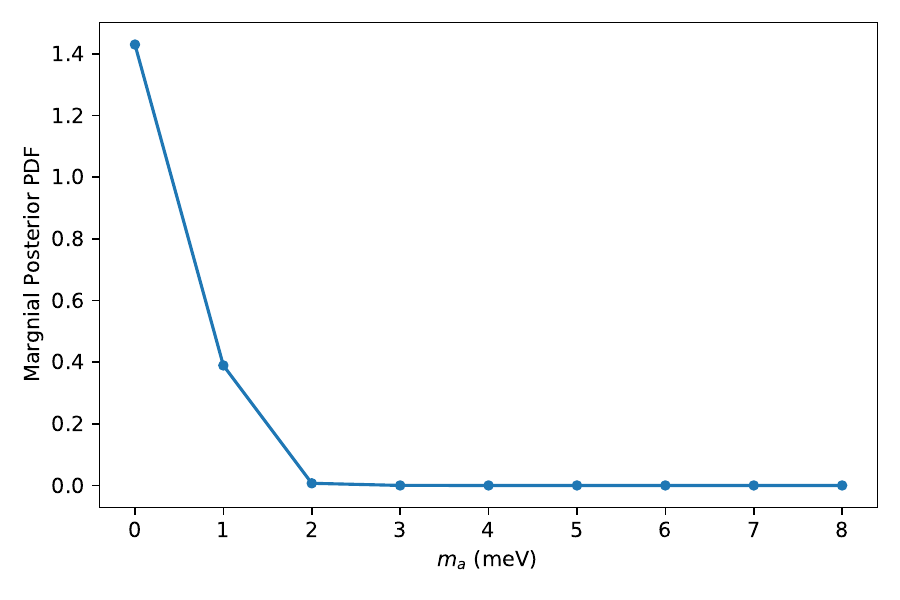}
        \caption{Axion mass}
        \label{fig:app_47tuc_axions_1d_marg_dist_ma_acs_case2}
    \end{subfigure}
    \\[1.5ex]
    \caption{ACS, case 2: one-dimensional posterior density distributions for each parameter after marginalising over all other model parameters.}
    \label{fig:app_47tuc_axions_1d_marginal_distributions_acs_case2}
\end{figure}
\clearpage


\begin{figure}[!h]
    \centering
    \setlength{\figwidth}{0.625\textwidth}
    \includegraphics[width=\figwidth]{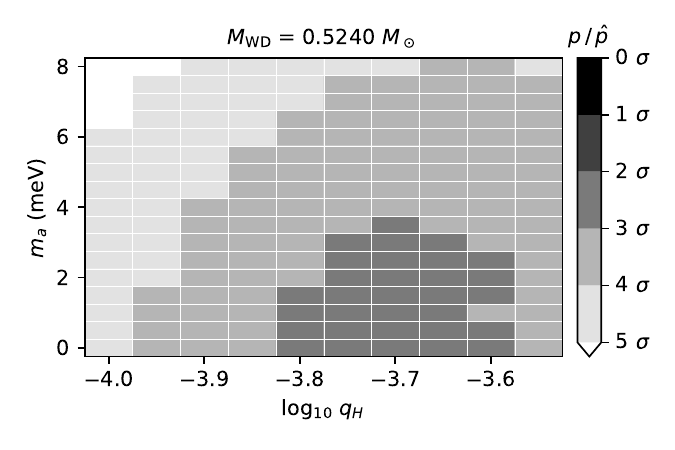}\\
    \includegraphics[width=\figwidth]{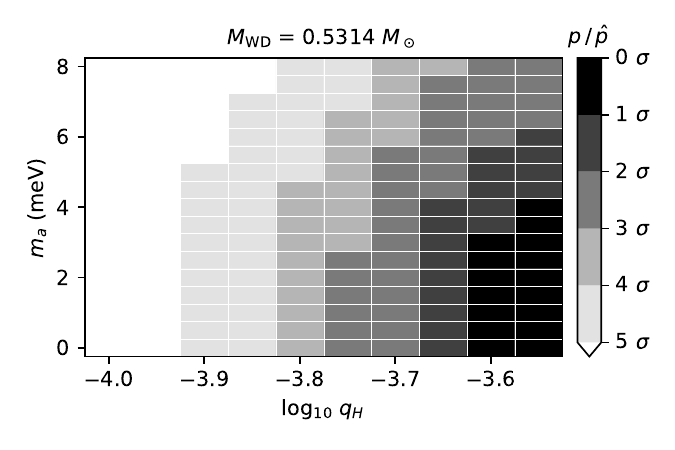}\\
    \includegraphics[width=\figwidth]{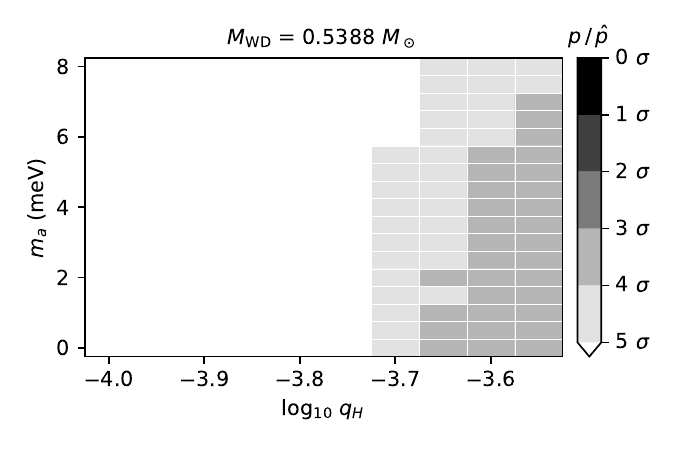}\\
    \caption{ACS, case 3: joint posterior probability density distribution after marginalising over the birthrate.}
    \label{fig:app_47tuc_axions_joint_density_acs_case3}
\end{figure}
\clearpage

\begin{figure}
    \centering
    \includegraphics[width=0.75\textwidth]{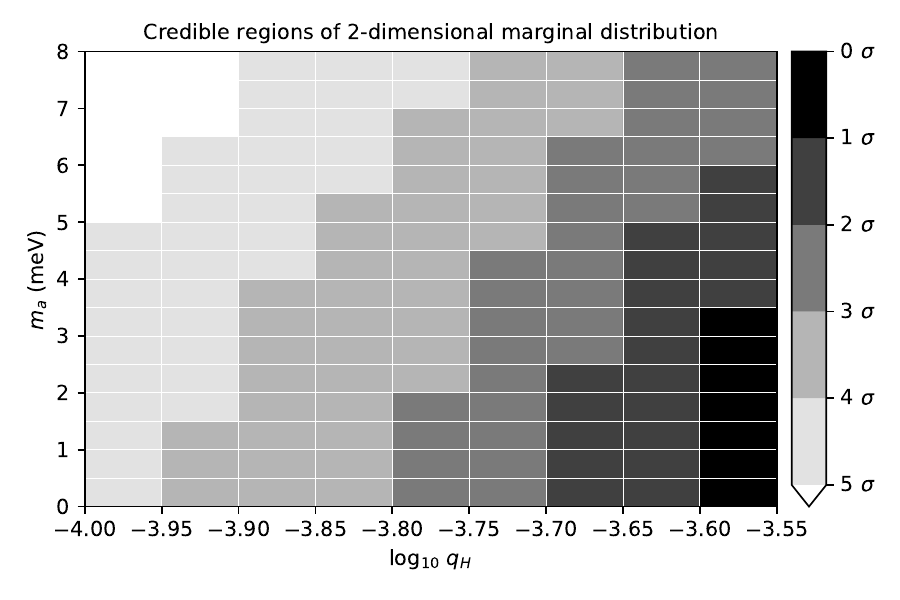}
    \caption{ACS, case 3: two-dimensional joint credible regions of axion mass ($m_a$) and envelope thickness ($q_H$) after marginalising over the other parameters.}
    \label{fig:app_47tuc_axions_2d_CRs_acs_case3}
\end{figure}

\begin{figure}
    \centering
    \begin{subfigure}[t]{0.49\textwidth}
        \includegraphics[width=\textwidth]{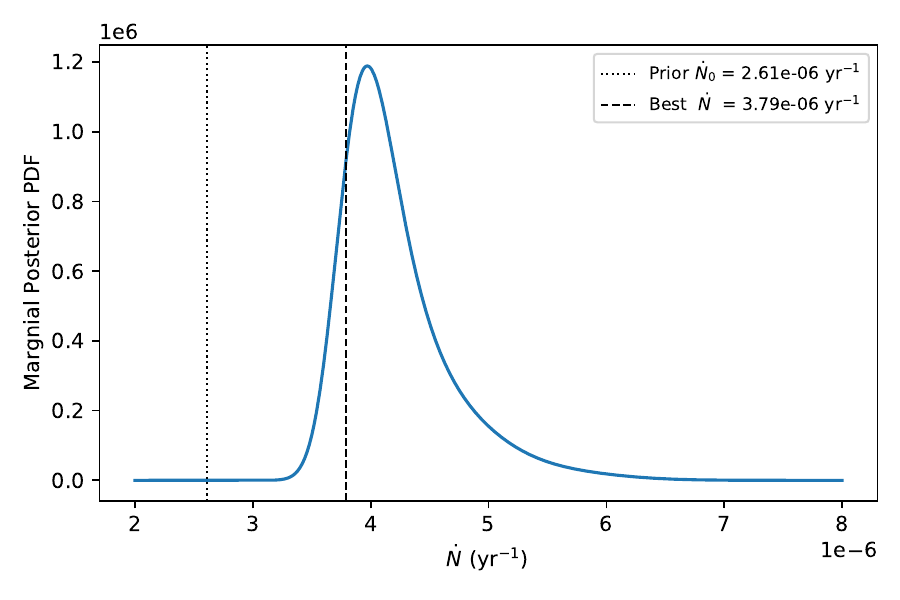}
        \caption{Birthrate}
        \label{fig:app_47tuc_axions_1d_marg_dist_Ndot_acs_case3}
    \end{subfigure}
    \hfill
    \begin{subfigure}[t]{0.49\textwidth}
        \includegraphics[width=\textwidth]{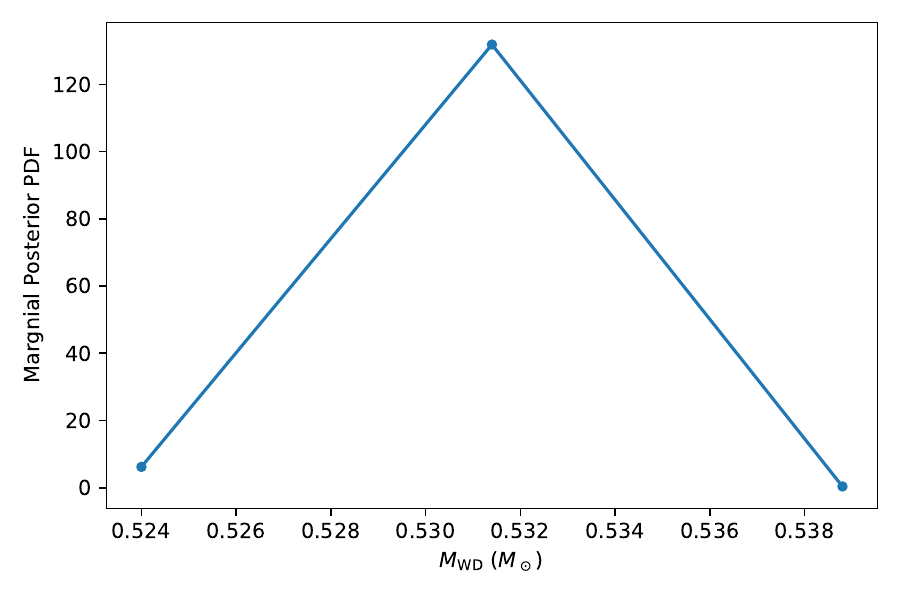}
        \caption{White dwarf mass}
        \label{fig:app_47tuc_axions_1d_marg_dist_MWD_acs_case3}
    \end{subfigure}
    \\[1.5ex]
    \begin{subfigure}[t]{0.49\textwidth}
        \includegraphics[width=\textwidth]{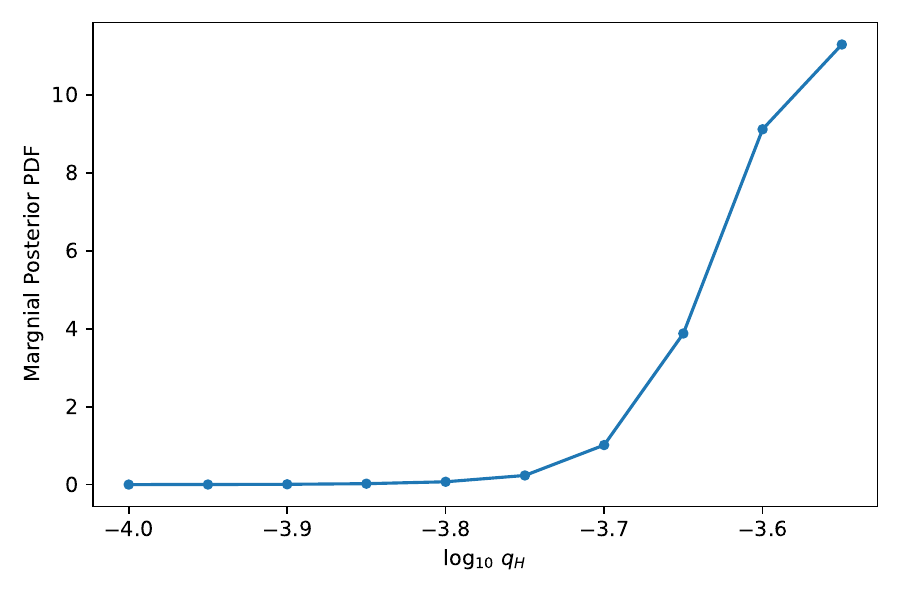}
        \caption{Envelope thickness}
        \label{fig:app_47tuc_axions_1d_marg_dist_lqh_acs_case3}
    \end{subfigure}
    \hfill
    \begin{subfigure}[t]{0.49\textwidth}
        \includegraphics[width=\textwidth]{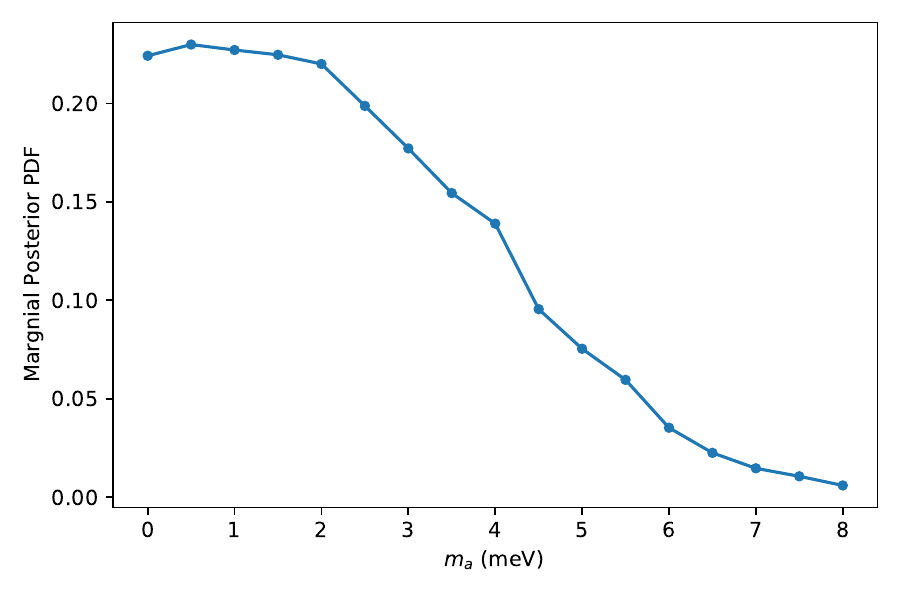}
        \caption{Axion mass}
        \label{fig:app_47tuc_axions_1d_marg_dist_ma_acs_case3}
    \end{subfigure}
    \\[1.5ex]
    \caption{ACS, case 3: one-dimensional posterior density distributions for each parameter after marginalising over all other model parameters.}
    \label{fig:app_47tuc_axions_1d_marginal_distributions_acs_case3}
\end{figure}
\clearpage

\subsubsection{Best-Fitting Models Comparison} \label[subsubappendix]{sec:appendix_47tuc_axions_cumdists_acs_cases}

\begin{figure}[!h]
    \centering
    \setlength{\figwidth}{0.35\textwidth} 
    \begin{subfigure}[t]{0.49\textwidth}
        \centering
        \includegraphics[width=\figwidth]{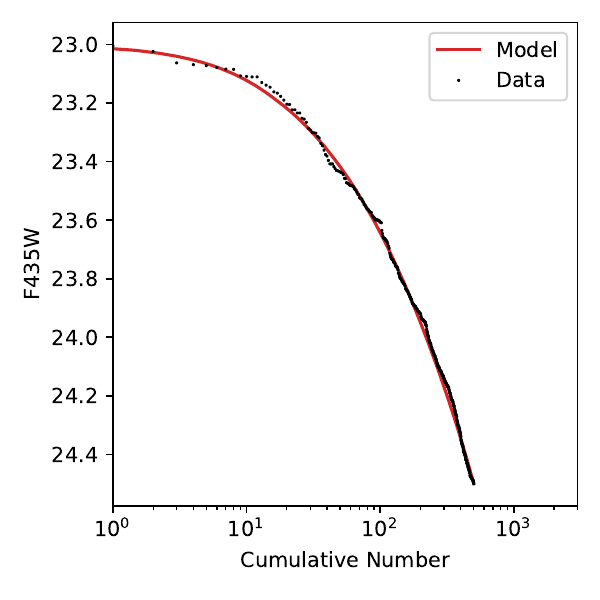}
        \caption{Case 1, F435W}
        \label{fig:app_47tuc_axions_invLFs_acs_f435w_case1}
    \end{subfigure}
    \hfill
    \begin{subfigure}[t]{0.49\textwidth}
        \centering
        \includegraphics[width=\figwidth]{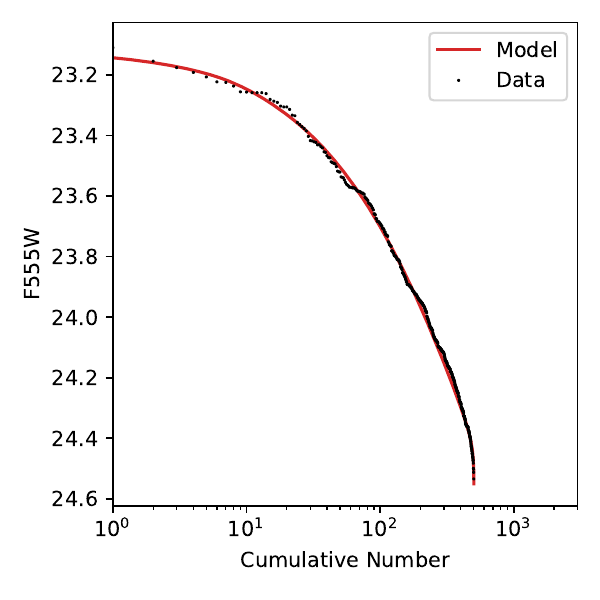}
        \caption{Case 1, F555W}
        \label{fig:app_47tuc_axions_invLFs_acs_f555w_case1}
    \end{subfigure}\\[1.5ex]
    \begin{subfigure}[t]{0.49\textwidth}
        \centering
        \includegraphics[width=\figwidth]{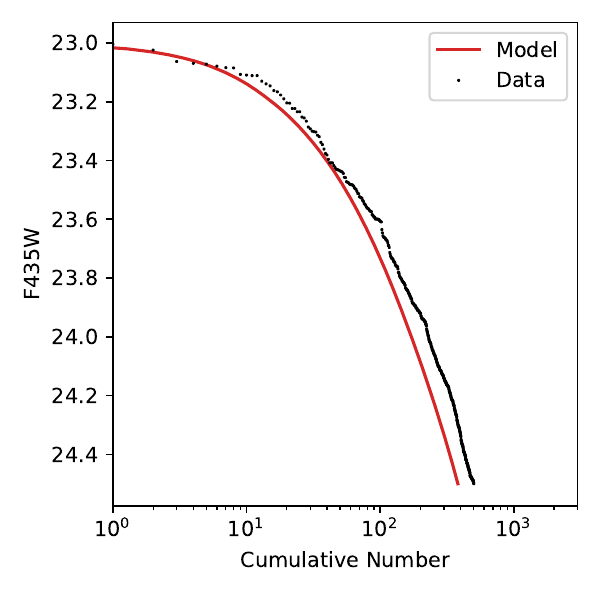}
        \caption{Case 2, F435W}
        \label{fig:app_47tuc_axions_invLFs_acs_f435w_case2}
    \end{subfigure}
    \hfill
    \begin{subfigure}[t]{0.49\textwidth}
        \centering
        \includegraphics[width=\figwidth]{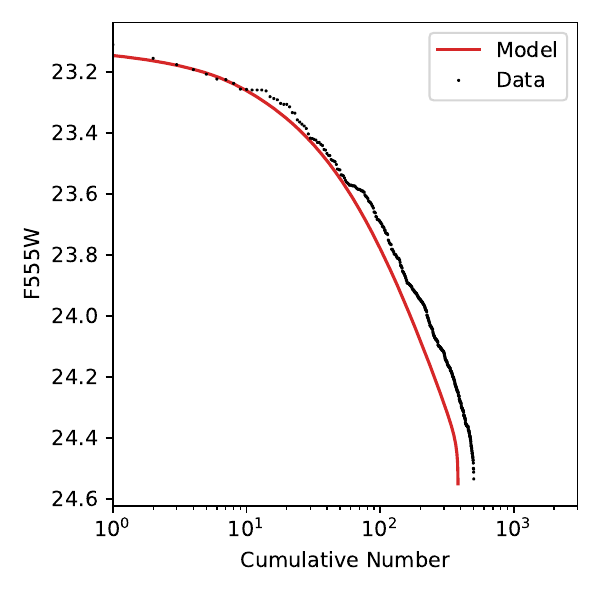}
        \caption{Case 2, F555W}
        \label{fig:app_47tuc_axions_invLFs_acs_f555w_case2}
    \end{subfigure}\\[1.5ex]
    \begin{subfigure}[t]{0.49\textwidth}
        \centering
        \includegraphics[width=\figwidth]{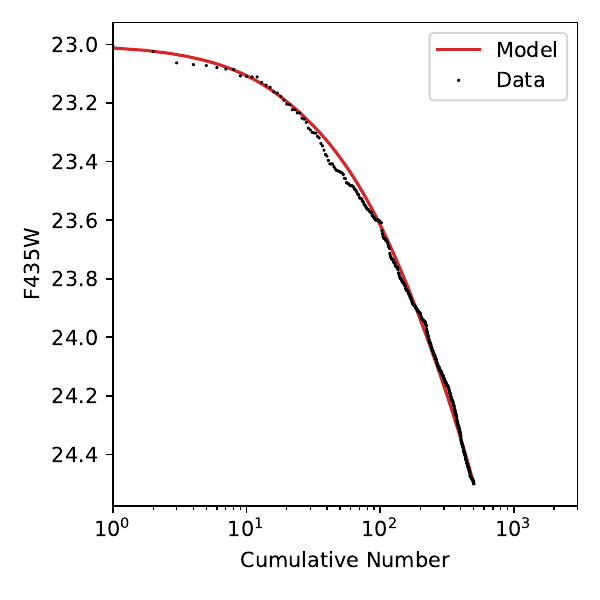}
        \caption{Case 3, F435W}
        \label{fig:app_47tuc_axions_invLFs_acs_f435w_case3}
    \end{subfigure}
    \hfill
    \begin{subfigure}[t]{0.49\textwidth}
        \centering
        \includegraphics[width=\figwidth]{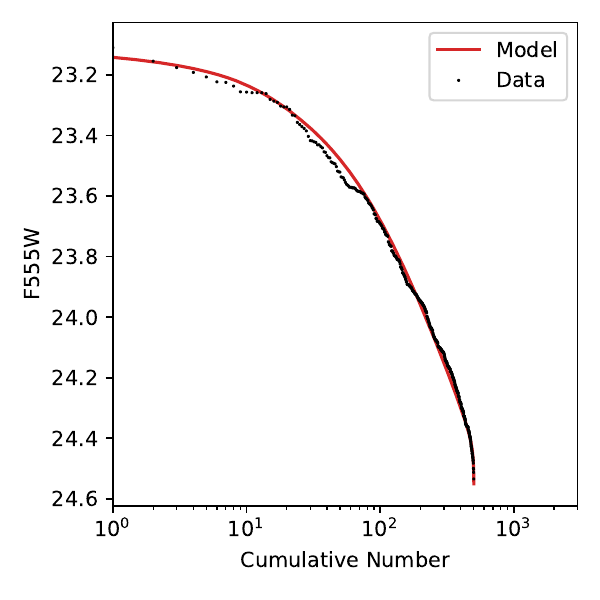}
        \caption{Case 3, F555W}
        \label{fig:app_47tuc_axions_invLFs_acs_f555w_case3}
    \end{subfigure}\\[1.5ex]
    \caption{ACS, all cases: inverse cumulative luminosity function of optimal model (red curve) compared to the data (black points) for both F435W (left column) and F555W (right column).
    Each row shows a different case: case 1 (top), case 2 (middle), and case 3 (bottom).}
    \label{fig:app_47tuc_axions_invLFs_acs_cases}
\end{figure}
\clearpage

\begin{table}
    \centering
    \begin{tabular}{c c c c c}
        \toprule
        ~ & $\dot{N}~(\mathrm{Myrs}^{-1})$ & $M_\mathrm{WD}~(M_\odot)$ & $\log_{10} q_H$ & $m_a~(\mathrm{meV})$ \\
        \midrule
        Case 1 & $3.94$ & $0.5240$ & $-3.65$ & $0.0$ \\
        Case 2 & $2.82$ & $0.5240$ & $-3.55$ & $0.0$ \\
        Case 3 & $3.79$ & $0.5314$ & $-3.55$ & $1.5$ \\
        \bottomrule
    \end{tabular}
    \caption{Parameter values of the optimal model for the different ACS cases. This is the combination of values that maximises the joint posterior distribution on the parameter grid.}
    \label{tab:app_47tuc_axions_best_model_params_acs_cases}
\end{table}

\begin{table}
    \centering
    \begin{tabular}{c c c}
    \toprule
        ~ & F435W & F555W \\
    \midrule
        Case 1 & 0.2098 & 0.2979 \\
        Case 2 & 0.3490 & 0.4693 \\
        Case 3 & 0.2660 & 0.2787 \\
    \bottomrule
    \end{tabular}
    \caption{Results of KS tests for the different ACS cases. The \textit{p}-values are reported for KS tests comparing the one-dimensional marginal cumulative probability distribution functions predicted by the optimal model for each case to the corresponding empirical distribution.}
    \label{tab:app_47tuc_axions_KS_test_results_acs_cases}
\end{table}

It can be seen from \cref{fig:app_47tuc_axions_invLFs_acs_cases} that the best-fitting models for both case 1 and case 3 fit the data well (for both magnitudes), as was also found for these cases when analysing the WFC3 data alone in \cref{sec:appendix_47tuc_axions_wfc3_only}.
This is expected because the optimal combination of parameter values is similar for case 1 and case 3.
The benefit of case 3, which uses the \citet{47tuc_deep_acs} priors for $\mwd$ and $\lqh$, over case 1, which uses uniform priors for $\mwd$ and $\lqh$, is that case 3 provides tighter credible regions.
To see this, compare \cref{fig:app_47tuc_axions_2d_CRs_acs_case1} and \cref{fig:app_47tuc_axions_2d_CRs_acs_case3} (and likewise \cref{fig:app_47tuc_axions_2d_CRs_wfc3_case1} and \cref{fig:app_47tuc_axions_2d_CRs_wfc3_case3} for the WFC3 only analysis) and note that \cref{fig:app_47tuc_axions_2d_CRs_acs_case3} (and \cref{fig:app_47tuc_axions_2d_CRs_wfc3_case3}) are plotted over a smaller range of $\lqh$ values (with the plot truncated at a lower limit of $\lqh = -4.0$ instead of $\lqh = -4.5$).
The joint posterior and credible region plots for case 3 are shown over a smaller range of $\lqh$ values than case 1 or case 2 for the sake of visualisation because the posterior distribution for case 3 drops below the 5-$\sigma$ level at larger values of $\lqh$ than the other cases; in other words, the posterior distribution (and credible regions) for case 1 (and case 2) are more extended in $\lqh$ than for case 3.

It can also be seen from \cref{fig:app_47tuc_axions_invLFs_acs_cases} that the best-fitting model for case 2 (using the Gaussian prior of \citet{Goldsbury2016} for the birthrate) does not fit the data as well as the best-fitting models of case 1 or case 3 (using a uniform birthrate prior).
The distributions predicted by the best-fitting model for case 2 in \cref{fig:app_47tuc_axions_invLFs_acs_cases} are shifted to the left (to smaller cumulative number values) compared to the corresponding empirical distributions, indicating that the birthrate is too small.
Aside from the birthrate being too small (which just affects the overall normalisation of the distribution), the morphology of each of the case 2 model distributions appears reasonable.
This is confirmed by the results of the KS tests (see \cref{tab:app_47tuc_axions_KS_test_results_acs_cases}), which assess the morphology but not overall normalisation of the distributions.
A reasonable morphology for the distributions for ACS case 2 should be expected because the best-fitting parameter values on the cooling model parameter grid are similar for all three cases in the ACS only analysis. 
In all three cases, the posterior probability is concentrated at the small $m_a$ and large $\lqh$ limits of the parameter grid (which are also physical limits, as discussed in \cref{sec:47tuc_axions_postdists} of the main text).
It can be seen from \cref{fig:app_47tuc_axions_joint_density_acs_case2} (and \cref{fig:app_47tuc_axions_2d_CRs_acs_case2}) that imposing a prior that underestimates the value of the birthrate drives the most likely parameter values to smaller $m_a$ and larger $\lqh$ values, but as the posterior distributions for the uniform birthrate cases are already concentrated near the limiting values of $m_a$ and $\lqh$ (\cref{fig:app_47tuc_axions_joint_density_acs_case1} and \cref{fig:app_47tuc_axions_joint_density_acs_case3}), 
the posterior distribution for case 2 simply ends up being very tightly concentrated in the low $m_a$, high $\lqh$ corner of the plots while being optimised at a similar location in parameter space.

This is in contrast to what is seen for case 2 of the WFC3 only analysis, where the WFC3 birthrate prior from \citet{Goldsbury2016} seems to overestimate the birthrate (instead of underestimating it like for the ACS birthrate). 
Imposing this likely overestimated value of the WFC3 birthrate prior drives the most likely parameter values to smaller $\lqh$ values, with some degeneracy between $\lqh$, $\mwd$, and $m_a$.
This resulted in the $\lqh$ value of the best-fitting model for WFC3 case 2 being much lower than for the other WFC3 (and ACS) cases, so the cumulative number distribution functions for WFC3 case 2 had a notably different morphology than the other cases in additional to a notably different birthrate value (see \cref{fig:app_47tuc_axions_invLFs_wfc3_cases}).
This different morphology is why the \textit{p}-values of the KS tests for WFC3 case 2 were much smaller than the values for the other cases (especially for the magnitude variables).

\section{Comparison of WFC3, ACS, and Combined Analyses} \label[appendix]{sec:appendix_47tuc_axions_comparison}

\Cref{tab:app_47tuc_axions_CRs_comparison,tab:app_47tuc_axions_best_model_params_comparison,tab:app_47tuc_axions_KS_test_results_comparison} summarise and compare the results of the combined analysis of the WFC3 and ACS data from \cref{sec:47tuc_axions_results} of the main text to the results of the individual analyses of the WFC3 data alone and the ACS data alone for the same choice of priors as were used in the combined analysis. This choice of priors consisted of uniform priors for the white dwarf birthrates and the prior from the results of \citet{47tuc_deep_acs} for $\mwd$ and $\lqh$, which corresponds to case 3 for both the WFC3 only analysis in \cref{sec:appendix_47tuc_axions_wfc3_only} and the ACS only analysis in \cref{sec:appendix_47tuc_axions_acs_only}.

The 95\% credible regions are reported directly in \cref{tab:app_47tuc_axions_CRs_comparison}.
These credible regions were calculated from the one-dimensional marginal posterior distribution of each parameter.
In \cref{tab:app_47tuc_axions_best_model_params_comparison}, these same credible regions are given as errors on the best-fitting combination of parameters on the discrete parameter grid for $\mwd$, $\lqh$, and $m_a$.
These best-fitting values are the combination of values that optimise the full joint posterior distribution.

These results show that analysing the WFC3 and ACS data individually gives similar best-fitting values.
The combined analysis likewise gives a similar result to analysing either of the data sets individually, but with tighter credible regions.
Compared to the WFC3 only credible regions, the combined analysis gives only a minor improvement.
The tightening of the credible region is much more dramatic when comparing the result of the ACS only analysis to the combined analysis.
This is to be expected because there are less white dwarfs in the ACS data space than the WFC3 data space due to the ACS data space being shorter (for reasons discussed in \cref{sec:47tuc_axions_analysis} of the main text), so the WFC3 data has more weight in the combined analysis.
The KS test results to check the goodness-of-fit are given in \cref{tab:app_47tuc_axions_KS_test_results_comparison} and have reasonable values in all cases.

\begin{table}
    \centering
    \setlength\tabcolsep{4pt} 
    \setlength\extratabcolsep{5.5\tabcolsep}
    \setlength\pad{12.5pt} 
    \begin{tabularx}{\columnwidth}{X @{\hspace*{\extratabcolsep}} 
    r @{\hspace*{\pad}} c d{2.4} @{\hspace*{\extratabcolsep}} 
    r @{\hspace*{\pad}} c d{2.4} @{\hspace*{\extratabcolsep}} 
    r @{\hspace*{\pad}} c d{2.4}}
        \toprule
        ~\hspace*{\extratabcolsep} 
        & \multicolumn{3}{c}{WFC3}\hspace*{\extratabcolsep} 
        & \multicolumn{3}{c}{ACS}\hspace*{\extratabcolsep} 
        & \multicolumn{3}{c}{Combined} \\
        \midrule
        $\dot{N}_\mathrm{WFC3}~(\mathrm{Myr}^{-1})$ & $6.66$ & $-$ & 7.60 & ~ & $-$ & ~ & $6.68$ & $-$ & 7.73 \\
        $\dot{N}_\mathrm{ACS}~(\mathrm{Myr}^{-1})$ & ~ & $-$ & ~ & $3.48$ & $-$ & 5.22 & $3.49$ & $-$ & 4.35 \\
        $M_\mathrm{WD}~(M_\odot)$ & $0.5327$ & $-$ & 0.5388 & $0.5254$ & $-$ & 0.5371 & $0.5282$ & $-$ & 0.5388 \\
        $\log_{10} q_H$ & $-3.65$ & $-$ & -3.55 & $-3.69$ & $-$ & -3.55 & $-3.67$ & $-$ & -3.55 \\
        $m_a~(\mathrm{meV})$ & $0.00$ & $-$ & 2.93 & $0.00$ & $-$ & 5.64 & $0.00$ & $-$ & 2.85 \\
        \addlinespace[1ex]
        \midrule
        \addlinespace[1ex]
        $g_{aee} \, / \, 10^{-13}$ & $0.00$ & $-$ & 0.83 & $0.00$ & $-$ & 1.60 & $0.00$ & $-$ & 0.81\\
        \bottomrule
    \end{tabularx}
    \caption{Comparison of 95\% credible regions given by the analysis of only WFC3, only ACS, and combined WFC3 and ACS data. The same priors and parameter grid were used in all three cases (where applicable).
    These priors correspond to case 3 for the WFC3 only and ACS only analyses and correspond to the main results presented in \cref{sec:47tuc_axions_results} of the main text for the combined analysis.}
    \label{tab:app_47tuc_axions_CRs_comparison}
\end{table}

\begin{table}
    \centering
    \begin{tabular}{l D{,}{.}{2.10} D{,}{.}{2.10} D{,}{.}{2.10}}
        \toprule
        ~ & \multicolumn{1}{l}{~WFC3} & \multicolumn{1}{l}{~ACS} & \multicolumn{1}{l}{~Combined} \\
        \midrule
        $\dot{N}_\mathrm{WFC3}~(\mathrm{Myr}^{-1})$ & 6,91_{-0.25}^{+0.69} & \multicolumn{1}{l}{~~~$-$} & 6,91_{-0.23}^{+0.82} \\[1ex]
        $\dot{N}_\mathrm{ACS}~(\mathrm{Myr}^{-1})$ & \multicolumn{1}{l}{~~~~$-$} & 3,79_{-0.31}^{+1.43} & 3,73_{-0.24}^{+0.62} \\[1ex]
        $M_\mathrm{WD}~(M_\odot)$ & 0,5388_{-0.0061}^{+0.0000} & 0,5314_{-0.0060}^{+0.0057} & 0,5388_{-0.0106}^{+0.0000} \\[1ex]
        $\log_{10} q_H$ & -3,55_{-0.10}^{+0.00} & -3,55_{-0.14}^{+0.00} & -3,55_{-0.12}^{+0.00} \\[1ex]
        $m_a~(\mathrm{meV})$ & 0,00_{-0.00}^{+2.93} & 1,50_{-1.50}^{+4.14} & 0,00_{-0.00}^{+2.85} \\
        \addlinespace[1.5ex] \midrule \addlinespace[1.5ex]
        $g_{aee} \, / \, 10^{-13}$ & 0,00_{-0.00}^{+0.83} & 0,42_{-0.42}^{+1.17} & 0,00_{-0.00}^{+0.81} \\
        \bottomrule
    \end{tabular}
    \caption{Comparison of best-fitting model parameters for the WFC3 only (case 3), ACS only (case 3), and combined (main text) analyses.
    The 95\% credible regions from \cref{tab:app_47tuc_axions_CRs_comparison} are reported as errors on the best-fitting combination of parameter values on the parameter grid.}
    \label{tab:app_47tuc_axions_best_model_params_comparison}
\end{table}

\begin{table}
    \centering
    \begin{tabular}{l c c c}
    \toprule
        ~ & WFC3 & ACS & Combined \\
    \midrule
        R & 0.0156 & $-$ & 0.0156 \\
        F225W & 0.0362 & $-$ & 0.0362 \\
        F336W & 0.0132 & $-$ & 0.0132 \\
        F435W & $-$ & 0.2660 & 0.3527 \\
        F555W & $-$ & 0.2787 & 0.2739 \\
    \bottomrule
    \end{tabular}
    \caption{Comparison of KS test results for WFC3 only (case 3), ACS only (case 3), and combined analyses.}
    \label{tab:app_47tuc_axions_KS_test_results_comparison}
\end{table}

\clearpage



\acknowledgments

This research is based on observations made with the NASA/ESA \textit{Hubble Space Telescope} obtained from the Space Telescope Science Institute, which is operated by the Association of Universities for Research in Astronomy, Inc., under NASA contract NAS 5–26555. These observations are associated with proposal GO-12971 (PI: H. Richer).
This work has been supported by the Natural Sciences and Engineering Research Council of Canada through the Discovery Grants program and Compute Canada.
This research was also supported in part through computational resources and services provided by Advanced Research Computing at the University of British Columbia.

We thank Pierre Bergeron for providing bolometric correction data upon request for the HST photometric system. 
These data correspond to the synthetic colours from the website ``Synthetic Colors and Evolutionary Sequences of Hydrogen- and Helium-Atmosphere White Dwarfs'' at \url{http://www.astro.umontreal.ca/~bergeron/CoolingModels/} but calculated for the relevant HST filters.
We retrieved the extinction data from the \texttt{PARSEC} database through the website \url{https://stev.oapd.inaf.it}.

\section*{Data Availability}

The raw data from the HST observations analysed in this work are publicly available through the Mikulski Archive for Space Telescopes (MAST; \url{https://archive.stsci.edu/}).
More specifically, the data from the individual orbits comprising the HST observations can be retrieved through the MAST Portal at \url{https://mast.stsci.edu/portal/Mashup/Clients/Mast/Portal.html}.
The extinction data used in this work are available through the website \url{https://stev.oapd.inaf.it}.


\bibliographystyle{JHEP_modified} 
\bibliography{main}

@ARTICLE{47tuc_deep_acs,
       author = {{Fleury}, Leesa and {Richer}, Harvey and {Heyl}, Jeremy},
        title = "{The Cooling of Old White Dwarfs in 47 Tucanae}",
      journal = {\mnras},
         year = 2026,
        month = jun,
          doi = {10.1093/mnras/stag1024},
archivePrefix = {arXiv},
       eprint = {2507.22046},
 primaryClass = {astro-ph.SR},
       adsurl = {https://ui.adsabs.harvard.edu/abs/2026MNRAS.tmp..974F}
}

@ARTICLE{Goldsbury2016,
       author = {{Goldsbury}, R. and {Heyl}, J. and {Richer}, H.~B. and {Kalirai}, J.~S. and
         {Tremblay}, P.~E.},
        title = "{Constraining White Dwarf Structure and Neutrino Physics in 47 Tucanae}",
      journal = {\apj},
         year = 2016,
        month = apr,
       volume = {821},
       number = {1},
          eid = {27},
        pages = {27},
          doi = {10.3847/0004-637X/821/1/27},
archivePrefix = {arXiv},
       eprint = {1602.06286},
 primaryClass = {astro-ph.SR},
       adsurl = {https://ui.adsabs.harvard.edu/abs/2016ApJ...821...27G}
}

@ARTICLE{Dine:1981,
       author = {{Dine}, Michael and {Fischler}, Willy and {Srednicki}, Mark},
        title = "{A simple solution to the strong CP problem with a harmless axion}",
      journal = {Phys. Lett. B},
         year = 1981,
        month = aug,
       volume = {104},
       number = {3},
        pages = {199-202},
          doi = {10.1016/0370-2693(81)90590-6},
       adsurl = {https://ui.adsabs.harvard.edu/abs/1981PhLB..104..199D}
}

@ARTICLE{Zhitnitsky:1980,
       author = {{Zhitnitsky}, A.~R.},
        title = "{On Possible Suppression of the Axion Hadron Interactions. (In Russian)}",
      journal = {Sov. J. Nucl. Phys.},
         year = 1980,
       volume = {31},
        pages = {260},
         note = {[Yad. Fiz. 31, 497 (1980)]}
}

@ARTICLE{2019PhRvL.123f1104D,
       author = {{Dessert}, Christopher and {Long}, Andrew J. and {Safdi}, Benjamin R.},
        title = "{X-Ray Signatures of Axion Conversion in Magnetic White Dwarf Stars}",
      journal = {\prl},
         year = 2019,
        month = aug,
       volume = {123},
       number = {6},
          eid = {061104},
        pages = {061104},
          doi = {10.1103/PhysRevLett.123.061104},
archivePrefix = {arXiv},
       eprint = {1903.05088},
 primaryClass = {hep-ph},
       adsurl = {https://ui.adsabs.harvard.edu/abs/2019PhRvL.123f1104D}
}

@ARTICLE{2022PhRvL.128g1102D,
       author = {{Dessert}, Christopher and {Long}, Andrew J. and {Safdi}, Benjamin R.},
        title = "{No Evidence for Axions from Chandra Observation of the Magnetic White Dwarf RE J0317-853}",
      journal = {\prl},
         year = 2022,
        month = feb,
       volume = {128},
       number = {7},
          eid = {071102},
        pages = {071102},
          doi = {10.1103/PhysRevLett.128.071102},
archivePrefix = {arXiv},
       eprint = {2104.12772},
 primaryClass = {hep-ph},
       adsurl = {https://ui.adsabs.harvard.edu/abs/2022PhRvL.128g1102D}
}

@ARTICLE{2025PhRvD.111j3002N,
       author = {{Ning}, Orion and {Dessert}, Christopher and {Hong}, Vi and {Safdi}, Benjamin R.},
        title = "{Search for axions from magnetic white dwarfs with Chandra x-ray observations}",
      journal = {\prd},
         year = 2025,
        month = may,
       volume = {111},
       number = {10},
          eid = {103002},
        pages = {103002},
          doi = {10.1103/PhysRevD.111.103002},
archivePrefix = {arXiv},
       eprint = {2411.05041},
 primaryClass = {astro-ph.HE},
       adsurl = {https://ui.adsabs.harvard.edu/abs/2025PhRvD.111j3002N}
}

@ARTICLE{Munn:2017,
       author = {{Munn}, Jeffrey A. and {Harris}, Hugh C. and {von Hippel}, Ted and {Kilic}, Mukremin and {Liebert}, James W. and {Williams}, Kurtis A. and {DeGennaro}, Steven and {Jeffery}, Elizabeth and {Dame}, Kyra and {Gianninas}, A. and {Brown}, Warren R.},
        title = "{A Deep Proper Motion Catalog Within the Sloan Digital Sky Survey Footprint. II. The White Dwarf Luminosity Function}",
      journal = {\aj},
         year = 2017,
        month = jan,
       volume = {153},
       number = {1},
          eid = {10},
        pages = {10},
          doi = {10.3847/1538-3881/153/1/10},
archivePrefix = {arXiv},
       eprint = {1611.06275},
 primaryClass = {astro-ph.SR},
       adsurl = {https://ui.adsabs.harvard.edu/abs/2017AJ....153...10M}
}

@ARTICLE{Kilic:2017,
       author = {{Kilic}, Mukremin and {Munn}, Jeffrey A. and {Harris}, Hugh C. and {von Hippel}, Ted and {Liebert}, James W. and {Williams}, Kurtis A. and {Jeffery}, Elizabeth and {DeGennaro}, Steven},
        title = "{The Ages of the Thin Disk, Thick Disk, and the Halo from Nearby White Dwarfs}",
      journal = {\apj},
         year = 2017,
        month = mar,
       volume = {837},
       number = {2},
          eid = {162},
        pages = {162},
          doi = {10.3847/1538-4357/aa62a5},
archivePrefix = {arXiv},
       eprint = {1702.06984},
 primaryClass = {astro-ph.SR},
       adsurl = {https://ui.adsabs.harvard.edu/abs/2017ApJ...837..162K}
}

@ARTICLE{Rowell:2011,
       author = {{Rowell}, N. and {Hambly}, N.~C.},
        title = "{White dwarfs in the SuperCOSMOS Sky Survey: the thin disc, thick disc and spheroid luminosity functions}",
      journal = {\mnras},
         year = 2011,
        month = oct,
       volume = {417},
       number = {1},
          eid = {93},
        pages = {93-113},
          doi = {10.1111/j.1365-2966.2011.18976.x},
archivePrefix = {arXiv},
       eprint = {1102.3193},
 primaryClass = {astro-ph.GA},
       adsurl = {https://ui.adsabs.harvard.edu/abs/2011MNRAS.417...93R}
}

@ARTICLE{2018MNRAS.478.2569I,
       author = {{Isern}, J. and {Garc{\'\i}a-Berro}, E. and {Torres}, S. and {Cojocaru}, R. and {Catal{\'a}n}, S.},
        title = "{Axions and the luminosity function of white dwarfs: the thin and thick discs, and the halo}",
      journal = {\mnras},
         year = 2018,
        month = aug,
       volume = {478},
       number = {2},
        pages = {2569-2575},
          doi = {10.1093/mnras/sty1162},
archivePrefix = {arXiv},
       eprint = {1805.00135},
 primaryClass = {astro-ph.SR},
       adsurl = {https://ui.adsabs.harvard.edu/abs/2018MNRAS.478.2569I}
}

@ARTICLE{2008ApJ...682L.109I,
       author = {{Isern}, J. and {Garc{\'\i}a-Berro}, E. and {Torres}, S. and {Catal{\'a}n}, S.},
        title = "{Axions and the Cooling of White Dwarf Stars}",
      journal = {\apjl},
         year = 2008,
        month = aug,
       volume = {682},
       number = {2},
        pages = {L109},
          doi = {10.1086/591042},
archivePrefix = {arXiv},
       eprint = {0806.2807},
 primaryClass = {astro-ph},
       adsurl = {https://ui.adsabs.harvard.edu/abs/2008ApJ...682L.109I}
}

@ARTICLE{2014JCAP...10..069M,
       author = {{Miller Bertolami}, M.~M. and {Melendez}, B.~E. and {Althaus}, L.~G. and {Isern}, J.},
        title = "{Revisiting the axion bounds from the Galactic white dwarf luminosity function}",
      journal = {\jcap},
         year = 2014,
        month = oct,
       volume = {10},
       number = {10},
          eid = {069},
        pages = {069},
          doi = {10.1088/1475-7516/2014/10/069},
archivePrefix = {arXiv},
       eprint = {1406.7712},
 primaryClass = {hep-ph},
       adsurl = {https://ui.adsabs.harvard.edu/abs/2014JCAP...10..069M}
}

@ARTICLE{DiLuzio:2020wdo,
       author = {{Di Luzio}, Luca and {Giannotti}, Maurizio and {Nardi}, Enrico and {Visinelli}, Luca},
        title = "{The landscape of QCD axion models}",
      journal = {\physrep},
         year = 2020,
       volume = {870},
        pages = {1-117},
          doi = {10.1016/j.physrep.2020.06.002},
archivePrefix = {arXiv},
       eprint = {2003.01100},
 primaryClass = {hep-ph},
       adsurl = {https://ui.adsabs.harvard.edu/abs/2020PhR...870....1D}
}

@ARTICLE{2022JCAP...02..035D,
       author = {{Di Luzio}, Luca and {Fedele}, Marco and {Giannotti}, Maurizio and {Mescia}, Federico and {Nardi}, Enrico},
        title = "{Stellar evolution confronts axion models}",
      journal = {\jcap},
         year = 2022,
        month = feb,
       volume = {02},
       number = {02},
          eid = {035},
        pages = {035},
          doi = {10.1088/1475-7516/2022/02/035},
archivePrefix = {arXiv},
       eprint = {2109.10368},
 primaryClass = {hep-ph},
       adsurl = {https://ui.adsabs.harvard.edu/abs/2022JCAP...02..035D}
}

@ARTICLE{1990PhR...198....1R,
       author = {{Raffelt}, Georg G.},
        title = "{Astrophysical methods to constrain axions and other novel particle phenomena}",
      journal = {\physrep},
         year = 1990,
        month = dec,
       volume = {198},
       number = {1-2},
        pages = {1-113},
          doi = {10.1016/0370-1573(90)90054-6},
       adsurl = {https://ui.adsabs.harvard.edu/abs/1990PhR...198....1R}
}

@ARTICLE{2025PhyR.1117....1C,
       author = {{Carenza}, Pierluca and {Giannotti}, Maurizio and {Isern}, Jordi and {Mirizzi}, Alessandro and {Straniero}, Oscar},
        title = "{Axion astrophysics}",
      journal = {\physrep},
         year = 2025,
        month = apr,
       volume = {1117},
        pages = {1-102},
          doi = {10.1016/j.physrep.2025.02.002},
archivePrefix = {arXiv},
       eprint = {2411.02492},
 primaryClass = {hep-ph},
       adsurl = {https://ui.adsabs.harvard.edu/abs/2025PhyR.1117....1C}
}

@ARTICLE{2020PhRvD.102h3007C,
       author = {{Capozzi}, Francesco and {Raffelt}, Georg},
        title = "{Axion and neutrino bounds improved with new calibrations of the tip of the red-giant branch using geometric distance determinations}",
      journal = {\prd},
         year = 2020,
        month = oct,
       volume = {102},
       number = {8},
          eid = {083007},
        pages = {083007},
          doi = {10.1103/PhysRevD.102.083007},
archivePrefix = {arXiv},
       eprint = {2007.03694},
 primaryClass = {astro-ph.SR},
       adsurl = {https://ui.adsabs.harvard.edu/abs/2020PhRvD.102h3007C}
}

@ARTICLE{2020A&A...644A.166S,
       author = {{Straniero}, O. and {Pallanca}, C. and {Dalessandro}, E. and {Dom{\'\i}nguez}, I. and {Ferraro}, F.~R. and {Giannotti}, M. and {Mirizzi}, A. and {Piersanti}, L.},
        title = "{The RGB tip of galactic globular clusters and the revision of the axion-electron coupling bound}",
      journal = {\aap},
         year = 2020,
        month = dec,
       volume = {644},
          eid = {A166},
        pages = {A166},
          doi = {10.1051/0004-6361/202038775},
archivePrefix = {arXiv},
       eprint = {2010.03833},
 primaryClass = {astro-ph.SR},
       adsurl = {https://ui.adsabs.harvard.edu/abs/2020A&A...644A.166S}
}

@ARTICLE{2025JETPL.121..159T,
       author = {{Troitsky}, S.~V.},
        title = "{Stellar Evolution and Axion-Like Particles: New Constraints and Hints from Globular Clusters in the GAIA DR3 Data}",
      journal = {Journal of Experimental and Theoretical Physics Letters},
         year = 2025,
        month = feb,
       volume = {121},
       number = {3},
        pages = {159-165},
          doi = {10.1134/S0021364024603798},
archivePrefix = {arXiv},
       eprint = {2410.02266},
 primaryClass = {hep-ph},
       adsurl = {https://ui.adsabs.harvard.edu/abs/2025JETPL.121..159T}
}

@ARTICLE{2013PhRvL.111w1301V,
       author = {{Viaux}, N. and {Catelan}, M. and {Stetson}, P.~B. and {Raffelt}, G.~G. and {Redondo}, J. and {Valcarce}, A.~A.~R. and {Weiss}, A.},
        title = "{Neutrino and Axion Bounds from the Globular Cluster M5 (NGC 5904)}",
      journal = {\prl},
         year = 2013,
        month = dec,
       volume = {111},
       number = {23},
          eid = {231301},
        pages = {231301},
          doi = {10.1103/PhysRevLett.111.231301},
archivePrefix = {arXiv},
       eprint = {1311.1669},
 primaryClass = {astro-ph.SR},
       adsurl = {https://ui.adsabs.harvard.edu/abs/2013PhRvL.111w1301V}
}

@ARTICLE{2013A&A...558A..12V,
       author = {{Viaux}, N. and {Catelan}, M. and {Stetson}, P.~B. and {Raffelt}, G.~G. and {Redondo}, J. and {Valcarce}, A.~A.~R. and {Weiss}, A.},
        title = "{Particle-physics constraints from the globular cluster M5: neutrino dipole moments}",
      journal = {\aap},
         year = 2013,
        month = oct,
       volume = {558},
          eid = {A12},
        pages = {A12},
          doi = {10.1051/0004-6361/201322004},
archivePrefix = {arXiv},
       eprint = {1308.4627},
 primaryClass = {astro-ph.SR},
       adsurl = {https://ui.adsabs.harvard.edu/abs/2013A&A...558A..12V}
}

@ARTICLE{2015ApJ...809..141H,
       author = {{Hansen}, Brad M.~S. and {Richer}, Harvey and {Kalirai}, Jason and {Goldsbury}, Ryan and {Frewen}, Shane and {Heyl}, Jeremy},
        title = "{Constraining Neutrino Cooling Using the Hot White Dwarf Luminosity Function in the Globular Cluster 47 Tucanae}",
      journal = {\apj},
         year = 2015,
        month = aug,
       volume = {809},
       number = {2},
          eid = {141},
        pages = {141},
          doi = {10.1088/0004-637X/809/2/141},
archivePrefix = {arXiv},
       eprint = {1507.05665},
 primaryClass = {astro-ph.SR},
       adsurl = {https://ui.adsabs.harvard.edu/abs/2015ApJ...809..141H}
}

@ARTICLE{Nakagawa:1987,
       author = {{Nakagawa}, Masayuki and {Kohyama}, Yasuharu and {Itoh}, Naoki},
        title = "{Axion Bremsstrahlung in Dense Stars}",
      journal = {\apj},
         year = 1987,
        month = nov,
       volume = {322},
        pages = {291},
          doi = {10.1086/165724},
       adsurl = {https://ui.adsabs.harvard.edu/abs/1987ApJ...322..291N}
}

@ARTICLE{Nakagawa:1988,
       author = {{Nakagawa}, Masayuki and {Adachi}, Tomoo and {Kohyama}, Yasuharu and {Itoh}, Naoki},
        title = "{Axion Bremsstrahlung in Dense Stars. II. Phonon Contributions}",
      journal = {\apj},
         year = 1988,
        month = mar,
       volume = {326},
        pages = {241},
          doi = {10.1086/166085},
       adsurl = {https://ui.adsabs.harvard.edu/abs/1988ApJ...326..241N}
}

@ARTICLE{Friedland:2012,
       author = {{Friedland}, Alexander and {Giannotti}, Maurizio and {Wise}, Michael},
        title = "{Constraining the Axion-Photon Coupling with Massive Stars}",
      journal = {\prl},
         year = 2013,
        month = feb,
       volume = {110},
       number = {6},
          eid = {061101},
        pages = {061101},
          doi = {10.1103/PhysRevLett.110.061101},
archivePrefix = {arXiv},
       eprint = {1210.1271},
 primaryClass = {hep-ph},
       adsurl = {https://ui.adsabs.harvard.edu/abs/2013PhRvL.110f1101F}
}

@ARTICLE{mesa1,
  author = {{Paxton}, B. and {Bildsten}, L. and {Dotter}, A. and {Herwig}, F. and {Lesaffre}, P. and {Timmes}, F.},
  title = {{Modules for Experiments in Stellar Astrophysics (MESA)}},
  journal = {\apjs},
  archivePrefix = {arXiv},
  eprint = {1009.1622},
  primaryClass = {astro-ph.SR},
  year = {2011},
  month = {jan},
  volume = {192},
  eid = {3},
  pages = {3},
  doi = {10.1088/0067-0049/192/1/3},
  adsurl = {https://ui.adsabs.harvard.edu/abs/2011ApJS..192....3P},
}

@ARTICLE{mesa2,
  author = {{Paxton}, B. and {Cantiello}, M. and {Arras}, P. and {Bildsten}, L. and {Brown}, E.~F. and {Dotter}, A. and {Mankovich}, C. and {Montgomery}, M.~H. and {Stello}, D. and {Timmes}, F.~X. and {Townsend}, R.},
  title = {{Modules for Experiments in Stellar Astrophysics (MESA): Planets, Oscillations, Rotation, and Massive Stars}},
  journal = {\apjs},
  archivePrefix = {arXiv},
  eprint = {1301.0319},
  primaryClass = {astro-ph.SR},
  year = {2013},
  month = {sep},
  volume = {208},
  eid = {4},
  pages = {4},
  doi = {10.1088/0067-0049/208/1/4},
  adsurl = {https://ui.adsabs.harvard.edu/abs/2013ApJS..208....4P},
}

@ARTICLE{mesa3,
  author = {{Paxton}, B. and {Marchant}, P. and {Schwab}, J. and {Bauer}, E.~B. and {Bildsten}, L. and {Cantiello}, M. and {Dessart}, L. and {Farmer}, R. and {Hu}, H. and {Langer}, N. and {Townsend}, R.~H.~D. and {Townsley}, D.~M. and {Timmes}, F.~X.},
  title = {{Modules for Experiments in Stellar Astrophysics (MESA): Binaries, Pulsations, and Explosions}},
  journal = {\apjs},
  archivePrefix = {arXiv},
  eprint = {1506.03146},
  primaryClass = {astro-ph.SR},
  year = {2015},
  month = {sep},
  volume = {220},
  eid = {15},
  pages = {15},
  doi = {10.1088/0067-0049/220/1/15},
  adsurl = {https://ui.adsabs.harvard.edu/abs/2015ApJS..220...15P},
}

@ARTICLE{mesa4,
  author = {{Paxton}, B. and {Schwab}, J. and {Bauer}, E.~B. and {Bildsten}, L. and {Blinnikov}, S. and {Duffell}, P. and {Farmer}, R. and {Goldberg}, J.~A. and {Marchant}, P. and {Sorokina}, E. and {Thoul}, A. and {Townsend}, R.~H.~D. and {Timmes}, F.~X.},
  title = {{Modules for Experiments in Stellar Astrophysics (MESA): Convective Boundaries, Element Diffusion, and Massive Star Explosions}},
  journal = {\apjs},
  archivePrefix = {arXiv},
  eprint = {1710.08424},
  primaryClass = {astro-ph.SR},
  year = {2018},
  month = {feb},
  volume = {234},
  eid = {34},
  pages = {34},
  doi = {10.3847/1538-4365/aaa5a8},
  adsurl = {https://ui.adsabs.harvard.edu/abs/2018ApJS..234...34P},
}

@ARTICLE{mesa5,
       author = {{Paxton}, Bill and {Smolec}, R. and {Schwab}, Josiah and {Gautschy}, A. and
         {Bildsten}, Lars and {Cantiello}, Matteo and {Dotter}, Aaron and
         {Farmer}, R. and {Goldberg}, Jared A. and {Jermyn}, Adam S. and
         {Kanbur}, S.~M. and {Marchant}, Pablo and {Thoul}, Anne and
         {Townsend}, Richard H.~D. and {Wolf}, William M. and {Zhang}, Michael and
         {Timmes}, F.~X.},
        title = "{Modules for Experiments in Stellar Astrophysics (MESA): Pulsating Variable Stars, Rotation, Convective Boundaries, and Energy Conservation}",
      journal = {\apjs},
         year = "2019",
        month = "Jul",
       volume = {243},
       number = {1},
          eid = {10},
        pages = {10},
          doi = {10.3847/1538-4365/ab2241},
archivePrefix = {arXiv},
       eprint = {1903.01426},
 primaryClass = {astro-ph.SR},
       adsurl = {https://ui.adsabs.harvard.edu/abs/2019ApJS..243...10P}
}

@ARTICLE{1963MNRAS.126..499M,
       author = {{Michie}, R.~W.},
        title = "{The dynamics of spherical stellar systems, IV}",
      journal = {\mnras},
         year = 1963,
        month = jan,
       volume = {126},
        pages = {499},
          doi = {10.1093/mnras/126.6.499},
       adsurl = {https://ui.adsabs.harvard.edu/abs/1963MNRAS.126..499M}
}

@ARTICLE{1966AJ.....71...64K,
       author = {{King}, Ivan R.},
        title = "{The structure of star clusters. III. Some simple dynamical models}",
      journal = {\aj},
         year = 1966,
        month = feb,
       volume = {71},
        pages = {64},
          doi = {10.1086/109857},
       adsurl = {https://ui.adsabs.harvard.edu/abs/1966AJ.....71...64K}
}

@ARTICLE{Goldsbury2013,
       author = {{Goldsbury}, Ryan and {Heyl}, Jeremy and {Richer}, Harvey},
        title = "{Quantifying Mass Segregation and New Core Radii for 54 Milky Way Globular Clusters}",
      journal = {\apj},
         year = 2013,
        month = nov,
       volume = {778},
       number = {1},
          eid = {57},
        pages = {57},
          doi = {10.1088/0004-637X/778/1/57},
archivePrefix = {arXiv},
       eprint = {1308.3706},
 primaryClass = {astro-ph.GA},
       adsurl = {https://ui.adsabs.harvard.edu/abs/2013ApJ...778...57G}
}

@ARTICLE{Chen2018,
       author = {{Chen}, Seery and {Richer}, Harvey and {Caiazzo}, Ilaria and {Heyl}, Jeremy},
        title = "{Distances to the Globular Clusters 47 Tucanae and NGC 362 Using Gaia DR2 Parallaxes}",
      journal = {\apj},
         year = 2018,
        month = nov,
       volume = {867},
       number = {2},
          eid = {132},
        pages = {132},
          doi = {10.3847/1538-4357/aae089},
archivePrefix = {arXiv},
       eprint = {1807.07089},
 primaryClass = {astro-ph.SR},
       adsurl = {https://ui.adsabs.harvard.edu/abs/2018ApJ...867..132C}
}

@ARTICLE{1996AJ....112.1487H,
       author = {{Harris}, William E.},
        title = "{A Catalog of Parameters for Globular Clusters in the Milky Way}",
      journal = {\aj},
         year = 1996,
        month = oct,
       volume = {112},
        pages = {1487},
          doi = {10.1086/118116},
       adsurl = {https://ui.adsabs.harvard.edu/abs/1996AJ....112.1487H}
}

@ARTICLE{1989ApJ...345..245C,
       author = {{Cardelli}, Jason A. and {Clayton}, Geoffrey C. and {Mathis}, John S.},
        title = "{The Relationship between Infrared, Optical, and Ultraviolet Extinction}",
      journal = {\apj},
         year = 1989,
        month = oct,
       volume = {345},
        pages = {245},
          doi = {10.1086/167900},
       adsurl = {https://ui.adsabs.harvard.edu/abs/1989ApJ...345..245C}
}

@ARTICLE{1994ApJ...422..158O,
       author = {{O'Donnell}, James E.},
        title = "{$R_\nu$-dependent Optical and Near-Ultraviolet Extinction}",
      journal = {\apj},
         year = 1994,
        month = feb,
       volume = {422},
        pages = {158},
          doi = {10.1086/173713},
       adsurl = {https://ui.adsabs.harvard.edu/abs/1994ApJ...422..158O}
}

@ARTICLE{2015ApJ...804...53H,
       author = {{Heyl}, Jeremy and {Richer}, Harvey B. and {Antolini}, Elisa and {Goldsbury}, Ryan and {Kalirai}, Jason and {Parada}, Javiera and {Tremblay}, Pier-Emmanuel},
        title = "{A Measurement of Diffusion in 47 Tucanae}",
      journal = {\apj},
         year = 2015,
        month = may,
       volume = {804},
       number = {1},
          eid = {53},
        pages = {53},
          doi = {10.1088/0004-637X/804/1/53},
archivePrefix = {arXiv},
       eprint = {1502.01890},
 primaryClass = {astro-ph.SR},
       adsurl = {https://ui.adsabs.harvard.edu/abs/2015ApJ...804...53H}
}

@ARTICLE{Heyl2017,
       author = {{Heyl}, J. and {Caiazzo}, I. and {Richer}, H. and {Anderson}, J. and {Kalirai}, J. and {Parada}, J.},
        title = "{Deep HST Imaging in 47 Tucanae: A Global Dynamical Model}",
      journal = {\apj},
         year = 2017,
        month = dec,
       volume = {850},
       number = {2},
          eid = {186},
        pages = {186},
          doi = {10.3847/1538-4357/aa974f},
archivePrefix = {arXiv},
       eprint = {1710.10666},
 primaryClass = {astro-ph.GA},
       adsurl = {https://ui.adsabs.harvard.edu/abs/2017ApJ...850..186H}
}

@MANUAL{WFC3DataHandbook,
author = {{K. C. Sahu et al.}},
title = "{WFC3 Data Handbook}",
version = {5.0},
year = 2021,
organization = {STScI},
address = {Baltimore},
note = {Documentation available at \url{https://hst-docs.stsci.edu/wfc3dhb}}
}

@MANUAL{WFC3InstrumentHandbook,
author = {{Marinelli}, M. and {Dressel}, L.},
title = "{WFC3 Instrument Handbook}",
version = {16.0},
year = 2024,
organization = {STScI},
address = {Baltimore},
note = {Documentation available at \url{https://hst-docs.stsci.edu/wfc3ihb}}
}

@ARTICLE{2021MNRAS.505.5957B,
       author = {{Baumgardt}, H. and {Vasiliev}, E.},
        title = "{Accurate distances to Galactic globular clusters through a combination of Gaia EDR3, HST, and literature data}",
      journal = {\mnras},
         year = 2021,
        month = aug,
       volume = {505},
       number = {4},
        pages = {5957-5977},
          doi = {10.1093/mnras/stab1474},
archivePrefix = {arXiv},
       eprint = {2105.09526},
 primaryClass = {astro-ph.GA},
       adsurl = {https://ui.adsabs.harvard.edu/abs/2021MNRAS.505.5957B}
}

@ARTICLE{2012AJ....143...11K,
       author = {{Kalirai}, Jason S. and {Richer}, Harvey B. and {Anderson}, Jay and {Dotter}, Aaron and {Fahlman}, Gregory G. and {Hansen}, Brad M.~S. and {Hurley}, Jarrod and {King}, Ivan R. and {Reitzel}, David and {Rich}, R.~M. and {Shara}, Michael M. and {Stetson}, Peter B. and {Woodley}, Kristin A.},
        title = "{A Deep, Wide-field, and Panchromatic View of 47 Tuc and the SMC with HST: Observations and Data Analysis Methods}",
      journal = {\aj},
         year = 2012,
        month = jan,
       volume = {143},
       number = {1},
          eid = {11},
        pages = {11},
          doi = {10.1088/0004-6256/143/1/11},
archivePrefix = {arXiv},
       eprint = {1112.1426},
 primaryClass = {astro-ph.SR},
       adsurl = {https://ui.adsabs.harvard.edu/abs/2012AJ....143...11K}
}

@ARTICLE{2025MNRAS.541.1390S,
       author = {{Salaris}, M. and {Scalco}, M. and {Bedin}, L. and {Cassisi}, S.},
        title = "{James Webb Space Telescope observations of the white dwarf cooling sequence of 47 Tucan{\ae}}",
      journal = {\mnras},
         year = 2025,
        month = aug,
       volume = {541},
       number = {2},
        pages = {1390-1402},
          doi = {10.1093/mnras/staf1039},
archivePrefix = {arXiv},
       eprint = {2506.20732},
 primaryClass = {astro-ph.SR},
       adsurl = {https://ui.adsabs.harvard.edu/abs/2025MNRAS.541.1390S}
}

@ARTICLE{2009ApJ...692.1013S,
       author = {{Salaris}, Maurizio and {Serenelli}, Aldo and {Weiss}, Achim and {Miller Bertolami}, Marcelo},
        title = "{Semi-empirical White Dwarf Initial-Final Mass Relationships: A Thorough Analysis of Systematic Uncertainties Due to Stellar Evolution Models}",
      journal = {\apj},
         year = 2009,
        month = feb,
       volume = {692},
       number = {2},
        pages = {1013-1032},
          doi = {10.1088/0004-637X/692/2/1013},
archivePrefix = {arXiv},
       eprint = {0807.3567},
 primaryClass = {astro-ph},
       adsurl = {https://ui.adsabs.harvard.edu/abs/2009ApJ...692.1013S}
}

@ARTICLE{2026ApJS..283...41B,
       author = {{Bauer}, Evan B. and {Dotter}, Aaron and {Conroy}, Charlie and {Cunningham}, Tim and {Park}, Minjung and {Tremblay}, Pier-Emmanuel},
        title = "{MESA Isochrones and Stellar Tracks (MIST). III. The White Dwarf Cooling Sequence}",
      journal = {\apjs},
         year = 2026,
        month = mar,
       volume = {283},
       number = {1},
          eid = {41},
        pages = {41},
          doi = {10.3847/1538-4365/ae401e},
archivePrefix = {arXiv},
       eprint = {2509.21717},
 primaryClass = {astro-ph.SR},
       adsurl = {https://ui.adsabs.harvard.edu/abs/2026ApJS..283...41B}
}

@ARTICLE{2013A&A...555A..96S,
       author = {{Salaris}, M. and {Althaus}, L.~G. and {Garc{\'\i}a-Berro}, E.},
        title = "{Comparison of theoretical white dwarf cooling timescales}",
      journal = {\aap},
         year = 2013,
        month = jul,
       volume = {555},
          eid = {A96},
        pages = {A96},
          doi = {10.1051/0004-6361/201220622},
archivePrefix = {arXiv},
       eprint = {1306.2575},
 primaryClass = {astro-ph.SR},
       adsurl = {https://ui.adsabs.harvard.edu/abs/2013A&A...555A..96S}
}

@ARTICLE{2021A&A...654A.149C,
       author = {{Cassisi}, Santi and {Potekhin}, Alexander Y. and {Salaris}, Maurizio and {Pietrinferni}, Adriano},
        title = "{Electron conduction opacities at the transition between moderate and strong degeneracy: Uncertainties and impacts on stellar models}",
      journal = {\aap},
         year = 2021,
        month = oct,
       volume = {654},
          eid = {A149},
        pages = {A149},
          doi = {10.1051/0004-6361/202141425},
archivePrefix = {arXiv},
       eprint = {2108.11653},
 primaryClass = {astro-ph.SR},
       adsurl = {https://ui.adsabs.harvard.edu/abs/2021A&A...654A.149C}
}

@ARTICLE{2020ApJ...899...46B,
       author = {{Blouin}, Simon and {Shaffer}, Nathaniel R. and {Saumon}, Didier and {Starrett}, Charles E.},
        title = "{New Conductive Opacities for White Dwarf Envelopes}",
      journal = {\apj},
         year = 2020,
        month = aug,
       volume = {899},
       number = {1},
          eid = {46},
        pages = {46},
          doi = {10.3847/1538-4357/ab9e75},
archivePrefix = {arXiv},
       eprint = {2006.16390},
 primaryClass = {astro-ph.SR},
       adsurl = {https://ui.adsabs.harvard.edu/abs/2020ApJ...899...46B}
}

@ARTICLE{2016ApJ...826...88P,
       author = {{Parada}, Javiera and {Richer}, Harvey and {Heyl}, Jeremy and {Kalirai}, Jason and {Goldsbury}, Ryan},
        title = "{Dynamical Estimate of Post-main-sequence Stellar Masses in 47 Tucanae}",
      journal = {\apj},
         year = 2016,
        month = jul,
       volume = {826},
       number = {1},
          eid = {88},
        pages = {88},
          doi = {10.3847/0004-637X/826/1/88},
archivePrefix = {arXiv},
       eprint = {1605.05740},
 primaryClass = {astro-ph.SR},
       adsurl = {https://ui.adsabs.harvard.edu/abs/2016ApJ...826...88P}
}

@ARTICLE{2016ApJ...830..139P,
       author = {{Parada}, Javiera and {Richer}, Harvey and {Heyl}, Jeremy and {Kalirai}, Jason and {Goldsbury}, Ryan},
        title = "{Formation and Evolution of Blue Stragglers in 47 Tucanae}",
      journal = {\apj},
         year = 2016,
        month = oct,
       volume = {830},
       number = {2},
          eid = {139},
        pages = {139},
          doi = {10.3847/0004-637X/830/2/139},
archivePrefix = {arXiv},
       eprint = {1609.02115},
 primaryClass = {astro-ph.SR},
       adsurl = {https://ui.adsabs.harvard.edu/abs/2016ApJ...830..139P}
}

@ARTICLE{2015ApJ...810..127H,
       author = {{Heyl}, Jeremy and {Kalirai}, Jason and {Richer}, Harvey B. and {Marigo}, Paola and {Antolini}, Elisa and {Goldsbury}, Ryan and {Parada}, Javiera},
        title = "{When do stars in 47 Tucanae lose their mass?}",
      journal = {\apj},
         year = 2015,
        month = sep,
       volume = {810},
       number = {2},
          eid = {127},
        pages = {127},
          doi = {10.1088/0004-637X/810/2/127},
archivePrefix = {arXiv},
       eprint = {1502.07306},
 primaryClass = {astro-ph.SR},
       adsurl = {https://ui.adsabs.harvard.edu/abs/2015ApJ...810..127H}
}

\end{document}